\documentclass[11pt]{article}
\pdfoutput=1
\usepackage[centertags]{amsmath}
\usepackage[square, comma, sort&compress,numbers]{natbib}
\usepackage{array,multirow}
\numberwithin{equation}{section}
\usepackage{amssymb,amsfonts}
\usepackage{graphicx}
\usepackage{color}
\usepackage{mathtools,bm}
\usepackage{accents}

\usepackage{epsfig}
\usepackage{bbold}
\newcommand{\pref}{C}
\usepackage{wrapfig}
\usepackage{float}
\usepackage{soul}
\usepackage{tkz-euclide}
\usepackage{braket}
\usepackage{tikz,pgf}
\usetikzlibrary{shapes}
\usetikzlibrary{calc}
\usetikzlibrary{decorations.pathmorphing}
\usetikzlibrary{decorations.pathreplacing,shapes.misc}
\usetikzlibrary{positioning}
\usetikzlibrary{arrows}
\usetikzlibrary{decorations.markings}
\usetikzlibrary{shadings}

\usetikzlibrary{intersections}

\newcommand{\be}{\begin{equation}}
\newcommand{\ee}{\end{equation}}
\newcommand{\ba}{\begin{array}}
\newcommand{\ea}{\end{array}}

\newcommand{\dps}{\displaystyle}
\newcommand{\half}{\frac{1}{2}}

\newcommand{\F}{\;_2F_1}

\newtheorem{prop}{Proposition}
\newtheorem{lemma}[prop]{Lemma}

\newcommand{\bref}[1]{\textbf{\ref{#1}}}

\newcommand{\RR}{\mathbb{R}}

\newcommand{\cD}{\mathcal{D}}

\newcommand{\cF}{\mathcal{F}}

\newcommand{\cJ}{\mathcal{J}}

\newcommand{\cL}{\mathcal{L}}

\newcommand{\cO}{\mathcal{O}}

\newcommand{\cR}{\mathcal{R}}

\newcommand{\cV}{\mathcal{V}}
\newcommand{\cW}{\mathcal{W}}

\numberwithin{equation}{section} \makeatletter
\@addtoreset{equation}{section}

\newcommand{\ads}{AdS$_2\;$}

\newcommand{\cft}{CFT$_1$ }

\newcommand{\pt}{w}
\newcommand{\h}{h}
\newcommand{\cross}{\eta}

\newcommand{\sltwo}{sl(2,\mathbb{R})}

\newcommand{\sJ}{J}

\usepackage{jheppub}
\makeatletter
\def\@fpheader{\vspace{-.1cm}}
\makeatother

\title{\centering{Holographic reconstruction for AdS Wilson line networks and  scalar Witten diagrams }}

\author[a,b]{Konstantin\ Alkalaev}  
\author[a]{and Vladimir\ Khiteev}

\affiliation[a]{I.E. Tamm Department of Theoretical Physics, \\P.N. Lebedev Physical
Institute, 119991 Moscow, Russia}
\affiliation[b]{Institute for Theoretical and Mathematical Physics,\\
Lomonosov Moscow State University,
119991 Moscow, Russia}

\emailAdd{alkalaev@lpi.ru}
\emailAdd{khiteev@lpi.ru}

\abstract{We find a holographic reconstruction formula  for gravitational Wilson line network operators in AdS$_2$  evaluated between Ishibashi states of the algebra $\sltwo$. It is given in integral form where the integrand is the global conformal block multiplied by a smearing function which is the product of the scalar bulk-to-boundary  propagators. The integral can be explicitly calculated  as multidimensional series of which arguments are rational functions of endpoint coordinates. In the case of two and three endpoints the resulting expressions   allow one to establish a number of relations between the gravitational Wilson line networks and Witten diagrams for massive scalar fields in AdS$_2$.}

\begin{document}

\maketitle
\flushbottom

\section{Introduction}

The Wilson line  networks in low-dimensional AdS  gravities formulated either as $3d$ Chern-Simons \cite{Achucarro:1987vz,Witten:1988hc}  or $2d$ BF model \cite{Fukuyama:1985gg}  are simple topological objects which are built as invariant compositions  of   Wilson lines carrying different representations  of the space-time isometry algebra. Thus, a general $n$-point gravitational Wilson line  network has $n-2$ cubic  vertices and $n$ endpoints in the bulk and/or on the conformal boundary. The key property here is that the gravitational connection, which  satisfies the zero-curvature condition being the equation of motion, is flat. This implies  that the respective Wilson line is independent of a path connecting two given points thereby demonstrating a topological behaviour.\footnote{The Wilson line networks are  labelled graphs and have much in common with the Penrose spin networks if viewed in the context of the gauge theory of gravitation \cite{Penrose}. On the other hand, the same construction of labelled graphs is used to build   conformal blocks in a CFT  \cite{Moore:1988qv}.} Such gravitational  networks have various properties assigned to the AdS/CFT correspondence \cite{Ammon:2013hba,deBoer:2013vca,deBoer:2014sna,Hegde:2015dqh,Bhatta:2016hpz,Besken:2016ooo,Besken:2017fsj,Hikida:2017ehf,Anand:2017dav,Hikida:2018eih,Hikida:2018dxe,Besken:2018zro,Bhatta:2018gjb,DHoker:2019clx,Castro:2018srf,Kraus:2018zrn,Blommaert:2018oro,Hulik:2018dpl,Hung:2018mcn,Castro:2020smu,Alkalaev:2020yvq,Belavin:2022bib,Belavin:2023orw,Castro:2023bvo,Alkalaev:2023axo}. E.g. their near-the-boundary asymptotics reproduce conformal blocks of a boundary CFT. In the  bulk they are interesting as probes of the geometry though this usually applies to single Wilson lines or loops, not Wilson line networks.

Following our previous work \cite{Alkalaev:2023axo}, in this paper we continue to study the Wilson line networks in two-dimensional AdS space and focus on them as functions in the bulk. The $n$-point Wilson line network in this case is built as an $n$-valent $\sltwo$ intertwiner of external weights $\h_1, ..., \h_n$ and intermediate weights $\tilde \h_1, ..., \tilde \h_{n-3}$ (see Fig. \bref{fig:comb}) contracted with $n$ Wilson line operators of weights $h_1,..., h_n$ between the point $0$ and points $x_1, ... , x_n$ in the bulk. Let $\cV_{\h_1...\h_n \tilde \h_1 ... \tilde \h_{n-3}}(x_1,...,x_n)$  be a particular matrix element of the Wilson line network associated to infinite-dimensional $\sltwo$ modules and evaluated  between $n$ cap states which are the Ishibashi states in given  $\sltwo$ modules of weights $h_1, ... , h_n$. We call these matrix elements the $n$-point AdS vertex functions \cite{Alkalaev:2023axo}.  The main technical problem   is that such matrix contractions which involve $3n-3$ infinite summations  coming from each $3j$ symbol of the $n$-valent intertwiner and the Ishibashi states are extremely complicated functions of the bulk points $x_1, ..., x_n \in$ AdS$_2$. Thus, our goal here is twofold: first, we pursue to find a convenient integral representation of $\cV_{\h_1...\h_n\tilde \h_1 ... \tilde \h_{n-3}}(x_1,...,x_n)$ that considerably simplifies the whole analysis; second, we aim to   calculate  lower-point  AdS vertex functions explicitly and study their properties.

We found that the AdS vertex functions can have the HKLL-type integral representation \cite{Hamilton:2005ju} which reconstructs the Wilson line networks from the global conformal blocks\footnote{By global conformal blocks we mean that the Virasoro conformal block is calculated in the regime when all conformal dimensions (external and intermediate) are fixed while the central charge goes to infinity. In this case the Virasoro algebra is truncated to the projective subalgebra $\sltwo$. The $n$-point global conformal blocks in CFT$_2$ (here we consider just a chiral copy) were extensively studied in Refs. \cite{Alkalaev:2015fbw,Rosenhaus:2018zqn,Fortin:2020zxw,Fortin:2023xqq}. The issue of $1/c$ corrections  to global blocks within the Wilson line approach was considered in \cite{Fitzpatrick:2016mtp,Besken:2017fsj,Hikida:2017ehf,Hikida:2018eih,Hikida:2018dxe}.} on the boundary by integrating them with a smearing function given by the bulk-to-boundary propagators  over the Pochhammer contours (such a contour is a curve on the doubly punctured plane which  cannot be shrunk to a single point, see Fig. \bref{fig:poch}). In this form the $n$-point AdS vertex functions can be calculated explicitly as multidimensional series. In particular cases such series can be represented in terms of known special functions. On the other hand, having such a holographic reconstruction formula one immediately sees that the extrapolate dictionary relation reproduces the $n$-point global conformal blocks as the asymptotic values of the $n$-point AdS vertex functions. It should be noted that both the Wilson line networks and the global conformal blocks are considered in the comb channel.

At $n=2,3$ there is an  interesting connection with  massive scalar dynamics in AdS$_2$. Namely, in the 2-point case one can show that   \cite{Castro:2018srf,Alkalaev:2023axo}
\be
\cV_{\h_1\h_2}(x_1,x_2) \sim \delta_{h_1h_2}G(x_1,x_2|h_1)\;, 
\ee
where the right-hand side is the bulk-to-bulk propagator of a free scalar field of mass $m^2 =h_1(h_1-1)$ in AdS$_2$. As shown by Castro, Iqbal, and Llabrés  \cite{Castro:2018srf} such a relation is due to that the Wilson line operator acting on the Ishibashi state in a given $\sltwo$ module  can be interpreted as a wave function realizing one-particle states in \ads massive scalar theory\footnote{These authors considered the Chern-Simons gravity in three dimensions and the respective  algebra is $\sltwo \oplus \sltwo$. Their results also apply to a chiral copy $\sltwo$ and two-dimensional Wilson line operators in AdS$_2$.} (see also related discussions in \cite{Nakayama:2015mva,Nakayama:2016xvw,Bhatta:2018gjb,Alkalaev:2023axo}). This important observation begs the question: can  the $3$-point AdS vertex function also be   related to some $3$-point diagrams in a scalar field theory? Indeed, in this paper we show that the 3-point scalar Witten diagram decomposes into the 3-point AdS vertex functions of running weights:\footnote{In the 2-point and 3-point cases there are no intermediate weights and the respective intertwiners are reduced to an identity operator and 3j symbol, respectively.} 
\be 
\label{summary}
\ba{c}
\dps
\int_{\text{AdS}_2}d^2x\, G(x,x_1|\h_1) G(x,x_2|\h_2)G(x,x_3|\h_3) 
\sim \cV_{\h_1\h_2\h_3}(x_1,x_2,x_3)
\vspace{3mm}
\\
\dps
+\sum_{n=0}^{\infty} \alpha_{\h_1 \h_2 \h_3}\,\cV_{\h_2+\h_3+2n\  \h_2\h_3}(x_1,x_2,x_3)
\vspace{3mm}
\\
\dps
+\sum_{n=0}^{\infty} \beta_{\h_1 \h_2 \h_3}\,\cV_{\h_1\  \h_1+\h_3+2n\ \h_3}(x_1,x_2,x_3)
\vspace{3mm}
\\
\dps
+\sum_{n=0}^{\infty}\gamma_{\h_1 \h_2 \h_3}\,\cV_{\h_1 \h_2\ \h_1+\h_2+2n}(x_1,x_2,x_3)\,,
\ea
\ee 
where $\alpha, \beta, \gamma$ are some coefficients dependent on conformal dimensions $\h_i$. The essential ingredient here is an exact expression for the 3-point Witten diagram calculated by Jepsen and Parikh in \cite{Jepsen:2019svc}\footnote{The 3-point Witten diagram with one bulk point was previously    calculated by Zhou et al in \cite{Zhou:2018sfz,Giombi:2020xah} using a method   introduced by D'Hoker et al in \cite{DHoker:1999mqo}.} which allowed us to identify the right-hand side as structured linear combinations of the 3-point AdS vertex functions.

The general formula \eqref{summary}  can be drastically simplified if some of points are on the conformal boundary. E.g. we show that in the case of two boundary points $x_2, x_3 \to z_2, z_3$ the respective 3-point AdS vertex function calculates the {\it geodesic} Witten diagram \cite{Hijano:2015zsa}:
\be
\label{summary_G}
\int_{\gamma_{23}}d\lambda \,G(x(\lambda),x_1|\h_1)\, K(x(\lambda),z_2|\h_2) \, K(x(\lambda),z_3|\h_3) 
\sim \cV_{\h_1\h_2\h_3}(x_1,z_2,z_3)\,,
\ee 
where $\gamma_{23}$ is a geodesic $x = x(\lambda)$ connecting two boundary points $z_2$ and $z_3$, and $K(x,z|\h)$ is the bulk-to-boundary propagator. Fewer points on the boundary extend the integration domain from the geodesic to the whole \ads and, as a consequence, this  revives  infinite summation tails over the AdS vertex functions with running weights present in  \eqref{summary}.

The paper is organized as follows. In section \bref{sec:wilson} we review  the Wilson line networks and  the AdS vertex functions built as the corresponding matrix elements of intertwined Wilson lines averaged between the cap states. We reconsider  the $\sltwo$ spacetime invariance conditions imposed on the AdS vertex functions and discuss their solutions. In section \bref{sec:higher} we formulate the HKLL-type  representation for the Wilson line networks. In particular, in section \bref{sec:exact} we  find exact expression for the $n$-point AdS vertex function as multidimensional series. Section \bref{sec:near} studies the conformal boundary asymptotics.   In section \bref{sec:lower} we consider the 2-point and 3-point AdS vertex functions in more detail and find for them a few analytic expressions. In section \bref{sec:points} we consider 3-point Witten diagrams and, in particular,  explicitly derive  expressions \eqref{summary} and \eqref{summary_G} described  above. In the concluding section \bref{sec:conclusion} we summarize our results and discuss  some further directions. Appendix  \bref{app:int} contains various  special functions and their identities used throughout the paper. Appendix \bref{app:n_point} describes  explicit calculation of the $n$-point AdS vertex function. Appendix \bref{app:bb} shortly reviews the \ads propagators. Appendices \bref{app:details} contain  calculation details of obtaining relations between Witten diagrams and AdS vertex functions. In Appendix \bref{app:special} we collect various expressions of  the AdS vertex functions which we managed to express in terms of known special functions.               

\section{Gravitational AdS Wilson line networks}
\label{sec:wilson}

In this section we  review the gravitational Wilson line networks in two-dimensional Euclidean AdS spacetime in the comb channel and the corresponding  AdS vertex functions \cite{Alkalaev:2023axo}. Let us briefly recap the main constructions. 

We consider $\sltwo$ connections $A$ on two-dimensional manifold which are subject to the zero curvature condition $dA + A\wedge A = 0$. The last relation is one of the equations of motion in the BF formulation \cite{Fukuyama:1985gg,Chamseddine:1989wn,Isler:1989hq} of the Jackiw-Teitelboim gravity \cite{Teitelboim:1983ux,Jackiw:1984je}. Introducing the Wilson line operators associated  to flat connections in given $\sltwo$ modules and using $\sltwo$ intertwiners one  can construct  Wilson line networks with arbitrary number of endpoints. By averaging them over particular cap states one obtains the AdS vertex functions, the main object of our interest.    
    
\begin{enumerate}
\item We fix a particular solution of the zero-curvature condition which describes \ads space\-time. The metric  in Poincare coordinates is given by  
\be 
\label{metric}
ds^2 = e^{2\rho}dz^2+d\rho^2\,,
\qquad
\rho, z \in \mathbb{R}\,.
\ee
\item We are interested in two types of infinite-dimensional $\sltwo$ modules $\cR_\h$ to be carried by Wilson lines. Below they are  described in the ladder basis. The $\sltwo$ basis elements  $J_0, J_{\pm 1}$ are subject to the commutation relations $[J_k,J_l] = (k-l)J_{k+l}$, $k,l = 0, \pm1$. 

{\it Negative(positive) discrete series}  $\cR_\h = \cD_\h^{\mp}$ with weights  $\h\in\mathbb{R}$, $\dim \cD^{\mp}_{\h} = \infty$. The basis is given by 
\be
\label{-+D_basis}
\ba{l}
\dps
\{\cD_\h^- \ni  \ket{\h,m}:\, J_0\ket{\h,m} = m \ket{\h,m},\, m = -\h, -\h-1, -\h-2, ..., -\infty\}\,,
\vspace{3mm}
\\
\dps
\{\cD_\h^+ \ni  \ket{\h,m}:\, J_0\ket{\h,m} = m \ket{\h,m},\, m = \h, \h+1, \h+2, ..., \infty\}\,,
\ea
\ee
where the highest weight (HW) vector $\ket{\h,-\h}$ and the lowest weight (LW) vector $\ket{\h,\h}$ are defined by
\be
\label{hw1}
\ba{l}
\text{HW}:\qquad J_0\ket{\h,-\h} = -\h \ket{\h,-\h}\;,
\qquad
J_{-1} \ket{\h,-\h} = 0\;,
\vspace{3mm}
\\
\dps
\text{LW}:\qquad J_0\ket{\h,\h} = \h \ket{\h,\h}\;,
\qquad
J_{1}\ket{\h,\h} = 0\;;
\ea
\ee 
a magnetic number $m$ is generally non-integer.

For particular weights $\h\in\mathbb{Z}_-$, modules $\cD_\h^{\pm}$ possess singular vectors. Factoring out the respective submodules yields finite-dimensional modules. Later in this section, we show that the AdS vertex function calculated in infinite-dimensional modules with weights $\h_i\in\mathbb{Z}_-$ coincide with that one calculated in finite-dimensional modules. Also, one may consider the infinite-dimensional continuous series which is much harder to deal with since  the explicit form of the intertwiners is quite  complicated \cite{HOLMAN19661} and requires a separate study. Moreover, due to the operator-state correspondence in a boundary CFT we prefer to have HW/LW representations of the conformal algebra in order to operate with bounded spectra. For these reasons, we restrict ourselves to examining simpler cases of negative/positive discrete series.

\item
The Wilson line operator $W_\h[\gamma] = \mathbb{P}e^{-\int_{\gamma}A}$, where $\gamma$ is a path from  $x_1$ to $x_2$ in \ads and the flat $\sltwo$ connection takes values  in  $\cD_\h^-$ or $\cD_\h^{+}$ is given by \cite{Banados:1994tn} 
\be 
\label{Wilson_operator}
W_\h^-[x_1,x_2]=e^{-\rho_2 J_0}e^{z_{_{12}} J_1}e^{\rho_1 J_0}\,,
\ee 
\be 
\label{Wilson_operator_positive}
\;\,W^+_\h[x_1,x_2]=e^{\rho_2 J_0}e^{z_{_{12}} J_{-1}}e^{-\rho_1 J_0}\,,
\ee 
where $x_i=(\rho_i,z_i)$. The property of a flat connection is that the respective Wilson line depends only on the endpoints of $\gamma$.    

\item Cap states in modules $\cD^{\mp}_h$ of the weight $\h\notin\mathbb{Z}_-$ are the Ishibashi state \cite{Ishibashi:1988kg},  
\be 
\label{cap_states}
\ket{a^\mp} = \sum_{n=0}^\infty \prod_{k=1}^n \frac{(-)^n}{4k \h +4k^2 -2k} \, (J_{\pm1})^{2n}\ket{\h,\mp \h}\,,
\ee 
where $\ket{\h,\mp \h}$ are H(L)W vectors in  $\cD_\h^\mp$. For  $\h\in\mathbb{Z}_-$  the cap states are linear combinations of two Ishibashi states 
\be 
\label{cap_states_int}
\ba{l}
\dps
\ket{a^\mp} = \alpha\sum_{n=0}^{-\h} \prod_{k=1}^n \frac{(-)^n}{4k \h +4k^2 -2k} \, (J_{\pm1})^{2n}\ket{\h,\mp \h} 
\vspace{2mm}
\\
\dps
\hspace{9mm}  + \beta\sum_{n=0}^\infty \prod_{k=1}^n \frac{(-)^n}{-4k \h +4k^2 +2k} \, (J_{\pm1})^{2n}\ket{\h,\pm \h\mp1}\,,
\ea
\ee 
where $\alpha,\beta\in\mathbb{R}$ are arbitrary coefficients. Factoring out singular submodules corresponds to dropping out the second term in \eqref{cap_states_int} that produces the Ishibashi states for finite-dimensional modules \cite{Alkalaev:2023axo}.   

\item An intertwiner is the 3-valent $\sltwo$ invariant operator which projects a tensor product of   two $\sltwo$ modules onto a third one, 
\be 
\label{intertwiner}
I_{\h_1 \h_2 \h_3}:\quad\mathcal{R}_{\h_2}\otimes\mathcal{R}_{\h_3}\xrightarrow{}\mathcal{R}_{\h_1}\,.
\ee
The corresponding matrix element $[I_{\h_1\h_2 \h_3}]^{m_1}{}_{m_2 m_3}$ is the Clebsch-Gordan coefficient  \eqref{intertwiner_integral}. Note that the intertwiner restricts  possible values of  weights $\h_{1,2,3}$ depending on the choice of modules $\cR_{\h_{1,2,3}}$. We consider the following cases:
\be 
\label{infinite_restrictions1}
\cD^\pm_{\h_2}\otimes\cD^\pm_{\h_3}\xrightarrow{}\cD^\pm_{\h_1}\,, \qquad \h_1\geq \h_2+\h_3\,,
\ee
\be
\label{infinite_restrictions2}
\cD^\mp_{\h_2}\otimes\cD^\pm_{\h_3}\xrightarrow{}\cD^\mp_{\h_1}\,, \qquad \h_1<\h_2+\h_3\,.
\ee
\end{enumerate}

\paragraph{Wilson line matrix elements.} Using the Wilson line operator \eqref{Wilson_operator} and the Ishibashi state \eqref{cap_states}, one can calculate the Wilson line matrix elements for $\cD^-_{\h}$ at $\h\notin\mathbb{Z}_-$ \cite{Alkalaev:2023axo}:\footnote{Similar expressions for the Wilson matrix elements hold  in the case of AdS$_3$ flat connections \cite{Bhatta:2016hpz}.}
\be 
\label{closed_left0}
\ba{l}
\dps
\braket{a^-|W^-_\h[0,x]|\h,m} =\, T_{\h m}
e^{-\rho m}\,(q+i)^{-\h}(q-i)^{m}\,{}_2F_1\left(\h, m+\h; m+1\big|\frac{q-i}{q+i}\right),
\vspace{3mm}
\\
\dps
 \braket{\h,m|W^-_\h[x,0]|a^-}= (-)^{-m} \, T_{\h m}\, e^{\rho m}\,(q+i)^{-\h-m}\,{}_2F_1\left(\h,m+\h; m+1\big|\frac{q-i}{q+i}\right),
\ea
\ee 
where 
\be
\label{q_and_Tjm}
q = -ze^{\rho}\,, \qquad T_{\h m} = (-)^{-\h} \frac{(-\h)!}{m!}\left[\frac{(m-\h)!}{(-\h-m)!(-2\h)!}\right]^\half.
\ee
For $\cD^-_\h$ at  $\h\in \mathbb{Z}_-$  the Wilson matrix elements are given by 
\be 
\label{closed_left_int}
\ba{l}
\dps
\braket{a^-|W^-_\h[0,x]|\h,m} = 
\begin{cases} 
\alpha\braket{a^-|W^-_\h[0,x]|\h,m} \;\text{eq. } \eqref{closed_left0},& m\geq\h\\
\beta\braket{a^-|W^-_{1-\h}[0,x]|1-\h,m} \;\text{eq. }\eqref{closed_left0},& m<\h
\end{cases}\;,
\vspace{3mm}
\\
\dps
 \braket{\h,m|W^-_\h[x,0]|a^-} =
\begin{cases} 
\alpha \braket{\h,m|W^-_\h[x,0]|a^-} \;\text{eq. }\eqref{closed_left0},&m\geq\h\\
\beta \braket{1-\h,m|W^-_{1-\h}[x,0]|a^-}\;\text{eq. } \eqref{closed_left0},& m<\h
\end{cases}\;,
\ea
\ee
where the cap states are given by \eqref{cap_states_int}. For the sake of simplicity, one fixes   $\alpha=1$ and $\beta=0$ which correspond to factoring out the singular submodule and, hence,  considering a finite-dimensional module \cite{Alkalaev:2023axo}. 

Finally, one shows that  the Wilson matrix elements for $\cD_\h^+$ and $\cD_\h^-$ are related as 
\be 
\label{positive_negative_rel}
\ba{l}
\dps
\braket{a^+|W_\h^+[0,x]|\h,m} = \braket{a^-|W^-_\h[0,x]|\h,-m}\,,
\vspace{3mm}
\\
\dps
\braket{\h,m|W^+_\h[x,0]|a^+} =  \braket{\h,-m|W^-_\h[x,0]|a^-}\,.
\ea
\ee

\paragraph{AdS vertex functions.} An $n$-point AdS vertex function associated with a Wilson network in the comb channel with endpoints ${\bf x} = (x_1, ..., x_n)\in\text{AdS}_2$, where all edges carry infinite-dimensional modules is defined as 
\be 
\label{n-p_def}
\ba{c}
\dps
\cV_{ \h\tilde{\h}}({\bf x})= \hspace{-5mm}\sum_{\substack{\{m_i\in \sJ_{i}^{\pm}\}_{i=1,..., n}
\\ 
\{p_i\in\tilde{J}_{i}^{\pm}\}_{i=1,..., n-3}}}[I_{\h_1\h_2\tilde{\h}_1}]^{m_1}{}_{m_2p_1}\dots[I_{\tilde{\h}_{n-4}\h_{n-2} \tilde{\h}_{n-3}}]^{p_{n-4}}{}_{m_{n-2}p_{n-3}}[I_{\tilde{\h}_{n-3}\h_{n-1} \h_n}]^{p_{n-3}}{}_{m_{n-1}m_n}
\vspace{-3mm}
\\
\hspace{20mm}\times\,\braket{a^{\pm}_1|W^{\pm}_{\h_1}[0,x_1]|\h_1,m_1}\dots  \braket{\h_n,m_n|W^{\pm}_{\h_n}[x_n,0]|a^{\pm}_n}\,,
\ea
\ee
where $\sJ_i^-= [\![-\infty,-\h_i]\!]$, $\sJ_i^+= [\![\h_i,\infty]\!]$ and $\ket{a_i^\pm}$ are the cap states associated to each external leg. Here,  the notation $[\![-n,n]\!]$ means that $k = -n,-n+1, ...\,, n-1, n$, a tilde in $\tilde{\sJ}_i$ stands for a weight $\tilde{\h}_i$. The first line in \eqref{n-p_def} is the $n$-valent intertwiner shown in Fig. \bref{fig:comb}, the second line is the product of  localized cap states obtained from the cap states by acting with the corresponding Wilson line operators. Note that the choice of particular $\sltwo$ modules  restricts the possible values of the weights $\h_i$ and $\tilde{\h}_i$ as follows from the intertwiners \eqref{infinite_restrictions1}, \eqref{infinite_restrictions2}.
\begin{figure}
\centering
\begin{tikzpicture}[scale = 0.7]
{\draw (-2,0) -- (0,0) -- (0,2) -- (0,0) -- (2,0) -- (2,2) -- (2,0);
\draw[dashed] (2,0)-- (4,0);
\draw(4,0) -- (4,2) -- (4,0)-- (6,0)--(6,2)--(6,0)--(8,0) (-2.3, 0) node {$\h_1$} (0, 2.3) node {$\h_2$} (1, -0.4) node {$\tilde{\h}_1$} (2, 2.3) node {$\h_3$} (6, 2.3) node {$\h_{n-1}$} (5, -0.4) node {$\tilde{\h}_{n-3}$} (8.3, 0) node {$\h_n$} (4, 2.3) node {$\h_{n-2}$};}
\end{tikzpicture}
\caption{An  $n$-valent $\sltwo$ intertwiner $I_{\h\tilde \h} \equiv I_{\h_1...\h_n|\tilde \h_1 ... \tilde \h_{n-3}}$ in the comb channel. External edges  correspond to modules $\cR_{\h_i}$, $i=1,...,n$, inner edges correspond  to modules $\cR_{\tilde \h_i}$, $i=1,...,n-3$. Each vertex is given by a 3-valent intertwiner.}    
\label{fig:comb}
\end{figure}

As such the formula \eqref{n-p_def} defines a number of AdS vertex functions depending on which particular $\sltwo$ modules are chosen for external and intermediate edges. Nonetheless, one can show that all possible AdS vertex functions are reduced to a single AdS vertex function which contains  $\sltwo$ modules from the  negative discrete series only. Indeed, the AdS vertex function built from the  $n$-valent intertwiner involving a number of $\cD^+$-s can be obtained by analytic continuation of the AdS vertex function built from the $n$-valent intertwiner for all modules being $\cD^-$-s. To show this, one makes the change $m_i\to-m_i$ and $p_i\to-p_i$ in \eqref{n-p_def} for every summation variable corresponding to $\cD^+$-s, i.e. for those in the summation domains $\sJ_i^+$ or $\tilde{\sJ}_i^+$:
\be 
\label{n-p_interm}
\ba{c}
\dps
\cV_{ \h\tilde{\h}}({\bf x})= \hspace{-5mm}\sum_{\substack{\{m_i\in \sJ_{i}^{-}\}_{i=1,..., n}
\\ 
\{p_i\in\tilde{J}_{i}^{-}\}_{i=1,..., n-3}}}\hspace{-5mm}[I_{\h_1\h_2\tilde{\h}_1}]^{m_1}{}_{m_2p_1}\dots[I_{\tilde{\h}_{i-2}\h_{i} \tilde{\h}_{i-1}}]^{-p_{i-2}}{}_{-m_{i}-p_{i-1}}\dots[I_{\tilde{\h}_{n-3}\h_{n-1} \h_n}]^{p_{n-3}}{}_{m_{n-1}m_n}
\vspace{3mm}
\\
\times\,\braket{a^{-}_1|W^{-}_{\h_1}[0,x_1]|\h_1,m_1}\dots\braket{\h_i,-m_i|W^{+}_{\h_i}[x_i,0]|a^{+}_i}\dots  \braket{\h_n,m_n|W^{-}_{\h_n}[x_n,0]|a^{-}_n}\,,
\ea
\ee
whereupon all $\sJ_i^+$, $\tilde{\sJ}_i^+$ go to $\sJ_i^-$, $\tilde{\sJ}_i^-$. After the change,  the Wilson matrix elements for $\cD^+_\h$  become the Wilson matrix elements for $\cD^-_\h$ due to the relation \eqref{positive_negative_rel}, which, at the same time, substitutes $\ket{a^{+}_i}$ by $\ket{a^{-}_i}$.    The intertwiner with at least one module from the  positive discrete series coincide after the change $m_i\to-m_i$ and $p_i\to-p_i$  with the analytically continued intertwiner for three modules from the  negative discrete series (up to an overall phase factor dependent on the weights) \cite{HOLMAN19661}. Applying these observations in  \eqref{n-p_interm} one obtains the analytically continued AdS vertex function, where all (external and intermediate) modules  belong to the negative discrete series. 

Note that in the case $h_i\in\mathbb{Z}_-$, $\tilde{h}_i\in\mathbb{Z}_-$ the sums in \eqref{n-p_def} become finite due to our choice of $\beta=0$ in \eqref{closed_left_int}. Despite that  the intertwiners in \eqref{n-p_def} are taken in  infinite-dimensional modules, for such summation domains they coincide with the intertwiners for finite-dimensional modules. Thus,  \eqref{n-p_def} also defines the AdS vertex function for  finite-dimensional modules.

\paragraph{The extrapolate  dictionary.} The near-the-boundary expansion of the $n$-point AdS vertex function is related to the  {\it global}  conformal block of  a boundary CFT \cite{Bhatta:2016hpz,Besken:2016ooo,Bhatta:2018gjb,Alkalaev:2020yvq,Alkalaev:2023axo}. In the $n$-point case the exact relation called the extrapolate dictionary reads \cite{Alkalaev:2023axo} 
\be
\label{extrapol}
\lim_{\rho\to\infty}e^{\rho\sum_{i=1}^n\h_i}\,\cV_{\h \tilde{\h}}(\rho,{\bf z})= \pref_{\h\tilde{\h}}\,\cF_{h \tilde{h}}({\bf z})\,,
\ee
where all points are placed on a hypersurface $\rho  = const$, the normalization coefficients $\pref_{{\h}\tilde{{\h}}} \equiv \pref_{\h_1... \h_n\tilde{\h}_1...\tilde{\h}_{n-3}}$ are given by \be
\label{prefactor}
\ba{l}
\dps
n=2:\quad \pref_{\h_1\h_2} = \frac{\delta_{\h_1\h_2}}{(-2\h_1+1)^{\half}} \,;  
\qquad 
n=3:\quad\pref_{\h_1\h_2\h_3} = \left[\frac{(-2\h_1)!(-2\h_2)!(-2\h_3)!}{\Delta(\h_1,\h_2,\h_3)}\right]^{\half};
\vspace{3mm}\\ \dps
n>3:\quad\pref_{{\h}\tilde{{\h}}} = \pref_{\h_1\h_2\tilde{\h}_1}\Big[\prod_{i=1}^{n-4}\pref_{\tilde{\h}_i\h_{i+2}\tilde{\h}_{i+1}}\Big]\pref_{\tilde{\h}_{n-3}\h_{n-1}\h_n}\,,
\ea
\ee
where $\Delta(a,b,c) = (-a-b-c+1)!(c-a-b)!(b-a-c)!(a-b-c)!$ is the modified triangle coefficient, and $\cF_{h \tilde{h}}({\bf z})$ is the conformal block function in the comb channel (see Fig. \bref{fig:comb}).

For future convenience, we  define the $n$-point regularized AdS vertex function with $k$ points on the boundary 
\be
\label{regular}
\ba{l}
\dps
\cV_{ \h\tilde{\h}}^{\text{reg}}(z_1,...,z_k,x_{k+1},...,x_{n}):=\lim_{\rho\to\infty}e^{\rho \sum_{i=1}^k \h_i}\cV_{ \h\tilde{\h}}({\bf x})\big|_{\rho_1=...=\rho_k=\rho}\,.
\ea
\ee
Indeed, the AdS vertex function is divergent or equals zero as $\rho_i\to\infty$ because of  $e^{-\rho_i \h_i}$ in the leading term of the expansion near $\rho_i=\infty$. In fact, this regularization has been used in the extrapolate dictionary relation \eqref{extrapol}.

\paragraph{Solving the Ward identities.} Following the extrapolate dictionary \eqref{extrapol} the AdS vertex functions are required to satisfy the $sl(2, \RR)$ Ward identities \cite{Alkalaev:2023axo}
\be
\label{WA_wo}
\sum_{i= 1}^{n} \cJ_{m}^{(i)}\, \cV_{ \h\tilde{\h}}(x_1,...,x_i, ...,x_n) = 0\;, 
\qquad
m = 0, \pm 1\;,
\ee
where $x_i=(\rho_i,z_i)$ are local coordinates and the Killing vector fields $\xi_m(x)$   of  metric \eqref{metric} are given by  
$\cJ_{-1} = \partial_z$,  $\cJ_{0} = z\partial_z-\partial_\rho$, 
$\cJ_{1} = z^2\partial_z-2z\partial_\rho - e^{-2\rho}\partial_z$.  
One can show that the system of the first-order PDE  \eqref{WA_wo} has $2n-3$ independent $\sltwo$ invariant variables
\be
\label{integrals_of_motion}
\ba{c}
\dps
c_{k\,k+1}=e^{r_{k\,k+1}}+e^{-r_{k\,k+1}}+v_{k\,k+1}^2-2\,,  \qquad k = 1,..., n-1\,,
\vspace{2mm}
\\
\dps
c_{m\,m+2}=e^{r_{m\,m+2}}+e^{-r_{m\,m+2}}+v_{m\,m+2}^2-2\,, \qquad m = 1,..., n-2\,,
\ea
\ee
where  
\be
v_{ij}=(z_i-z_{j})e^{\frac{\rho_i+\rho_{j}}{2}}\,,
\qquad
r_{ij}=\rho_i-\rho_{j}\,.
\ee 
Thus, the AdS vertex functions can  be parameterized as
\be 
\label{general_dependence}
\cV_{ \h\tilde{\h}}({\bf x})  = \cV_{ \h\tilde{\h}}(c_{12},...,c_{n-1\,n},c_{13},...,c_{n-2\,n})\,.
\ee
In view of the extrapolate dictionary  relation \eqref{extrapol} one finds that  on the $\rho = const$ hypersurface the invariant variables \eqref{integrals_of_motion} are simplified to become 
\be
\label{integrals_of_motion_rho_const}
\ba{l}
\dps
c_{k\,k+1}|_{\rho_k=\rho_{k+1}=\rho}=q_{k\,k+1}^2\,,  \qquad k = 1,..., n-1\,,
\vspace{2mm}
\\
\dps
c_{m\,m+2}|_{\rho_m=\rho_{m+2}=\rho}=q_{m\,m+2}^2\,, \qquad m = 1,..., n-2\,,
\ea
\ee
where using the variable $q$ from \eqref{q_and_Tjm} one  defines 
\be 
\label{q_def}
q_{i j}:=q_j-q_i = e^\rho(z_i-z_j), \qquad i,j = 1,..., n\,.
\ee 
Obviously,  $n-2$ invariant variables in  \eqref{integrals_of_motion_rho_const} depend on the other $n-1$ variables so that the AdS vertex functions on the hypersurface have only $n-1$ arguments:
\be 
\cV_{ \h\tilde{\h}}({\bf z},\rho)  = \cV_{ \h\tilde{\h}}(q_{1 2},..., q_{n-1\,n})\,.
\ee

\section{AdS vertex functions}
\label{sec:higher}

In order to calculate the $n$-point AdS vertex function explicitly one substitutes the Wilson matrix elements \eqref{cap_states} and the intertwiners \eqref{general-intertwiner} into the defining  matrix relation \eqref{n-p_def}. However, the resulting expression is extremely complicated and requires  many transformations such as Pfaff transformations, generalized Newton binomials and hypergeometric function reductions (see Appendix \bref{app:int})  aimed to simplify or reduce a part of initial  $3n-3$ summations.  The way to get around this is to represent the Wilson matrix elements \eqref{closed_left0} in integral form. This entails the HKLL-type representation of the AdS vertex functions which reverses the extrapolate dictionary relation \eqref{extrapol}. In other words, there is a holographic reconstruction formula for the Wilson line networks.    

\subsection{Wilson matrix elements in integral form}
\label{sec:matrix}

\begin{figure}
\centering
\includegraphics[scale=1]{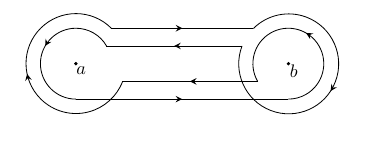}
\caption{The Pochhammer contour $P[a,b]$ in $\mathbb{C}\symbol{92} \{a,b\}$. The integration over the Pochhammer contour arises when considering the AdS vertex functions with arbitrary conformal weights. For positive real weights, these contours can be reduced to the standard line intervals connecting points $a$ and $b$. However, this cannot be done in the case of non-positve real weights as the integrals over the line contours diverge.}
\label{fig:poch}
\end{figure}
In Appendix \bref{app:wils_integ} we  derive the following integral representation \eqref{integral_rep_inf} for the Wilson  matrix elements:
\be
\label{integral_rep_inf_v2}
\ba{c}
\dps
\braket{a^-|W^-_\h[0,x]|\h,m} =  (-)^{m-\h}K_\h \oint_{P[\pt,\bar{\pt}]}du \, ((u-z)^2e^\rho+e^{-\rho})^{\h-1}\left[\frac{(-2\h)!}{(m-\h)!(-\h-m)!}\right]^\half u^{m-\h}\, ,
\vspace{8mm}
\\
\dps
\braket{\h,m|W^-_\h[x,0]|a^-} = K_\h 
\oint_{P[\pt,\bar{\pt}]}dv \,((v-z)^2e^\rho+e^{-\rho})^{\h-1}\left[\frac{(-2\h)!}{(m-\h)!(-\h-m)!}\right]^\half v^{-\h-m}\, ,
\ea
\ee 
where  $P[w, \bar w]$ is the Pochhammer contour around points $\pt$ and $\bar{\pt}$, see Fig. \bref{fig:poch},  the normalization factor and the branch points are  
\be
\label{w_var}
K_\h = \frac{(-2i)^{-2\h+1}}{\pi(1-e^{-2\pi i \h})}\frac{(-\h)!^2}{(-2\h)!} \,,
\qquad
\pt = z+ie^{-\rho}\,, \quad \bar \pt = z-ie^{-\rho}\,.
\ee
The first factor in the integrand is recognized as the scalar bulk-to-boundary propagator of weight $1-\h$ in the Poincare coordinates \eqref{metric}  (see Appendix \bref{app:bb}):
\be 
K(x,u|1-\h) = \left(\frac{e^{-\rho}}{e^{-2\rho} + (z-u)^2}\right)^{1-\h}\,.
\ee 
The last two factors are the leading terms of the asymptotic expansion of the corresponding Wilson matrix element in the points $y_1= (u, \rho)$ and $y_2= (v, \rho)$ near $\rho = \infty$: 
\be 
\label{matrix_boundary}
\ba{l}
\dps
\braket{a^-|W^-_\h[0,y_1]|\h,m} = e^{-\rho \h}\braket{a^-|W^-_\h[0,y_1]|\h,m}_{\partial}  + O(e^{-\rho(\h+1)})\,,
\vspace{3mm}
\\
\dps
\braket{\h,m|W^-_\h[y_2,0]|a^-} = e^{-\rho \h} \braket{\h,m|W^-_\h[y_2,0]|a^-}_{\partial} + O(e^{-\rho(\h+1)})\,,
\ea
\ee
where the leading matrix contributions equipped with a subscript ${\partial}$ are given by 
\be 
\ba{l}
\dps
\braket{a^-|W^-_\h[0,y_1]|\h,m}_{\partial}\, :=\, (-)^{m-\h}\left[\frac{(-2\h)!}{(m-\h)!(-\h-m)!}\right]^\half\, u^{m-\h}\,,
\vspace{3mm}
\\
\dps
\braket{\h,m|W^-_\h[y_2,0]|a^-}_{\partial} \, :=  \,\left[\frac{(-2\h)!}{(m-\h)!(-\h-m)!}\right]^\half\, v^{-\h-m}\,.
\ea
\ee
Taking into account these observations the Wilson matrix elements \eqref{integral_rep_inf} can be rewritten as
\be 
\label{matrix_alternative}
\ba{c}
\dps
\braket{a^-|W^-_\h[0,x]|\h,m}  = K_\h\oint_{P[\pt,\bar{\pt}]}du \; \, K(x,u|1-\h)\braket{a^-|W^-_\h[0,y_1]|\h,m}_{\partial}\,,
\vspace{3mm}
\\
\dps
\braket{\h,m|W^-_\h[x,0]|a^-}  = K_\h\oint_{P[\pt,\bar{\pt}]}dv \; \, K(x,v|1-\h)\braket{\h,m|W^-_\h[y_2,0]|a^-}_{\partial}\,.
\ea
\ee 

\begin{figure}
\centering
\includegraphics[scale=0.8]{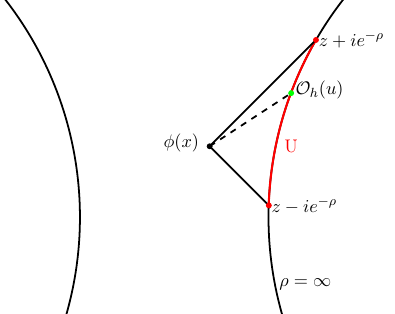}
\caption{A field $\phi(x)$  can be obtained by dragging  a boundary field $\cO_h(u)$ (the green point) into  the bulk by means of the bulk-to-boundary propagator (the dotted line). The \ads space has two conformal boundaries but the Poincare coordinates \eqref{metric} cover only one of them at $\rho = \infty$. Note that the integration domain $U$ is a complex contour which requires the complexified boundary space.}
\label{fig:HKLL_AdS2}
\end{figure}

The relations \eqref{matrix_alternative} resemble  the HKLL reconstruction \cite{Hamilton:2005ju},\footnote{The similar observation that the Wilson line operator acting on the Ishibashi state can be represented in terms of HKLL construction was made in \cite{Castro:2018srf}.} where a scalar field in  \ads with the mass $m^2 = \h(\h-1)$ can be obtained by integrating a primary operator of conformal weight $\h$ with the smearing function which is essentially the  bulk-to-boundary propagator over the part of the conformal boundary, 
\be 
\label{HKLL_rel}
\phi(x) = \int_{U} du\, K(x,u|1-\h)\cO_\h(u) \,.
\ee 
Here, the integration domain $U$ is the line contour\footnote{The integration contour $U$ in the Euclidean \ads is related to the integration contour $U_{\text{Lor}}$ in the Lorentzian \ads by  the Wick rotation $z\to iz$ and  $u \to iu$ in \eqref{HKLL_rel}. The  domain $U_{\text{Lor}} = [z-e^{-\rho}, z+e^{-\rho}]$ consists of boundary points which are spacelike separated from a bulk point $(z,\rho)$ in the Lorentzian \ads.} connecting points $z-i e^{-\rho}$ and $z+i e^{-\rho}$, see Fig. \bref{fig:HKLL_AdS2}. The appearance of the Pochhammer contour $P$ in \eqref{matrix_alternative} instead of the contour $U$ as in \eqref{HKLL_rel} is due to the divergence  of the integral \eqref{HKLL_rel} occurring in the case of arbitrary  real weights $\h$ (\eqref{HKLL_rel} is convergent only for $h>0$). A primary operator $\cO_\h(u)$ is the leading asymptotics of  a scalar field $\phi(x)$  expanded near $\rho=\infty$,  
\be 
\label{primary_rel}
\phi(\rho, u) = e^{-\h\rho} \cO_\h(u) + O(e^{-\rho(\h+1)})  \,,
\ee 
which is analogous to the relations \eqref{matrix_boundary}.

\subsection{HKLL-type representation of AdS vertex functions}
\label{sec:exact}

Using the HKLL-type  relations \eqref{matrix_alternative} one can obtain the following integral representation of the $n$-point AdS vertex function originally defined in matrix form \eqref{n-p_def}: 
\be 
\ba{l}
\dps
\hspace{5mm}\cV_{ \h\tilde{\h}}({\bf x})= \prod_{k = 1}^nK_{\h_k}\oint_{P[\pt_k,\bar{\pt}_k]}du_k\;K(x_k,u_k|1-\h_k)
\vspace{2mm}
\\
\dps
\hspace{20mm}\times\sum_{\substack{\{m_i\in \sJ_{i}^-\}_{i=1,..., n}\\\{p_i\in\tilde{J}_{i}^-\}_{i=1,..., n-3}}}
[I_{\h_1\h_2\tilde{\h}_1}]^{m_1}{}_{m_2p_1}\dots[I_{\tilde{\h}_{n-4}\h_{n-2} \tilde{\h}_{n-3}}]^{p_{n-4}}{}_{m_{n-2}p_{n-3}}
\vspace{2mm}
\\
\hspace{10mm}\times\,[I_{\tilde{\h}_{n-3}\h_{n-1} \h_n}]^{p_{n-3}}{}_{m_{n-1}m_n}\braket{a^-_1|W^-_{\h_1}[0,y_1]|\h_1,m_1}_{\partial}\dots \braket{\h_n,m_n|W^-_{\h_n}[y_n,0]|a^-_n}_{\partial}\,.
\ea
\ee
By virtue of the extrapolate dictionary \eqref{extrapol} the sum in the second and third lines which is the leading term of the asymptotic expansion of the $n$-point AdS vertex function at the points $y_i = (u_i,\rho_i)$ reduces to the $n$-point global conformal block. 
\begin{prop} 
\label{prop:HKLL}
The holographic reconstruction formula for the $n$-point AdS vertex functions is given by 
\be 
\label{HKLL_vertex}
\ba{l}
\dps
\hspace{-2mm}\cV_{ \h\tilde{\h}}({\bf x})=
\pref_{\h\tilde{\h}}  \prod_{k = 1}^nK_{\h_k} \oint_{P[\pt_k,\bar{\pt}_k]}du_k\;K(x_k,u_k|1-\h_k)\cF_{\h \tilde{\h}}({\bf u})\,,
\ea
\ee
where $\cF_{\h \tilde{\h}}({\bf u})$ is the $n$-point global conformal block in the comb channel which explicit form is given by 
\be
\label{comb_fin}
\ba{l}
\dps
\cF_{\h, \tilde{\h}}({\bf u})=\prod_{l=2}^{n-1}\Big(\frac{u_{_{l+1\,l-1}}}{u_{_{l+1\,l}}u_{_{l-1\,l}}}\Big)^{\h_l}\Big(\frac{u_{_{23}}}{u_{_{12}}u_{_{13}}}\Big)^{\h_1}\Big(\frac{u_{_{n-2\,n-1}}}{u_{_{n\,n-1}}u_{_{n\,n-2}}}\Big)^{\h_n}\prod_{k=1}^{n-3}\cross_k^{\tilde{\h}_k}
\vspace{2.5mm}  
\\
\dps
\times F_{K} \left[\begin{array}{cccc}
     \h_1+\tilde{\h}_1-\h_2,&\tilde{\h}_1+\tilde{\h}_2-\h_3,&...\,,&\h_n+\tilde{\h}_{n-3}-\h_{n-1} \\
     2\tilde{\h}_1, &2\tilde{\h}_2,&...\,, &2\tilde{\h}_{n-3} \end{array};\cross_1,...\,,\cross_{n-3}\right]\,,
\ea 
\ee
with arguments being the   cross-ratios
\be 
\label{cross-ratios}
\cross_i=\frac{u_{_{i\,i+1}}u_{_{i+2\,i+3}}}{u_{_{i\,i+2}}u_{_{i+1\,i+3}}}\,,
\qquad
u_{_{i\,j}} = u_i -u_j\,.
\ee
The function $F_{K}$  is the comb function \eqref{def_comb} which is a hypergeometric-type function  of $n-3$ variables  \cite{10.1007/BF02392525,Rosenhaus:2018zqn}.  
\end{prop}
The obtained HKLL-type  representation of the AdS vertex function along with  the extrapolate dictionary relation \eqref{extrapol} demonstrates that there is one-to-one holographic correspondence between the AdS vertex functions and the global conformal blocks in the comb channel.\footnote{The conformal block conveniently decomposes into a power-law leg-factor and a bare conformal block which is a conformally-invariant function of cross-ratios. In formula \eqref{comb_fin} these are the first and  second lines, respectively. Note that at $n=2,3$  the bare conformal blocks are trivial ($=1$) so that only the leg-factors contribute which are  $2$-point  and $3$-point  correlation functions in CFT.} Using the extrapolate relation \eqref{extrapol} one can obtain the global conformal block as a boundary value of the AdS vertex function and, vice versa,  using the HKLL-type representation \eqref{HKLL_vertex}  one can reconstruct the AdS vertex function from its boundary value, i.e. from the global conformal block.

The HKLL representation for correlation functions of scalar operators \eqref{HKLL_rel} in AdS/CFT  \cite{Hamilton:2005ju} is similar to our relation \eqref{HKLL_vertex}:  
\be 
\label{HKLL1}
\braket{\phi_1(x_1)\phi_2(x_2)...\phi_n(x_n)} = \left(\prod_i^n\int_{U_i}du\, K(x_i,u_i|1-\h_i)\right)\braket{\cO_{\h_1}(u_1)\cO_{\h_2}(u_2)...\cO_{\h_n}(u_n)}\,.
\ee
At $n=2,3$ the HKLL representation \eqref{HKLL1}  and that of the AdS vertex function \eqref{HKLL_vertex} coincide.\footnote{In the $n=2$ case,  the  expression similar to \eqref{HKLL_vertex} was obtained in  \cite{Castro:2018srf}. The novelty here is the integration contour which analytically continues the HKLL-type representation of the 2-point AdS vertex function to any real weights $\h$.}  At $n\geq4$,  \eqref{HKLL1} contains a correlation function of primary operators while  \eqref{HKLL_vertex} contains the global conformal block. Obviously, they can be related by virtue of the conformal block expansion. Adding the structure constants and summing the AdS vertex functions \eqref{HKLL_vertex} over intermediate weights $\tilde{\h}_i$ one obtains the  correlation function of \ads scalar operators. Thus, we conclude that the AdS vertex functions can be viewed as building blocks of \ads scalar correlation functions in the HKLL representation. In this way  one can think of a sort of AdS$_2$/\ads duality between topological \ads gravity and scalar field dynamics in AdS$_2$.   

The HKLL-type  integral formula \eqref{HKLL_vertex}  can be explicitly evaluated. Here is an exact expression  for the AdS vertex function.

\begin{prop}
\label{prop:expl}
The $n$-point AdS vertex function is  given by the following  multidimensional series: 
\be
\label{n-point_short}
\cV_{ \h\tilde{\h}}({\bf x})=\frac{\pref_{\h\tilde{\h}}\,\cL_{\h\tilde{\h}}({\bf x})}{(2i)^{\sum_{i=1}^n\h_i}}\hspace{-9mm}\sum_{\substack{m_1,...,m_{n-3}=0\\\{k_{i\,i-2},k_{i\,i-1},k_{i\,i+1},k_{i\,i+2}=0\}_{i=1,..., n}}}^{\infty}\hspace{-12mm}D_{\h\tilde{\h}}^{m_i\,k_{il}}\prod_{i=1}^{n-3}\chi_{i}^{m_i}\prod_{l=1}^{n-1}s_{l\, l+2}^{k_{l\,l+2}}s_{l+1\, l}^{k_{l+1\,l}}s_{l\, l+1}^{k_{l\,l+1}}s_{l+2\, l}^{k_{l+2\,l}}\;,
\ee 
where $\pref_{\h\tilde{\h}}$ is given by \eqref{prefactor}, the variables are defined as
\be
\label{npoint_variables}
\ba{l}
\dps 
\chi_i:=\frac{(\bar{\pt}_i-\bar{\pt}_{i+1})(\bar{\pt}_{i+2}-\bar{\pt}_{i+3})}{(\bar{\pt}_i-\bar{\pt}_{i+2})(\bar{\pt}_{i+1}-\bar{\pt}_{i+3})}\,,
\qquad 
s_{i\, j}:=\frac{\bar{\pt}_{i}-\pt_i}{\bar{\pt}_i-\bar{\pt}_j}\,,
\qquad
\pt_i = z_i+ie^{-\rho_i} \;;
\ea
\ee
we also introduced the $n$-point leg-factor
\be
\label{leg-factor}
\cL_{\h\tilde{\h}}({\bf x}) := \prod_{i=0}^{n-2}\chi_{i}^{\tilde{\h}_i}\prod_{l=1}^{n}(\bar{\pt}_l-\pt_l)^{\h_l}(\bar{\pt}_l-\bar{\pt}_{l+1})^{-\h_l-\h_{l+1}}
(\bar{\pt}_l-\bar{\pt}_{l+2})^{\h_{l+1}}\,,
\ee
and the $D$-coefficients
\be
\label{npoint_coeff}
\ba{l}
\dps 
D_{\h\tilde{\h}}^{m_i\,k_{il}}:=\prod_{s=1}^{n}\frac{(\h_s)_{k_{s\,s-2}+k_{s\,s-1}+k_{s\,s+1}+k_{s\,s+2}}}{k_{s\,s-2}!k_{s\,s-1}!k_{s\,s+1}!k_{s\,s+2}!m_s!(2\h_s)_{k_{s\,s-2}+k_{s\,s-1}+k_{s\,s+1}+k_{s\,s+2}}(2\tilde{\h}_s)_{m_s}}
\vspace{2.5mm}  
\\
\dps\times\prod_{t=1}^{n-1}(\h_t+\h_{t+1}-\tilde{\h}_{t-2}-\tilde{\h}_t-m_{t-2}-m_t)_{k_{t+1\,t}+k_{t\,t +1}}(\tilde{\h}_{t-1}+\tilde{\h}_t-\h_{t+1})_{m_{t-1}+m_t+k_{t+2\,t}+k_{t\,t +2}}\,.
\ea
\ee
\end{prop}
The proof is given in Appendix \bref{app:n_point}. In particular cases, the above multidimensional series can be given by special functions, see Appendix \bref{app:special}. To make the expression  \eqref{n-point_short} more compact we used the following conventions:
\begin{itemize}

\item if the indices of  $k_{ij}$ in  \eqref{n-point_short} satisfy one of the following four inequalities: $i\leq0$, $j\leq0$, $i>n$, $j>n$, then the sum over this variable is omitted and one sets $k_{ij}=0$. The same applies to the summation variable $m_i$ in the case $i>n-3$ or $i\leq0$;

\item if a  Pochhammer symbol or a power of a variable contain $\tilde{\h}_i$ with $i<0$ or $i>n-2$, then $\tilde{\h}_i=0$. If $i=0$ or $i=n-2$, then  $\tilde{\h}_0 = \h_1$ and $\tilde{\h}_{n-2} = \h_n$;

\item if the factor $(\bar{\pt}_i-\bar{\pt}_{j})$ with $i\leq0$, $j\leq0$, $i>n$ or $j>n$ appears in a sum, then it is replaced by $1$.

\end{itemize}

\noindent Note that despite the presence of the complex-valued arguments, the AdS vertex function is real (up to an overall phase factor which we do not fix) as it is constructed from the Wilson matrix elements and intertwiners which are real-valued functions.

The AdS vertex function in the form \eqref{n-point_short} explicitly depends on $5n-9$ variables but only $2n-2$ of them are independent. To show this, one notes that  $3n-7$ variables $\chi_i, s_{i+2\,i}$ and $s_{i\,i+2}$ in \eqref{n-point_short} can be expressed as functions of $2n-2$ variables $s_{i+1\,i}, s_{i\,i+1}$ as
\be
\label{functional_dependence}
\ba{c}
\dps
\chi_i = \frac{s_{i+2\,i+1}s_{i+1\,i+2}}{(s_{i+2\,i+3}-s_{i+2\,i+1})(s_{i+1\,i}- s_{i+1\,i+2})}\,,
\vspace{3mm}
\\
\dps
s_{i+2\,i}= \frac{s_{i+1\,i}s_{i+2\,i+1}}{s_{i+1\,i}-s_{i+1\,i+2}}\,,
\quad 
s_{i\,i+2}=\frac{s_{i\,i+1}s_{i+1\,i+2}}{ s_{i+1\,i+2}-s_{i+1\,i}}\,.
\ea 
\ee 
On the other hand, by examining the Jacobian matrix of $s_{i\,i+1}$ and $s_{i+1\,i}$ one shows that these variables are independent. From the Ward identities \eqref{WA_wo} it follows that the $n$-point AdS vertex function depends on  $2n-3$ invariant combinations  \eqref{integrals_of_motion} while the variables \eqref{npoint_variables} are not invariant. Finding the form of the $n$-point AdS vertex function which explicitly depends on invariant variables is technically difficult, so below we will consider only the simplest cases and obtain invariant parameterization of the $2$-point and $3$-point AdS vertex functions.

\subsection{Asymptotic  analysis} 
\label{sec:near}

To check  that the obtained $n$-point AdS vertex function \eqref{n-point_short} satisfies  the extrapolate dictionary relation \eqref{extrapol} one places all  coordinates $\rho_i$ on the same $\rho=const$ hypersurface and sends  $\rho\to\infty$. The asymptotic values of  variables  \eqref{npoint_variables} are given by
\be
\ba{l}
\dps
\chi_i\big|_{\rho_i=\rho}=\frac{z_{i\,i+1}z_{i+2\,i+3}}{z_{i\,i+2}z_{i+1\,i+3}}+O(e^{-\rho}) \equiv \cross_i+O(e^{-\rho})\,, 
\vspace{2.5mm}  
\\
\dps
s_{ij}|_{\rho_i=\rho} = e^{-\rho}\frac{-2i}{z_{ij}}+O(e^{-2\rho})\,,
\ea
\ee 
where $\cross_i$ are the cross-ratios \eqref{cross-ratios}.
Since $s_{ij}\to0$ the leading term in the $n$-point AdS vertex function \eqref{n-point_short} is given by the contribution with $k_{l\, l\pm1}=k_{l\,l\pm2}=0$. The $D$-coefficients \eqref{npoint_coeff} then become
\be
D_{\h\tilde{\h}}^{m_i\,0}=\prod_{s=1}^{n-3}\frac{(\tilde{\h}_{s-1}+\tilde{\h}_s-\h_{s+1})_{m_{s-1}+m_s}}{m_s!(2\tilde{\h}_s)_{m_s}}\,,
\ee
which are, in fact, the expansion coefficients in the $n$-point conformal block expression  \eqref{comb_fin}. The leg-factor \eqref{leg-factor} is expanded as 
\be
\cL_{\h\tilde{\h}}({\bf x})|_{\rho_i=\rho} = \frac{e^{-\rho\sum_{i=1}^n\h_i}}{(2i)^{-\sum_{i=1}^n\h_i}}\prod_{i=1}^{n-3}\cross_i^{\tilde{\h}_i}\prod_{l=1}^{n}z_{l\,l+1}^{-\h_l-\h_{l+1}}
z_{l\,l+2}^{\h_{l+1}}+O(e^{-\rho(\sum_{i=1}^n\h_i+1)})\,.
\ee
Substituting everything into the $n$-point AdS vertex function, taking into account the conventions listed below \eqref{npoint_coeff} and rearranging the terms one finds 
\be 
\ba{l}
\dps
e^{\rho\sum_{i=1}^n\h_i}\,\cV_{\h \tilde{\h}}(\rho,{\bf z}) = \pref_{\h\tilde{\h}}\prod_{i=1}^{n-3}\cross_i^{\tilde{\h}_i}\prod_{l=2}^{n-1}\Big(\frac{z_{l+1\,l-1}}{z_{l+1\,l}z_{l-1\,l}}\Big)^{\h_l}\Big(\frac{z_{23}}{z_{12}z_{13}}\Big)^{\h_1}\Big(\frac{z_{n-2\,n-1}}{z_{n\,n-1}z_{n\,n-2}}\Big)^{\h_n}
\vspace{2.5mm}  
\\
\dps
\times\hspace{-4mm}\sum_{m_1,...,m_{n-3}=0}^\infty\hspace{-2mm}\frac{(\h_1+\tilde{\h}_1-\h_2)_{m_1}}{m_1!(2\tilde{\h}_1)_{m_1}}\frac{(\h_n+\tilde{\h}_{n-3}-\h_{n-1})_{m_{n-3}}}{m_{n-3}!(2\tilde{\h}_{n-3})_{m_{n-3}}}\prod_{s=2}^{n-4}\frac{(\tilde{\h}_{s-1}+\tilde{\h}_s-\h_{s+1})_{m_{s-1}+m_s}}{m_s!(2\tilde{\h}_s)_{m_s}}\prod_{i=1}^{n-3}\cross_i^{m_i}+O(e^{-\rho})\,.
\ea
\ee
Using the comb function \eqref{def_comb} one obtains
\be 
\ba{l}
\dps
\lim_{\rho\to\infty}e^{\rho\sum_{i=1}^n\h_i}\,\cV_{\h \tilde{\h}}(\rho,{\bf z}) = \pref_{\h\tilde{\h}}\prod_{i=1}^{n-3}\cross_i^{\tilde{\h}_i}\prod_{l=2}^{n-1}\Big(\frac{z_{l+1\,l-1}}{z_{l+1\,l}z_{l-1\,l}}\Big)^{\h_l}\Big(\frac{z_{23}}{z_{12}z_{13}}\Big)^{\h_1}\Big(\frac{z_{n-2\,n-1}}{z_{n\,n-1}z_{n\,n-2}}\Big)^{\h_n}
\vspace{2.5mm}  
\\
\dps
\times F_{K} \left[\begin{array}{cccc}
     \h_1+\tilde{\h}_1-\h_2,&\tilde{\h}_2+\tilde{\h}_1-\h_{3},&...\,,&\h_n+\tilde{\h}_{n-3}-\h_{n-1} \\
     2\tilde{\h}_1, &2\tilde{\h}_2,&...\,, &2\tilde{\h}_{n-3} \end{array};\cross_1,...\,,\cross_{n-3}\right]\,,
\ea
\ee
which is the $n$-point global conformal block \eqref{comb_fin} with the overall coefficient $\pref_{\h\tilde{\h}}$. The resulting expression agrees with  the extrapolate dictionary  \eqref{extrapol}.

\section{Lower-point AdS vertex functions}
\label{sec:lower}

In this section the $2$-point and $3$-point AdS vertex functions \eqref{n-point_short} are expressed in terms of hypergeometric functions.    

\subsection{$2$-point AdS vertex function}
\label{sec:2pt}

In this case the AdS vertex function can be   simplified to  the Gauss hypergeometric function   (see Appendix \bref{app:2p_simpl})\footnote{Here, we generalize the calculation of the $2$-point AdS vertex function from \cite{Alkalaev:2023axo}, where it was done in the case $\rho_1=\rho_2$ and $\h\in\mathbb{Z_-}$.}
\be
\label{2pt_inf}
\cV_{\h_1 \h_2}({\bf x})= \pref_{\h_1\h_2}\, c_{12}^{-\h_1}\F\left(\h_1,\h_1;2\h_1\Big|-\frac{4}{c_{12}}\right).
\ee
As per the  discussion in the end of section \bref{sec:exact}, the obtained function  explicitly depends on invariant variables  \eqref{integrals_of_motion} which in the present case are reduced to   $c_{12}$. Now,  applying  the quadratic transformation  \eqref{quadratic} and using  the  relation 
$4(c_{12}+2)^{-2} =\xi(x_1,x_2)^2$,  
where $\xi(x_1,x_2)$ is the AdS invariant distance between points $x_1$ and $x_2$ defined in \eqref{AdS_invariant_dist}, one obtains the final expression 
\be 
\label{2pt_as_bb}
\cV_{\h_1 \h_2}({\bf x})= \pref_{\h_1\h_2}\left[\frac{\xi(x_1,x_2)}{2}\right]^{\h_1}\F\left(\frac{\h_1}{2},\frac{\h_1}{2}+\half;\h_1+\half\, \Big|\, \xi(x_1,x_2)^2\right)\,.
\ee 
Up to the normalization coefficient $\pref_{\h_1\h_2}$, the resulting $2$-point AdS vertex function coincides with the bulk-to-bulk propagator of a scalar field of mass $m^2 = \h_1(\h_1-1)$ in \ads \eqref{bulk-to-bulk}. Thus, we have 
\begin{prop}
\label{prop:2pt} 
The 2-point Witten diagram is proportional to the 2-point AdS vertex function
\be
\label{2pt_rel} 
G(x_1,x_2|\h)  = \pref_{\h_1\h_2}^{-1}\, \cV_{\h_1 \h_2}({\bf x})\,,
\qquad
\h_1 = \h_2 = \h \,.
\ee
\end{prop}
The same result was obtained in \cite{Castro:2018srf}. To examine the near-the-boundary behaviour one takes $\rho = const$  in which case $c_{12}=  q_{12}^2$ \eqref{integrals_of_motion_rho_const}: 
\be
\cV_{\h_1 \h_2}(\rho,{\bf z})= \frac{\pref_{\h_1\h_2}}{q_{12}^{\;2\h_1}}\,\F\left(\h_1,\h_1;2\h_1 \Big|-\frac{4}{q_{12}^2}\right)\;.
\ee
Then, the asymptotic expansion near  $\rho =\infty$  reads 
\be
\label{2pt_asmp}
e^{2\rho \h_1}\cV_{\h_1 \h_2}(\rho,{\bf z})= \frac{\pref_{\h_1\h_2}}{z_{12}^{\;2\h_1}} + O(e^{-\rho})\;.
\ee
In accordance with the extrapolation dictionary the leading term here is the $2$-point correlation function   of \cft primary operators of conformal dimensions $\h_1$.

\subsection{$3$-point AdS vertex function}

One can find two convenient representations  of the 3-point AdS vertex function. 

\paragraph{First representation.}  In this case the AdS vertex function \eqref{n-point_short}  is given by 
\be
\label{3p_first_rep}
\ba{c}
\dps
\cV_{\h_1 \h_2 \h_3}({\bf x})=  \frac{C_{\h_1 \h_2 \h_3}\,\cL_{\h_1 \h_2 \h_3}({\bf x})}{(2i)^{\h_1+\h_2+\h_3}}\hspace{-3mm}\sum_{k_{12},k_{21},k_{13},k_{31},k_{23},k_{32}=0}^{\infty}\hspace{-3mm}D_{\h_1\h_2\h_3}^{k_{il}}\,s_{12}^{k_{12}}s_{13}^{k_{13}}s_{21}^{k_{21}}s_{23}^{k_{23}}s_{31}^{k_{31}}s_{32}^{k_{32}}\,,
\ea
\ee 
where the leg-factor and the $D$-coefficients are  given by 
\be 
\label{3p_leg_factor}
\ba{l}
\dps
\cL_{\h_1 \h_2 \h_3}({\bf x}) = (\bar{\pt}_1-\pt_1)^{\h_1}(\bar{\pt}_2-\pt_2)^{\h_2}(\bar{\pt}_3-\pt_3)^{\h_3}
\vspace{3mm}
\\
\dps
\times(\bar{\pt}_1-\bar{\pt}_{2})^{\h_3-\h_1-\h_2}(\bar{\pt}_1-\bar{\pt}_{3})^{\h_2-\h_1-\h_3}(\bar{\pt}_2-\bar{\pt}_{3})^{\h_1-\h_2-\h_3}
\,,
\ea
\ee
\be 
\ba{c}
\dps
D_{\h_1\h_2\h_3}^{k_{il}}=\frac{(\h_1+\h_3-\h_2)_{k_{13}+k_{31}}(\h_2+\h_3-\h_1)_{k_{23}+k_{32}}(\h_1+\h_2-\h_3)_{k_{12}+k_{21}}}{k_{12}!k_{21}!k_{13}!k_{31}!k_{23}!k_{32}!}
\vspace{3mm}
\\
\dps
\times
\frac{(\h_1)_{k_{12}+k_{13}}(\h_2)_{k_{23}+k_{21}}(\h_3)_{k_{32}+k_{31}}}{(2\h_1)_{k_{12}+k_{13}}(2\h_2)_{k_{23}+k_{21}}(2\h_3)_{k_{32}+k_{31}}}\,.
\ea
\ee
This function has six arguments. As discussed below \eqref{functional_dependence}, variables $s_{13}$ and $s_{31}$ can be expressed through  the independent variables $s_{12},s_{21},s_{23}$, and $s_{32}$ as follows 
\be
\ba{c}
\dps
s_{31}= \frac{s_{32}s_{21}}{s_{21}-s_{23}}\,,
\qquad
s_{13}=\frac{s_{12}s_{23}}{ s_{23}-s_{21}}\,.
\ea 
\ee 
Despite that our analysis of the Ward identities in section \bref{sec:wilson} shows that the $3$-point AdS vertex function depends only on  three invariant variables  \eqref{general_dependence} the  function \eqref{3p_first_rep} has four independent  non-invariant arguments. It follows that this expression can be further simplified (see the next paragraph). Nonetheless, the representation \eqref{3p_first_rep} is convenient in the sense that it is (anti)symmetric  with respect to permutations of  indices of weights and coordinates. 
 
Finally, by placing all $\rho$ coordinates on the same $\rho = const$ hypersurface and taking $\rho\to\infty$ in the first representation \eqref{3p_first_rep} of the $3$-point AdS vertex function  one obtains 
\be
\label{asym3pt}
\ba{l}
\dps
e^{\rho(\h_1+\h_2+\h_3)}\cV_{\h_1 \h_2 \h_3}(\rho,{\bf z})\big|_{
\rho\to\infty } = \frac{C_{\h_1 \h_2 \h_3}}{
\,z_{23}^{\h_2+\h_3-\h_1}z_{31}^{\h_1+\h_3-\h_2}z_{21}^{\h_1+\h_2-\h_3}}+ O( e^{-\rho})\,.
\ea
\ee 
The $3$-point AdS vertex function is proportional to the $3$-point conformal correlation function thereby confirming the extrapolate dictionary relation \eqref{extrapol}.

\paragraph{Second representation.} In Appendix \bref{app:3p_second} we show that the first  representation can be further simplified: 
\be
\ba{l}
\dps
\cV_{\h_1 \h_2 \h_3}({\bf x})=  \frac{C_{\h_1 \h_2 \h_3}\, c_{13}^{-\h_3}\,c_{12}^{-\h_2}}{(2i)^{\h_1-\h_2-\h_3}}\sum_{\substack{k,s\geq0\vspace{1mm}\\ k<s}}\, \frac{(\h_1+\h_2-\h_3)_{k}(\h_1-\h_2-\h_3)_{k}(\h_2+\h_3-\h_1)_{s}(\h_1)_{k}}{k!(s-k)!(2\h_1)_{k}}
\vspace{3mm}
\\
\dps
\times y^{\h_3-\h_1+s}
{}_2F_1\left(k+\h_1-s,\h_3;2\h_3\big|\frac{-4}{c_{13}}\right){}_2F_1\left(\h_2+\h_3-\h_1-k+s,\h_2;2\h_2\big|\frac{-4}{c_{12}}\right),
\ea
\ee
where a new variable $y$ is introduced,   
\be 
\dps
y:=\frac{(\bar{\pt}_2-\pt_1)(\bar{\pt}_3-\bar{\pt}_1)}{(\bar{\pt}_2-\bar{\pt}_1)(\bar{\pt}_3-\pt_1)} \,.
\ee 
Note that $c_{23}$ is related to $y$, $c_{12}$, $c_{13}$ as\footnote{This equation has two complex conjugated roots. However, the $3$-point AdS vertex function does not depend on the choice of a root because it is real.}
\be
\dps
\label{c_as_y}
c_{23}=\frac{c_{13}c_{12}\, y}{(y-1)(y-1+c_{12}-y c_{13})}\,.
\ee 
The resulting representation of the $3$-point AdS vertex function depends only on invariant variables $c_{ij}$ and their combinations in agreement with the Ward identities \eqref{general_dependence}, but the explicit permutation (anti)symmetry of the first representation \eqref{3p_first_rep} is now lost. 

To summarize, there are two equivalent representations of the 3-point AdS vertex function, each of which is convenient in its own way: either  an explicit permutation symmetry or a minimal set of variables.

\section{Witten diagrams and AdS vertex functions}  
\label{sec:points}

\begin{figure}
\hspace{-11mm} \includegraphics[scale=1.1]{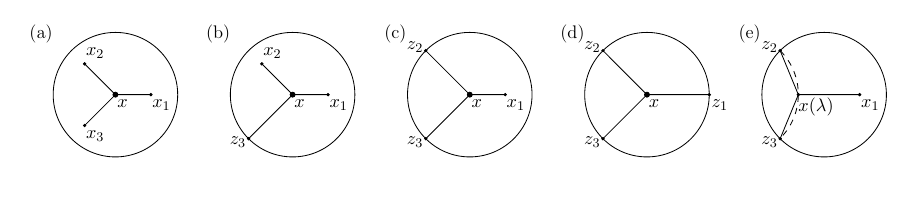}
\caption{$3$-point Witten diagrams with three (a), two (b), one (c), zero (d) bulk points and the geodesic Witten diagrams (e). The central dot in (a)-(d) denotes the integration over AdS. The dashed line is a geodesic $x(\lambda)$ between boundary points $z_2$ and $z_3$.}
\label{fig:geodesic}
\end{figure}

As we saw earlier, the $2$-point AdS vertex function coincides with the bulk-to-bulk propagator of free scalar fields in AdS$_2$.  In this section we show that the $3$-point (geodesic) scalar Witten diagram can be expressed in terms of the $3$-point AdS vertex functions, cf. \eqref{summary}.  Recall that the 3-point Witten diagrams \cite{Witten:1998qj} are given by 
\be
\label{WD}
\cW_{\h_1, \h_2, \h_3}[x_1,x_2,x_3] = \int_{\text{AdS}_2} d^2x\, G(x,x_1|\h_1) G(x,x_2|\h_2)G(x,x_3|\h_3)\,,
\ee
where $G(x,y|\h)$ is the bulk-to-bulk  propagator, $h$ parameterizes the mass $m^2 = h(h-1)$ of a scalar field in AdS$_2$. If  e.g. $\h_3=0$ and $\h_1 = \h_2 \equiv  \h$, then the 3-point Witten diagram  reduces to the 2-point Witten diagram which is the bulk-to-bulk propagator $G(x_1, x_2|\h)$.  For three points on the boundary the respective Witten diagram calculates the 3-point conformal correlation function of primary operators of conformal weights $\h_1$, $\h_2$, $\h_3$.  On the other hand, choosing two points on the boundary and  reducing  the integration domain yields the  geodesic Witten diagram \cite{Hijano:2015zsa}:   
\be
\label{WGD}
\widetilde \cW_{\h_1, \h_2, \h_3}[x_1,z_2,z_3] =\int_{\gamma_{23}}d\lambda \,G(x(\lambda),x_1|\h_1)\, K(x(\lambda),z_2|\h_2) \, K(x(\lambda),z_3|\h_3)\,,
\ee 
where $\gamma_{23}$ is a geodesic in the bulk, $x(\lambda)$, which connects two boundary points $z_2$ and $z_3$, $\lambda$ is the proper time,   see Fig. \bref{fig:geodesic}.

Below we analyze the form of the 3-point AdS vertex function depending on how many of three points are on the boundary. In the simplest case, when all three points lie on the boundary we obtain the 3-point conformal function in accordance with the extrapolate dictionary relation. The next simplest case is when two points are on the boundary and here we show that the 3-point AdS vertex function calculates the geodesic Witten diagram of three scalar fields of different masses. In the case of one point on the boundary  one obtains that the Witten diagram integral decomposes into 3-point AdS vertex functions of running weights. The case of three points in the bulk also inherits this pattern.     

\subsection{One or two points on the  boundary }  
\label{sec:12points}

\paragraph{Two boundary points.} Let two points $x_2$ and $x_3$ lie  on the boundary. In Appendix \bref{app:KGK_calc} we show that the respective AdS vertex function can be expressed in terms of the Gauss hypergeometric function with a complicated argument 
\be
\label{Wilson_KGK_t}
\ba{l}
\dps
\cV_{\h_1 \h_2 \h_3}^{\text{reg}}(x_1,z_2,z_3)=  \pref_{\h_1\h_2\h_3} \, z_{23}^{\h_1-\h_2-\h_3}\chi(x_1,z_2)^{\frac{\h_1+\h_2-\h_3}{2}}\chi(x_1,z_3)^{\frac{\h_1+\h_3-\h_2}{2}}
\vspace{3mm}
\\
\dps
\times{}_2F_1\left(\frac{\h_1+\h_3-\h_2}{2},\frac{\h_1+\h_2-\h_3}{2}; \h_1+\half\,\Big|\, z_{23}^2\,\chi(x_1,z_2)\,\chi(x_1,z_3)\right)\,,
\ea
\ee
where $\cV_{\h_1 \h_2 \h_3}^{\text{reg}}$ stands for  the regularized function  \eqref{regular} and $\chi(x,z)$ is the regularized invariant distance  between a bulk point $x$ and a boundary point $z$ \eqref{regul_distance}. Note that up to the normalization coefficient the obtained expression coincides with the correlation function of the one bulk and two boundary scalars calculated in \cite{Kabat:2012av} (see eq. (19) therein). This can be directly seen from comparing  the integral representation of the AdS vertex function \eqref{HKLL_vertex} and the integral used to calculate the mentioned correlation function since  these  two integrals are related by analytic continuation.  Moreover, it turns out that this 3-point AdS vertex function  is given by a geodesic Witten diagram with two boundary points (see Fig. \bref{fig:geodesic}).

\begin{prop}
\label{prop:geod} The 3-point AdS vertex functions with two boundary points  is proportional to the 3-point  geodesic Witten diagram, 
\be
\widetilde \cW_{\h_1, \h_2, \h_3}[x_1,z_2,z_3]  =\frac{A(\h_1,\h_2,\h_3)}{\pref_{\h_1 \h_2 \h_3}} \,\cV_{\h_1 \h_2 \h_3}^{\text{reg}}(x_1,z_2,z_3)\,,
\ee 
where 
\be 
\label{geodesic_coef}
A(\h_1,\h_2,\h_3):=\frac{\Gamma(\frac{\h_1+\h_2-\h_3}{2})\Gamma(\frac{\h_1+\h_3-\h_2}{2})}{2\Gamma(\h_1)}\,.
\ee 
\end{prop}
Indeed, this statement can be seen by comparing the 3-point AdS vertex function \eqref{Wilson_KGK_t} with the $3$-point geodesic Witten diagram \eqref{WGD} which  was explicitly calculated in \cite{Hijano:2015zsa}: 
\be
\label{Witten_3pt_geodesic}
\ba{c}
\dps
\int_{\gamma_{23}}d\lambda \,G(x(\lambda),x_1|\h_1)\, K(x(\lambda),z_2|\h_2) \, K(x(\lambda),z_3|\h_3) = A(\h_1,\h_2,\h_3)\, z_{23}^{\h_1-\h_2-\h_3}\chi(x_1,z_2)^{\frac{\h_1+\h_2-\h_3}{2}}
\vspace{3mm}
\\
\dps
\times\chi(x_1,z_3)^{\frac{\h_1-\h_2+\h_3}{2}}{}_2F_1\left(\frac{\h_1-\h_2+\h_3}{2},\frac{\h_1+\h_2-\h_3}{2}; \h_1+\half\,\Big| \,z_{23}^2\,\chi(x_1,z_2)\,\chi(x_1,z_3)\right),
\ea
\ee 
where the coefficient $A(\h_1,\h_2,\h_3)$ is given by  \eqref{geodesic_coef}.

Yet  another relation can be found by considering the following expression for the $3$-point Witten diagram \eqref{WD} with two boundary points  which was explicitly calculated  in Ref. \cite{Zhou:2018sfz}.\footnote{The same Witten diagram was also calculated in \cite{Jepsen:2019svc} by using a different method, see eq. (B.8) therein.} Therein we find formula (C.16) which by using the reflection formula for gamma functions and taking into account the difference in the normalization of the bulk-to-bulk propagators  can be cast into the form: 
\be
\label{KGK}
\ba{l}
\dps
\int_{\text{AdS}_2} d^2x\, G(x,x_1|\h_1) K(x,z_2|\h_2)K(x,z_3|\h_3)   
\vspace{3mm}
\\
\dps
=\alpha(\h_1,\h_2,\h_3)\, \sum_{k=0}^{\infty}\,\frac{\left(\frac{\h_1-\h_2+\h_3}{2}\right)_k\left(\frac{\h_1+\h_2-\h_3}{2}\right)_k}{k!(\h_1+\half)_k}\chi(x_1,z_2)^{\frac{\h_1+\h_2-\h_3}{2}+k}\chi(x_1,z_3)^{\frac{\h_1-\h_2+\h_3}{2}+k}z_{23}^{\h_1-\h_2-\h_3+2k}
\vspace{3mm}
\\
\dps
+\beta(\h_1;\h_2,\h_3)\,\sum_{k=0}^\infty \frac{(\h_2)_k(\h_3)_k}{(\frac{\h_2+\h_3-\h_1}{2}+1)_k(\frac{\h_1+\h_2+\h_3}{2}+\half)_k}\chi(x_1,z_2)^{k+\h_2}\chi(x_1,z_3)^{k+\h_3}z_{23}^{2k}
\vspace{4mm}
\\
\dps
\hspace{40mm}\equiv  GW^{(1)}[x_1,z_2,z_3]+GW^{(2)}[x_1,z_2,z_3]\;,
\ea
\ee 
where 
\be 
\label{alpha_beta}
\ba{l}
\dps
\alpha(\h_1,\h_2,\h_3) = \frac{\pi^{\half}}{2}\Gamma\left(\frac{\h_1+\h_2+\h_3}{2}-\half\right)\frac{\Gamma(\frac{\h_1+\h_2-\h_3}{2})\Gamma(\frac{\h_1-\h_2+\h_3}{2})\Gamma(\frac{\h_2-\h_1+\h_3}{2})}{\Gamma(\h_1)\Gamma(\h_2)\Gamma(\h_3)}\,,
\vspace{3mm}
\\
\dps
\beta(\h_1;\h_2,\h_3) = \frac{\pi^{\half}}{2}\frac{\Gamma(\frac{\h_1+\h_2+\h_3}{2}-\half)\Gamma\left(\frac{\h_1-\h_2-\h_3}{2}\right)\Gamma(\h_1+\half)}{\Gamma(\frac{\h_1+\h_2+\h_3}{2}+\half)\Gamma\left(\frac{\h_1-\h_2-\h_3}{2}+1\right)\Gamma(\h_1)}\,.
\ea
\ee 
$GW^{(1,2)}[x_1,z_2,z_3]$ stand for the second and third lines in \eqref{KGK}, respectively.    
Using the Gauss hypergeometric series one can directly see that $GW^{(1)}[x_1,z_2,z_3]$  is proportional to the $3$-point AdS vertex function \eqref{Wilson_KGK_t}. On the other hand, $GW^{(2)}[x_1,z_2,z_3]$ can be represented as a linear combination of the $3$-point (regularized) AdS vertex functions
\be
\label{KGK_2nd_term}
GW^{(2)}[x_1,z_2,z_3]=\sum_{n=0}^{\infty}\frac{a(\h_1;\h_2,\h_3;n)}{\pref_{\h_2+\h_3+2n\  \h_2\h_3}}\,\cV^{\text{reg}}_{\h_2+\h_3+2n\  \h_2\h_3}(x_1,z_2,z_3)\,,
\ee 
where  
\be 
\label{a_coef}
\ba{l}
a(\h_1;\h_2,\h_3;n)=
\vspace{3mm}
\\
\dps
\hspace{5mm}\frac{2\pi^{\half}}{\Gamma(\h_1)}\,\frac{(-)^n(\h_2)_n(\h_3)_n}{n!(\h_2+\h_3-\half+n)_n} \,\frac{\Gamma(\h_1+\half)}{\h_1(\h_1-1)-(\h_2+\h_3+2n)(\h_2+\h_3+2n-1)}\;.
\ea
\ee 
To prove  \eqref{KGK_2nd_term}  one substitutes the regularized $3$-point AdS vertex function \eqref{Wilson_KGK_t}  into \eqref{KGK_2nd_term}, then  converts the hypergeometric function into series with summation variable $k$, changes $k\to l=k+n$ and uses the identity \eqref{useful_identity} with $a=\h_1+\h_2$ and $b=\h_3$: 
\be
\ba{c}
\dps
GW^{(2)}[x_1,z_2,z_3]=\frac{\pi^{\half}}{2}\sum_{l=0}^{\infty}z_{23}^{2l}\,\chi(x_1,z_2)^{\h_2+l}\chi(x_1,z_3)^{\h_3+l}
\vspace{3mm}
\\
\dps
\times(\h_2)_{l}(\h_3)_{l}\frac{\Gamma(\frac{\h_1+\h_2-\h_3}{2})\Gamma(\frac{\h_1+\h_2+\h_3}{2}-\half)\Gamma(\h_1+\half)}{\Gamma(\frac{\h_1+\h_2-\h_3}{2}+l+1)\Gamma(\frac{\h_1+\h_2+\h_3}{2}+\half+l)\Gamma(\h_1)}\,.
\ea
\ee
Finally, rewriting  the gamma functions through the Pochhammer symbols, renaming $l\to n$ and gathering everything together one formulates the resulting statement: 

\begin{prop}
\label{prop:3pt_two} The conformal boundary asymptotics about two points  of the 3-point Witten diagram is expressed in terms of the 3-point AdS vertex functions as
\be
\label{KGK_rel}
\ba{r}
\dps
\int_{\text{AdS}_2}d^2x \,G(x,x_1|\h_1) K(x,z_2|\h_2)K(x,z_3|\h_3) =  \frac{\alpha(\h_1,\h_2,\h_3)}{\pref_{\h_1\h_2\h_3}}\cV^{\text{reg}}_{\h_1\h_2\h_3}(x_1,z_2,z_3)
\vspace{3mm}
\\
\dps
+\sum_{n=0}^{\infty}\frac{a(\h_1;\h_2,\h_3;n)}{\pref_{\h_2+\h_3+2n\ \h_2\h_3}}\cV^{\text{reg}}_{\h_2+\h_3+2n\ \h_2\h_3}(x_1,z_2,z_3)\,.
\ea
\ee
\end{prop}

\paragraph{One boundary point.} The relation between the $3$-point AdS vertex function and the $3$-point Witten diagram   \eqref{KGK_rel} can be straightforwardly generalized to the case of one boundary point. The proof is given in Appendix \bref{app:KGG}. 

\begin{prop}
\label{prop:3pt_one} 
The conformal boundary asymptotics about one point of the 3-point Witten diagram is expressed in terms of the 3-point AdS vertex functions as 
\be 
\label{KGG_rel}
\ba{r}
\dps
\int_{\text{AdS}_2}d^2x\, G(x,x_1|\h_1) G(x,x_2|\h_2)K(x,z_3|\h_3) = \frac{\alpha(\h_1,\h_2,\h_3)}{\pref_{\h_1\h_2\h_3}}\cV^{\text{reg}}_{\h_1\h_2\h_3}(x_1,x_2,z_3)
\vspace{3mm}
\\
\dps
+\sum_{n=0}^{\infty}\frac{a(\h_1;\h_2,\h_3;n)}{\pref_{\h_2+\h_3+2n\  \h_2\h_3}}\cV^{\text{reg}}_{\h_2+\h_3+2n\  \h_2\h_3}(x_1,x_2,z_3)
\vspace{3mm}
\\
\dps
+\sum_{n=0}^{\infty}\frac{a(\h_2;\h_1,\h_3;n)}{\pref_{\h_1\  \h_1+\h_3+2n\ \h_3}}\cV^{\text{reg}}_{\h_1\  \h_1+\h_3+2n\ \h_3}(x_1,x_2,z_3)\,,
\ea
\ee 
where the last two terms are symmetric with respect to the permutation of indices $1\leftrightarrow 2$ of weights and coordinates.
\end{prop}

\begin{figure}
\centering
\includegraphics[scale=1.1]{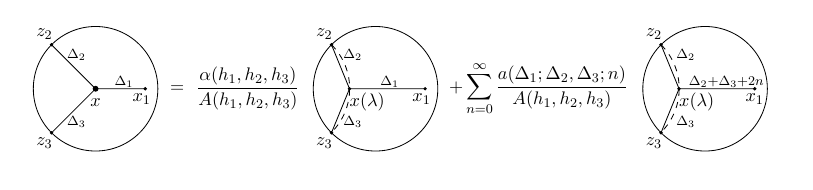}
\caption{The diagrammatic  representation of the relation \eqref{KGK_rel} with the geodesic $3$-point Witten integrals  \eqref{Witten_3pt_geodesic} instead of the 3-point AdS vertex functions. The dashed line represents the geodesic, the central dot on the left diagram  denotes the integral over AdS$_2$.}
\label{fig:KGK_rel}
\end{figure}

\paragraph{Addendum.} Let us replace  the $3$-point AdS vertex functions  in  \eqref{KGK_rel}  with  the geodesic $3$-point Witten diagrams \eqref{Witten_3pt_geodesic}. In this way one reveals a new propagator identity shown in Fig. \bref{fig:KGK_rel}. Another way to prove this identity   is to apply the  following two relations to the left-hand side of \eqref{KGK} (see \cite{Hijano:2015zsa} and  eqs. (4.1) and (4.8) therein)
\be 
\label{prop_identity_1}
\ba{c}
\dps
 K(x,z_1|\h_1)K(x,z_2|\h_2) = \frac{2}{A(\h_1,\h_2,\h_3)}\sum_{n=0}^\infty\frac{(-)^n\Gamma(\h_1+\h_2+2n)}{\Gamma(\h_1)\Gamma(\h_2)n!(\h_1+\h_2+n-\half)_n}
 \vspace{3mm}
\\
\dps
\times\int_{\gamma_{12}}d\lambda\, K(y(\lambda),z_1|\h_1)K(y(\lambda),z_2|\h_2)G(y(\lambda),x|\h_1+\h_2+2n)\,,
\ea
\ee
\be
\int_{\text{AdS}_2} d^2x\, G(x,x_1|\h_1)G(x,x_2|\h_2) = 2\pi^{\half}\frac{\frac{\Gamma(\h_2+\half)}{\Gamma(\h_2)}G(x_1,x_2|\h_1)-\frac{\Gamma(\h_1+\half)}{\Gamma(\h_1)}G(x_1,x_2|\h_2)}{\big(\h_1(\h_1-1)-\h_2(\h_2-1)\big)}\,,
\ee
where $A(\h_1,\h_2,\h_3)$ is given by  \eqref{geodesic_coef}.

\subsection{Three points in the bulk}  
\label{sec:3bulk}

One can further generalize the relation \eqref{KGG_rel} to the case of three points in the bulk by comparing the subleading terms in the boundary expansion of the Witten diagram and the AdS vertex function. It can be shown that the first few coefficients in both expansions coincide but at higher order the technical complexity drastically increases. In order to have a more tractable way to prove the desired relation we use a  different approach. We observe  that both the $3$-point AdS vertex function and a part of the $3$-point Witten diagram \eqref{WD} satisfy the homogeneous  Klein-Gordon equation with the same boundary condition which  uniquely fixes the solution. This allows us to identify these two functions. To this end, we recall that the $3$-point AdS vertex function satisfies the homogeneous  Klein-Gordon equations over each argument with respective  masses \cite{Alkalaev:2023axo}:
\be
\label{KG_vertex}
\left(\Box_{i}- m_i^2\right) \cV_{\h_1\h_2\h_3}(x_1, x_2, x_3) = 0\,,
\qquad
m_i^2 = \h_i(\h_i-1)\,,
\;\;
i = 1,2,3\,.
\ee
These equations hold by construction because each Wilson line matrix element \eqref{closed_left0} satisfies the homogeneous  Klein-Gordon equation \cite{Bhatta:2018gjb,Castro:2020smu,Alkalaev:2023axo}. 
 
Now, consider the $3$-point Witten diagram \eqref{WD} with three points in the bulk which was explicitly calculated in \cite{Jepsen:2019svc} (see eq. (B.56) therein which we restrict to two dimensions):
\be
\label{GGG}
\ba{l}
\dps
\int_{\text{AdS}_2}d^2x\, G(x,x_1|\h_1) G(x,x_2|\h_2)G(x,x_3|\h_3) =  \alpha(\h_1,\h_2,\h_3)\sum_{k_1,k_2,k_3=0}^{\infty}c^{\h_1;\h_2;\h_3}_{k_1;k_2;k_3}
\vspace{3mm}
\\
\dps
\times
\left[\frac{\xi(x_1,x_3)}{2}\right]^{\frac{\h_3+\h_1-\h_2}{2}+k_1-k_2+k_3}\left[\frac{\xi(x_2,x_3)}{2}\right]^{\frac{\h_3+\h_2-\h_1}{2}-k_1+k_2+k_3}\left[\frac{\xi(x_1,x_2)}{2}\right]^{\frac{\h_1+\h_2-\h_3}{2}+k_1+k_2-k_3}
\vspace{3mm}
\\
\dps
+\Bigg(\sum_{k_1,k_2,k_3=0}^{\infty}d^{\h_1;\h_2;\h_3}_{k_1;k_2;k_3}\left[\frac{\xi(x_1,x_3)}{2}\right]^{\h_3+2k_3+k_1}\left[\frac{\xi(x_2,x_3)}{2}\right]^{-k_1}\left[\frac{\xi(x_1,x_2)}{2}\right]^{\h_2+k_1+2k_2} +(1\leftrightarrow 2,3)\Bigg)  
\vspace{3mm}
\\
\equiv WF^{(1)}(x_1,x_2,x_3) + WF^{(2)}(x_1,x_2,x_3)+ WF^{(3)}(x_1,x_2,x_3)+ WF^{(4)}(x_1,x_2,x_3)\,,
\ea
\ee 
where $(1\leftrightarrow 2,3)$ stands for  interchanging indices $1$ and $2,3$ and  $WF^{(1,2,3,4)}(x_1,x_2,x_3)$ denote each sum in \eqref{GGG}, respectively, from  first to  fourth.    
The coefficients are given by 
$$
\ba{l}
\dps
c_{k_1 ; k_2;k_3}^{\h_1 ; \h_2; \h_3 } = \left(\frac{\h_1-\h_2+\h_3}{2}\right)_{k_1-k_2+k_3}\left(\frac{\h_1+\h_2-\h_3}{2}\right)_{k_1+k_2-k_3}\left(\frac{\h_2+\h_3-\h_1}{2}\right)_{k_2+k_3-k_1}
\vspace{3mm}
\\
\dps
\times
\frac{(-)^{k_1+k_2+k_3}}{k_{1}!k_{2}!k_3!}F^{(3)}_{A}\left[\begin{array}{cccc}
     -k_1,&-k_2,&-k_3, &\frac{\h_1+\h_2+\h_3}{2}-\half  \\
    & \h_1+\half, &\hspace{8mm}\h_2 +\half, &\h_3 +\half\end{array};1,1,1\right]\,,
\vspace{3mm}
\\
\dps
d_{k_1 ; k_2;k_3}^{\h_1 ; \h_2; \h_3 } =\frac{\pi^{\half}\Gamma(\frac{\h_1+\h_2+\h_3}{2}-\half)}{2\Gamma(\h_1)}(\h_3)_{2k_3+k_1}(\h_2)_{2k_2+k_1}\Gamma\left(\frac{\h_1-\h_2-\h_3}{2}-k_1-k_2-k_3\right)
\ea
$$
\be 
\label{GGG_coef}
\ba{l}
\dps
\times
\frac{(-)^{k_1+k_2+k_3}}{k_{1}!k_{2}!k_3!}F^{(3)}_{A}\left[\begin{array}{cccc}
     \frac{\h_1-\h_2-\h_3}{2}-k_1-k_2-k_3,&-k_2,&-k_3, &\frac{\h_1+\h_2+\h_3}{2}-\half  \\
      \h_1+\half, &\hspace{-14mm}\h_2 +\half, &\h_3 +\half\end{array};1,1,1\right]\,,
\ea
\ee 
where $F_A^{(3)}$ is the Lauricella function \eqref{def_lauricella_A}. 
\begin{lemma} 
\label{lem:KG} The first term $WF^{(1)}(x_1,x_2,x_3)$ satisfies the homogeneous  Klein-Gordon  equation with boundary condition:
\be
\label{KG_WF}
\ba{l}
\dps
\left(\Box_{i}- m_i^2\right) WF^{(1)}(x_1,x_2,x_3) = 0\,,
\qquad
m_i^2 = h_i(h_i-1)\,,
\;\;
i = 1,2,3\,;
\vspace{3mm}
\\
\dps
\lim_{\rho_i\to\infty}e^{\rho_i\h_i}WF^{(1)}(x_1,x_2,x_3) = \frac{\alpha(\h_1,\h_2,\h_3)}{\pref_{\h_1\h_2\h_3}}\lim_{\rho_i\to\infty}e^{\rho_i\h_i}\cV_{\h_1\h_2\h_3}(x_1,x_2,x_3)\,.
\ea
\ee
\end{lemma}

\begin{lemma}
\label{lem:Cauchy} 
Functions $WF^{(1)}(x_1,x_2,x_3)$ and $\cV_{\h_1\h_2\h_3}(x_1,x_2,x_3)$ coincide up to a factor. 

\end{lemma}
For explicit proofs of Lemmas see Appendix \bref{app:GGG}. Here we just outline the basic idea that  the two functions satisfy the same homogeneous Klein-Gordon equations and have the same boundary value on the conformal boundary that guarantees that they coincide. Indeed, one can consider one of three Klein-Gordon operators, e.g. $\Box_3-m_3^2$, and show that these two functions are in its kernel. On the other hand, they have the same boundary value (up to an overall factor dependent only on conformal dimensions) at $x_3 \to (\infty,z_3)$ because $\lim_{\rho_3\to\infty}e^{\rho_3\h_3}WF^{(1)}(x_1,x_2,x_3) \sim  \cV^{\text{reg}}_{\h_1\h_2\h_3}(x_1,x_2,z_3)$ as follows from  Lemma \bref{lem:KG}.

Other terms of the $3$-point Witten diagram can also be expressed in terms of the AdS vertex functions by analogy with the relation for one boundary point  \eqref{KGG_rel}. 

\begin{prop}
\label{prop:GGG} The 3-point Witten diagram is expressed in terms of the 3-point AdS vertex functions
\be 
\label{GGG_rel}
\ba{r}
\dps
\int_{\text{AdS}_2}d^2x\, G(x,x_1|\h_1) G(x,x_2|\h_2)G(x,x_3|\h_3) = \frac{\alpha(\h_1,\h_2,\h_3)}{\pref_{\h_1\h_2\h_3}}\cV_{\h_1\h_2\h_3}(x_1,x_2,x_3)
\vspace{3mm}
\\
\dps
+\sum_{n=0}^{\infty}\frac{a(\h_1;\h_2,\h_3;n)}{\pref_{\h_2+\h_3+2n\  \h_2\h_3}}\cV_{\h_2+\h_3+2n\  \h_2\h_3}(x_1,x_2,x_3)
\vspace{3mm}
\\
\dps
+\sum_{n=0}^{\infty}\frac{a(\h_2;\h_1,\h_3;n)}{\pref_{\h_1\   \h_1+\h_3+2n\ \h_3}}\cV_{\h_1 \  \h_1+\h_3+2n\ \h_3}(x_1,x_2,x_3)
\vspace{3mm}
\\
\dps
\;+\sum_{n=0}^{\infty}\frac{a(\h_3;\h_1,\h_2;n)}{\pref_{\h_1 \h_2\ \h_1+\h_2+2n}}\cV_{\h_1 \h_2\ \h_1+\h_2+2n}(x_1,x_2,x_3)\,,
\ea
\ee 
The coefficients are given by \eqref{prefactor}, \eqref{alpha_beta}, \eqref{a_coef}.
\end{prop}
The proof is given in Appendix \bref{app:GGG}. 

Let us  consider the boundary asymptotics of the $3$-point Witten diagram with the regularization factor similar to that one in the definition of the regularized AdS vertex function \eqref{regular}:
\be 
\label{KKK_asymp}
\ba{r}
\dps
\lim_{\rho_1,\rho_2,\rho_3\to\infty}e^{\rho_1 \h_1+\rho_2\h_2+\rho_3\h_3}\int_{\text{AdS}_2}d^2x\, G(x,x_1|\h_1) G(x,x_2|\h_2)G(x,x_3|\h_3) \hspace{10mm}
\vspace{3mm}
\\
\dps\equiv\int_{\text{AdS}_2}d^2x\, K(x,z_1|\h_1) K(x,z_2|\h_2)K(x,z_3|\h_3)= \frac{\alpha(\h_1,\h_2,\h_3)}{\pref_{\h_1\h_2\h_3}}\cV^{\text{reg}}_{\h_1\h_2\h_3}(z_1,z_2,z_3)
\vspace{3mm}
\\
\dps
+\sum_{n=0}^{\infty}\lim_{\rho_1\to\infty}e^{-\rho_1 (\h_2+\h_3-\h_1+2n)}\frac{a(\h_1;\h_2,\h_3;n)}{\pref_{\h_2+\h_3+2n\  \h_2\h_3}}\cV^{\text{reg}}_{\h_2+\h_3+2n\  \h_2\h_3}(z_1,z_2,z_3)
\vspace{3mm}
\\
\dps
+\sum_{n=0}^{\infty}\lim_{\rho_2\to\infty}e^{-\rho_2 (\h_1+\h_3-\h_2+2n)}\frac{a(\h_2;\h_1,\h_3;n)}{\pref_{\h_1 \  \h_1+\h_3+2n\ \h_3}}\cV^{\text{reg}}_{\h_1 \  \h_1+\h_3+2n\ \h_3}(z_1,z_2,z_3)
\vspace{3mm}
\\
\dps
\;+\sum_{n=0}^{\infty}\lim_{\rho_3\to\infty}e^{-\rho_3 (\h_1+\h_2-\h_3+2n)}\frac{a(\h_3;\h_1,\h_2;n)}{\pref_{\h_1 \h_2\ \h_1+\h_2+2n}}\cV^{\text{reg}}_{\h_1 \h_2\ \h_1+\h_2+2n}(z_1,z_2,z_3)\,.
\ea
\ee 
Note that the  boundary asymptotic of the $3$-point Witten diagram coincide with the conformal $3$-point function of three primary operators $\cO_{h_1}(z_1)$, $\cO_{h_2}(z_2)$, $\cO_{h_3}(z_3)$ (which is also the $3$-point AdS vertex function $\cV^{\text{reg}}_{\h_1\h_2\h_3}(z_1,z_2,z_3)$, see the extrapolate dictionary relation \eqref{asym3pt}) only when conformal dimensions  are subject to the triangle inequalities
\be 
\label{tringle_identity}
h_1< h_2+h_3\,,\quad
h_2< h_1+h_3\,,\quad
h_3< h_1+h_2\,.
\ee 
These restrictions also appear in the analysis of the convergence of the integral of three bulk-to-boundary propagators (see e.g. \cite{Castro:2024cmf} for a recent discussion). If the weights satisfy  other triangle identities, then the leading boundary asymptotics are given by other first terms from the three sums in \eqref{KKK_asymp}. The triangle identities \eqref{tringle_identity} correspond to choosing particular weights in the 3-point AdS vertex function which is built by intertwining the triples of $\sltwo$ modules $(\cD_{\h_1}^{\mp}, \cD_{\h_2}^{\mp}, \cD_{\h_3}^{\pm})$, see \eqref{infinite_restrictions2}.

Finally, we note that using \eqref{2pt_rel} the  relation \eqref{GGG_rel}  can be rewritten as 
\be 
\label{summary1}
\ba{c}
\dps
\pref_{\h_1\h_1}^{-1}\pref_{\h_2\h_2}^{-1}\pref_{\h_3\h_3}^{-1}\int_{\text{AdS}_2}d^2x\, \cV_{\h_1\h_1}(x,x_1) \cV_{\h_2\h_2}(x,x_2)\cV_{\h_3\h_3}(x,x_3) 
= \frac{\alpha(\h_1,\h_2,\h_3)}{\pref_{\h_1\h_2\h_3}}\cV_{\h_1\h_2\h_3}(x_1,x_2,x_3)
\vspace{3mm}
\\
\dps
+\sum_{n=0}^{\infty}\frac{a(\h_1;\h_2,\h_3;n)}{\pref_{\h_2+\h_3+2n\  \h_2\h_3}}\cV_{\h_2+\h_3+2n\  \h_2\h_3}(x_1,x_2,x_3)
\vspace{3mm}
\\
\dps
+\sum_{n=0}^{\infty}\frac{a(\h_2;\h_1,\h_3;n)}{\pref_{\h_1\ \h_1+\h_3+2n\ \h_3}}\cV_{\h_1\   \h_1+\h_3+2n\ \h_3}(x_1,x_2,x_3)
\vspace{3mm}
\\
\dps
\;+\sum_{n=0}^{\infty}\frac{a(\h_3;\h_1,\h_2;n)}{\pref_{\h_1 \h_2\ \h_1+\h_2+2n}}\cV_{\h_1 \h_2\ \h_1+\h_2+2n}(x_1,x_2,x_3)\,,
\ea
\ee
In this form it is about integral relations between the AdS vertex functions only. It would be interesting to understand this relation in inner group-theoretic  terms of the Wilson line networks.   

\section{Conclusion}
\label{sec:conclusion}

In this paper we have proven Proposition \bref{prop:HKLL} about the holographic reconstruction and then in Proposition \bref{prop:expl} explicitly calculated the $n$-point AdS vertex function as multidimensional series. Then, we have proven Propositions \bref{prop:2pt}-\bref{prop:GGG} which describe relations between the Witten diagrams and the AdS vertex functions depending on the number of points lying on the conformal  boundary. This last set of Propositions shows that there should be a relation between interacting scalar theory in \ads and Wilson line networks. We hope to consider this issue elsewhere.      

Our results can be extended in several directions. First of all, one should consider how  AdS vertex functions could be related to Witten diagrams with more endpoints. The first non-trivial case here is $n=4$ where the Wilson line network involves 4-valent intertwiner with one intermediate module while its  boundary value is given by the 4-point global conformal block which is known to admit the AdS bulk integral realization in terms of the geodesic Witten diagrams \cite{Hijano:2015zsa}. We expect that this relation can also  be extended to the corresponding AdS vertex functions and Witten diagrams in the bulk.   

On the other hand, having in mind this relation between Wilson line networks defined with respect to flat connections and a local dynamics in the bulk it would be interesting to generalize this construction to  any  theories formulated in terms of connections which equations of motion take the form of the zero-curvature conditions. In particular, since  $\sltwo$ is a chiral copy of $\sltwo\oplus \sltwo$  one can be sure that analogous  relations should hold between spinning (geodesic) Witten diagrams \cite{Nishida:2016vds,Dyer:2017zef,Sleight:2017fpc} and  Wilson line networks in $3d$ Chern-Simons gravity \cite{Bhatta:2016hpz,Besken:2016ooo,Castro:2018srf}, and, more generally, in $3d$ higher-spin Chern-Simons gravity \cite{Blencowe:1988gj}.

\vspace{3mm}

\noindent \textbf{Acknowledgements.}
We are grateful to A. Kanoda for collaboration at early stage of the project. We also thank  S.  Mandrygin, D. Ponomarev and A. Pribytok  for useful discussions. Our work was supported by the Foundation for the Advancement of Theoretical Physics and Mathematics “BASIS”. 

\appendix

\section{Hypergeometric-type functions and $3$-valent intertwiner}
\label{app:int}

Here we collect definitions of various special functions and useful identities used throughout the paper. 

\paragraph{Hypergeometric functions.}The integral representation of the Gauss hypergeometric function for $\text{Re}(c)>\text{Re}(b)>0$ is defined as \cite{Bateman:100233}
\be 
\label{hyper_integral}
\dps
{}_2F_1\left(a,b;c\big|z\right) = \frac{\Gamma(c)}{\Gamma(b)\Gamma(c-b)}\int_0^1 du\; u^{b-1}(1-u)^{c-b-1}\left(1-zu\right)^{-a}\,.
\ee 
The hypergeometric functions in different domains of convergence  can be related by the    Pfaff transformation
\be
\label{4kummer2f1}
{}_2F_1(a, b; c|z) =(1-z)^{-b}\,{}_2F_1\left(c-a, b; c| \frac{z}{z-1}\right)\,.
\ee
It allows one to derive the  Pfaff transformation in terms of series as  
\be 
\label{pfaff_series}
\sum_{k=0}^{\infty}\frac{(a)_k(b)_k}{k!(c)_k}z^k = (1-z)^{-b}\sum_{k=0}^{\infty}\frac{(c-a)_k(b)_k}{k!(c)_k}\left(\frac{z}{z-1}\right)^k \,,
\qquad \text{$|z|<\half$}\,,
\ee 
where $(p)_k={\Gamma(p+k)}/{\Gamma(p)}$ is the Pochhammer symbol. The hypergeometric function at unit argument simplifies to
\be 
\label{2f1_in_1}
\dps
{}_2F_1\left(a,b;c\big|1\right) = \frac{\Gamma(c)\Gamma(c-a-b)}{\Gamma(c-a)\Gamma(c-b)}\,,
\qquad \text{Re}(c-a-b)>0\,.
\ee 
At $c = 2a$ the hypergeometric function can be transformed as \cite{Bateman:100233} 
\be 
\label{quadratic}
\ba{l}
\dps
{}_2F_1(a, b; 2a|z)=\left(1-\frac{z}{2}\right)^{-b}{}_2F_1\left(\frac{b}{2},\frac{b}{2}+\half;a+\half\Big|\frac{z^2}{(2-z)^2}\right)\,.
\ea
\ee 
At $j\in\mathbb{Z}_-$ and $m = j-k$, where $k\in \mathbb{N}$, the following relation holds
\be 
\label{hyper_negative_int}
{}_2F_1\left(j+1,m+j+1;m+1\big|z\right) = \frac{\Gamma(-m-j)\Gamma(-j)}{\Gamma(-m)}\int_1^{\frac{1}{z}} du\; u^{m+j}(1-u)^{-j-1}\left(1-zu\right)^{-j-1}\,.
\ee 
To prove this relation one makes the following steps: 1) changes $v = \frac{(u-1)z}{(1-z)}$; 2) uses  \eqref{hyper_integral} and makes the Pfaff transformation \eqref{4kummer2f1}; 3) rewrites the obtained hypergeometric function as hypergeometric series and expands the argument $1-z$ using the Newton binomial to obtain
\be 
{}_2F_1\left(j+1,m+j+1;m+1\big|z\right) = \sum_{s=0}^\infty\,\sum_{k=2j+1+s}^\infty\,\frac{(-j+k-1)!(m-j+k-1)!(-j-1)!}{k!s!(k+2\h-1-s)!(m-j-1)!}z^s\,.
\ee 
Changing  $k = n+2j+1+s$, summing over $n$ and using  \eqref{2f1_in_1}, one finally obtains the relation \eqref{hyper_negative_int}. It can be analytically continued to any real $j$ using the Pochhammer contour around points $1$ and $\frac{1}{z}$ (see Fig. \bref{fig:poch}),
\be 
\label{hyper_negative_poch}
\ba{c}
\dps
{}_2F_1\left(j+1,m+j+1;m+1\big|z\right) = \frac{\Gamma(-m-j)\Gamma(-j)}{\Gamma(-m)}\frac{1}{(1-e^{2\pi i j})^2}
\vspace{2mm}
\\
\dps
\times\oint_{P[1,\frac{1}{z}]} \,du\; u^{m+j}(1-u)^{-j-1}\left(1-zu\right)^{-j-1}\,.
\ea
\ee 
Yet another useful representation of the hypergeometric function can be obtained by analytically  continuing  \eqref{hyper_integral} by changing $u = \frac{(x_1-v)(x_2-x_4)}{(x_1-x_2)(v-x_4)}$ and $z = \frac{(x_1-x_2)(x_3-x_4)}{(x_1-x_3)(x_2-x_4)}$, where $x_i\in\mathbb{R}$:
\be 
\label{hyper_integral_contin}
\ba{c}
\dps
{}_2F_1\left(a,b;c\big|z\right) = (-)^{b-a}\frac{(x_1-x_3)^{a} (x_1-x_2)^{1-c}(x_2-x_4)^{b}(x_1-x_4)^{c-a-b}}{(1-e^{2\pi i b})(1-e^{2\pi i (c-b)})}
\vspace{2mm}
\\
\dps\times
\frac{\Gamma(c)}{\Gamma(b)\Gamma(c-b)}\oint_{P[x_1,x_2]} du\;(v-x_1)^{b-1}(v-x_2)^{c-b-1}(v-x_3)^{-a}(v-x_4)^{a-c}\,.
\ea
\ee

The hypergeometric series ${}_3F_2(a_1, a_2, a_3; b_1, b_2| z)$ is defined as
\be
\label{def_hyper_3f2}
{}_3F_2(a_1, a_2, a_3; b_1, b_2| z)=\sum_{m=0}^{\infty}\frac{(a_1)_m (a_2)_m(a_3)_m }{(b_1)_m(b_2)_m}\frac{z^m}{m!}\,,
\qquad |z|<1\,.
\ee
At $a_1 = -k$, where $k\in\mathbb{N}$:
\be 
\label{3f2_idenity}
{}_3F_2(-k, a_2, a_3; b_1, b_2| 1) = \frac{(b_2-a_2)_k}{(b_2)_k}{}_3F_2(-k, a_2, b_1-a_3; b_1, a_2-k-b_2+1| 1)\,. 
\ee 

The series representation of the Bessel function of the first kind is given by
\be
\label{bes}
J_\alpha(z) = \sum_{k=0}^{\infty}\frac{(-)^k}{k!\Gamma(k+\alpha+1)}\left(\frac{z}{2}\right)^{2k+\alpha}\,,
\ee
where the series is convergent for any $z$.

\paragraph{Appell's functions.} The first Appell function is defined as
\be 
\label{def_appell}
F_1 \left[\begin{array}{ccc}
     a_1,&a_2, &b  \\
     &c \end{array};z_1,z_2\right] = \sum_{m_1,m_2 =0}^{\infty}\frac{(a_1)_{m_1}(a_2)_{m_2}(b)_{m_1+m_2}}{(c)_{m_1+m_2}}\frac{z_1^{m_1}}{m_1!}\frac{z_2^{m_2}}{m_2!}\,,
\ee 
where $|z_1|<1$, $|z_2|<1$ and $c\notin \mathbb{Z}_-$. 
It has the following analytically continued integral representation \cite{Bateman:100233}
\be 
\label{appell_poch}
\ba{c}
\dps
F_1 \left[\begin{array}{ccc}
     a_1,&a_2, &b  \\
     &c \end{array};z_1,z_2\right] = \frac{1}{[1-\exp(2\pi i b)]}\frac{1}{[1-\exp(2\pi i (c-b))}\frac{\Gamma(c)}{\Gamma(b)\Gamma(c-b)}
     \vspace{2mm}
     \\\dps
     \times\oint_{P[0,1]}\, du\; u^{b-1}(1-u)^{c-b-1}\left(1-z_1u\right)^{-a_1}(1-z_2u)^{-a_2}\,.
\ea
\ee 
At $c=2b$ the following relation holds
\be 
\label{Appell_1}
\ba{c}
\dps
F_1 \left[\begin{array}{ccc}
     a_1,&a_2, &a  \\
     &2a \end{array};z_1,z_2\right] = -\frac{(x_1-x_3)^{a-a_1-a_2}(x_2-x_3)^{a}(x_1-x_2)^{-2a+1}(x_1-x_4)^{a_1}(x_1-x_5)^{a_2}}{[1-\exp(2\pi i b)]^2}
\vspace{3mm}
\\
\dps
\times\,
\frac{\Gamma(2a)}{\Gamma(a)^2}\oint_{P[x_1,x_2]} dx\; (x-x_1)^{a-1}(x-x_2)^{a-1}(x-x_3)^{a_1+a_2-2a}(x-x_4)^{-a_1}(x-x_5)^{-a_2}\,,
\ea
\ee 
where arguments  $z_1$ and $z_2$ are given by 
\be 
z_1 = \frac{(x_1-x_2)(x_3-x_4)}{(x_1-x_4)(x_3-x_2)}\,,\quad
z_2 = \frac{(x_1-x_2)(x_3-x_5)}{(x_1-x_5)(x_3-x_2)}\,.
\ee 
The identity \eqref{Appell_1} can be proven by changing $u = -\frac{(x-x_1)(x_2-x_3)}{(x-x_3)(x_1-x_2)}$ in the integral representation \eqref{appell_poch}.

The second Appell function is defined as
\be 
\label{def_appell_f2}
F_2\left[\begin{array}{ccc}
     a_1,&a_2, &b  \\
     c_1, &c_2 \end{array};z_1,z_2\right] = \sum_{m_1,m_2 =0}^{\infty}\frac{(a_1)_{m_1}(a_2)_{m_2}(b)_{m_1+m_2}}{(c_1)_{m_1}(c_2)_{m_2}}\frac{z_1^{m_1}}{m_1!}\frac{z_2^{m_2}}{m_2!}\,,
\ee 
where $|z_1|+|z_2|<1$ and $c_1,c_2\notin \mathbb{Z}_-$. 

The fourth Appell function is defined as
\be 
\label{def_appell_f4}
 F_4\left[\begin{array}{cc}
 a,&b \\
  c_1, &c_2 \end{array};z_1,z_2\right] = \sum_{m_1,m_2 =0}^{\infty}\frac{(a)_{m_1+m_2}(b)_{m_1+m_2}}{(c_1)_{m_1}(c_2)_{m_2}}\frac{z_1^{m_1}}{m_1!}\frac{z_2^{m_2}}{m_2!}\,.
\ee 
The second and fourth Appell functions are related by the quadratic transformation \cite{Bailey:1938}
\be 
\label{f_2_to_f_4}
\ba{l}
\dps
F_2\left[\begin{array}{ccc}
     a_1,&a_2, &b  \\
     2a_1, &2a_2 \end{array};z_1,z_2\right] 
      \vspace{3mm}
\\
\dps
\hspace{15mm}=\left(1-\frac{z_1}{2}-\frac{z_2}{2}\right)^{-b}F_4\left[\begin{array}{cc}
 \frac{b}{2},&\frac{b}{2}+\half \\
  a_1+\half, &a_2 +\half\end{array};\frac{z_1^2}{(2-z_1-z_2)^2},\frac{z_2^2}{(2-z_1-z_2)^2}\right].
\ea
\ee 

\paragraph{Lauricella's  functions.} The direct generalization of the first Appell function is the Lauricella function D:
\be 
\label{def_lauricella}
F_D^{(k)} \left[\begin{array}{cccc}
     a_1,&... &a_k, &b  \\
     &c \end{array};z_1,...,z_k\right] = \sum_{m_1,...,m_k =0}^{\infty}\frac{(a_1)_{m_1}...(a_k)_{m_k}(b)_{m_1+...+m_k}}{(c)_{m_1+...+m_k}}\frac{z_1^{m_1}}{m_1!}...\frac{z_k^{m_k}}{m_k!}\,,
\ee 
where $|z_1|<1$,...,$|z_k|<1$. The integral representation is given by  \cite{CARLSON1963452}
\be 
\label{lauricella_poch}
\ba{c}
\dps
F_D^{(k)} \left[\begin{array}{cccc}
     a_1,&... &a_k, &b  \\
     &c \end{array};z_1,...,z_k\right] =\frac{1}{[1-\exp(2\pi i b)]}\frac{1}{[1-\exp(2\pi i (c-b))}\frac{\Gamma(c)}{\Gamma(b)\Gamma(c-b)}
\vspace{3mm}
\\
\dps
     \times\int_{P[0,1]} du\; u^{b-1}(1-u)^{c-b-1}\left(1-z_1u\right)^{-a_1}...(1-z_ku)^{-a_k}\,.
\ea
\ee 

The Lauricella function A generalizes the second Appell function:
\be 
\label{def_lauricella_A}
F_A^{(k)} \left[\begin{array}{cccc}
     a_1,&... &a_k, &b  \\
     c_1,&... &c_k \end{array};z_1,...,z_k\right] = \sum_{m_1,...,m_k =0}^{\infty}\frac{(a_1)_{m_1}...(a_k)_{m_k}(b)_{m_1+...+m_k}}{(c_1)_{m_1}...(c_k)_{m_k}}\frac{z_1^{m_1}}{m_1!}...\frac{z_k^{m_k}}{m_k!}\,,
\ee 
where $|z_1|+...+|z_k|<1$.

\paragraph{Comb function.} Both the hypergeometric function  and the second  Appell function can be viewed as  particular cases of the comb function \cite{10.1007/BF02392525,Rosenhaus:2018zqn}:
\be 
\label{def_comb}
\ba{l}
F_N \left[\begin{array}{ccccc}
     a_1,&b_1,&...\,,&b_{N-1}, &a_2 \\
     &c_1,&...\,, &c_N \end{array};z_1,...\,,z_N\right] 
\vspace{2mm}
\\ 
\dps
= \sum_{m_1,...,m_N=0}^{\infty}\frac{(a_1)_{m_1}(b_1)_{m_1+m_2}\dots(b_{N-1})_{m_{N-1}+m_N}(a_2)_{m_N}}{(c_1)_{m_1}\dots(c_N)_{m_N}}\frac{z_1^{m_1}}{m_1!}\cdots \frac{z_n^{m_N}}{m_N!}\,.
\ea 
\ee 
\paragraph{Useful identity:}  
\be
\label{useful_identity}
\ba{l}
\dps
\sum_{n=0}^{k}\frac{(-)^n\Gamma(a+n-\frac{1}{2})}{n!(k-n)!\Gamma(a+n+k+\half)}\,\frac{a+2n-\half}{b(b-1)-(a+2n)(a+2n-1)}
\vspace{2mm}
\\ 
\dps
\hspace{51mm}=-\frac{1}{4}\frac{\Gamma(\frac{a-b}{2})\Gamma(\frac{a+b-1}{2})}{\Gamma(\frac{a-b}{2}+k+1)\Gamma(\frac{a+b+1}{2}+k)}\,,
\qquad
k\in\mathbb{N}\,.
\ea
\ee 

\paragraph{$3$-valent intertwiner.}
The intertwiner for any three modules from the negative discrete series is given by 
$$ 
\ba{l}
\dps
[I_{\h_1\h_2\h_3}]^{m_1}{}_{m_2m_3}  = \delta_{m_1,m_2+m_3}(-)^{m_1-m_3-\h_2}
\vspace{2mm}
\\
\dps
\times\Bigg[\frac{\Gamma(1+\h_1-\h_2-\h_3)\Gamma(\h_3-\h_2-\h_1+1)\Gamma(1+\h_2-\h_1-\h_3)\Gamma(1+m_1-\h_1)\Gamma(1-m_1-\h_1)}{\Gamma(2-\h_1-\h_2-\h_3)}\Bigg]^{\half}
\ea
$$
\be
\label{general-intertwiner}
\ba{l}
\dps
\times\sum_{k\in \tilde{K}}\frac{\sqrt{\Gamma(-m_3-\h_3+1)\Gamma(-m_2-\h_2+1)\Gamma(1+m_2-\h_2)\Gamma(1+m_3-\h_3)}}{\Gamma(m_2+\h_1-\h_3+k+1)\Gamma(m_1-\h_3+\h_2+k+1)}\vspace{2mm}
\\
\dps
\times \frac{1}{k!\Gamma(1-\h_1-m_1-k)\Gamma(1+\h_2-m_2-k)\Gamma(\h_3-\h_2-\h_1-k+1)}\; ,
\ea
\ee 
where $\tilde{K} = [\![\max(0, -m_2-\h_1+\h_3),\min(-\h_2-m_2, -\h_1-m_1)]\!]$ \cite{HOLMAN19661}. The intertwiner for any (in)finite modules is defined as
$$ 
\ba{l}
\dps
[I_{\h_1\h_2\h_3}]^{m_1}{}_{m_2m_3}  = \delta_{m_1,m_2+m_3}(-)^{-\h_2-m_2}
\vspace{2mm}
\\
\dps
\times\Bigg[\frac{\Gamma(\h_1-\h_2+\h_3)\Gamma(\h_3-\h_2-\h_1+1)\Gamma(m_1-\h_1+1)\Gamma(m_1+\h_1)\Gamma(-m_2+\h_2)}{\Gamma(\h_1+\h_2+\h_3-2)\Gamma(-m_2-\h_2+1)\Gamma(1+m_3-\h_3)\Gamma(m_3+\h_3)\Gamma(\h_2+\h_3-\h_1)}\Bigg]^{\half}
\vspace{2mm}  
\\
\dps
\times\lim_{x\to1}e^{-i\pi(\h_1+\h_2+\h_3)}\left[(1-e^{-4\pi i\h_3})(1-e^{-2\pi i(\h_3+m_3)})(1-e^{2\pi i(m_1-\h_1)})(1-e^{-2\pi i(\h_3+m_3)})\right.
\ea
$$

\vspace{-5mm}

\be
\label{intertwiner_integral}
\ba{l}
\times\left.(1-e^{2\pi i(m_2-\h_1-\h_3)})\Gamma(\h_1+\h_3-\h_2)\Gamma(1-\h_3-m_3)\Gamma(1-\h_1+m_1)\right]^{-1}
\vspace{2mm}
\\
\dps
\times \oint_{P[0,1]}\,dv\, v^{m_2-\h_1-\h_3+1}(1-v)^{\h_2-\h_3+\h_1-1}[1-x(1-v)]^{\h_1+\h_3-\h_2-1}
\vspace{2mm}
\\
\dps
\times\oint_{P[0,1]}\,du\, u^{-\h_3-m_3}(1-u)^{m_1-\h_1}[1-uv]^{-\h_2-m_2}\,,
\ea
\ee 
where $P[0,1]$ is the Pochhammer contour shown in Fig. \bref{fig:poch} \cite{HOLMAN19661}. 
This general formula allows one to obtain the intertwiner for particular $\sltwo$ modules by  restricting  weights $\h_i$ and magnetic numbers $m_i$, integrating over $u$ and $v$ and then taking  the limit $x\to1$.

\section{Calculating the $n$-point HKLL-type integral}
\label{app:n_point}

In order to calculate the integral in the holographic reconstruction formula \eqref{HKLL_vertex} one substitutes the $n$-point global conformal block  \eqref{comb_fin}, expands the comb function into series by using \eqref{def_comb}, and then makes the change 
\be 
u_i \to v_i=\frac{u_i-\bar{\pt}_i}{\pt_i-\bar{\pt}_i}\,,
\ee
to obtain 
\be 
\label{start_np}
\ba{l}
\dps
\hspace{-2mm}\cV_{ \h\tilde{\h}}({\bf x})=\hspace{-5mm}\sum_{m_1,...,m_{n-3}=0}^{\infty}\hspace{-4mm}E_{\h\tilde{\h}}^{m_i}  \prod_{k = 1}^nK_{\h _k}\oint_{P[0,1]}dv_k\;v_k^{\h_k-1}(1-v_k)^{\h_k-1}\frac{(\pt_k-\bar{\pt}_k)^{\h_k}}{(2i)^{1-\h_k}}
\prod_{l=1}^{n-1}(v'_{l\,l+1})^{\sigma_l}(v'_{l\,l+2})^{\gamma_l},
\ea
\ee
where we defined the auxiliary variables: 
\be 
\label{npoint_aux}
\ba{l}
\dps
v'_{ij} := v_i(\pt_i-\bar{\pt}_i)-v_j(\pt_j-\bar{\pt}_j)+(\bar{\pt}_i-\bar{\pt}_j)\,,
\vspace{2.5mm}  
\\
\dps
\sigma_i:=\tilde{\h}_{i-2}+\tilde{\h}_{i}-\h_i-\h_{i+1}+m_{i-2}+m_{i}\,,
\vspace{2.5mm}  
\\
\dps
\gamma_i:=\h_{i+1}-\tilde{\h}_{i-1}-\tilde{\h}_{i}-m_{i-1}-m_{i}\,,
\vspace{2.5mm}  
\\
E_{\h\tilde{\h}}^{m_i} := \dps \pref_{\h\tilde{\h}}
\frac{( \h_1+\tilde{\h}_1-\h_2)_{m_1}(\tilde{\h}_1+\tilde{\h}_2-\h_3)_{m_1+m_2}\cdots(\h_n+\tilde{\h}_{n-3}-\h_{n-1})_{m_{n-3}}}{m_1!(2\tilde{\h}_1)_{m_1}\cdots m_{n-3}!(2\tilde{\h}_{n-3})_{m_{n-3}}}\,.
\ea
\ee 
To make  \eqref{start_np} more compact and take into account the fact that in the case $n=2,3$ the comb function in the expression of the global conformal block \eqref{comb_fin} equals $1$ we use the following conventions which are also used in further calculations:
\begin{itemize}
\item if the index of $m_i$ in  \eqref{start_np} satisfy $i\leq0$ or $i>n-3$, then one sets $m_i=0$ and omits the sum over this variable;

\item if a power of a variable contains $\tilde{\h}_i$ with $i<0$ or $i>n-2$, then $\tilde{\h}_i=0$. If $i=0$ or $i=n-2$, then $\tilde{\h}_0 = \h_1$ and $\tilde{\h}_{n-2} = \h_n$.
\end{itemize}

We sequentially evaluate  integrals over $v_i$ in  \eqref{start_np}. One  starts with the integral over $v_1$ and notes that it is the integral representation of the first Appell function \eqref{appell_poch}
\be 
\label{v_1_integral}
\ba{l}
\dps
K_{\h _1}\oint_{P[0,1]}dv_1\,v_1^{\h_1-1}(1-v_1)^{\h_1-1}(v'_{12})^{\sigma_1}(v'_{13})^{\gamma_1} 
\vspace{2.5mm}  
\\
\dps
\hspace{10mm}= (-)^{\h_1}(2i)^{-2\h_1+1}(s'_{12})^{\sigma_1}
(s'_{13})^{\gamma_1}F_1 \left[\begin{array}{ccc}
     -\sigma_1,&-\gamma_1, &\h_1  \\
     &2\h_1 \end{array};\frac{\bar{\pt}_1-\pt_1}{s'_{12}},\frac{\bar{\pt}_1-\pt_1}{s'_{13}}\right],
\ea
\ee 
where 
\be
s'_{ij} := -v_j(\pt_j-\bar{\pt}_j)+(\bar{\pt}_i-\bar{\pt}_j)\,.
\ee 
Then, using the first Appell series \eqref{def_appell} with $m_1 = k_{12}$ and $m_2=k_{13}$ one obtains
\be 
\ba{c}
\dps
\hspace{-2mm}\cV_{ \h\tilde{\h}}({\bf x})=
\hspace{-2mm}\sum_{m_1,...,m_{n-3},k_{12},k_{13}=0}^{\infty}{}_{(1)}E_{\h\tilde{\h}}^{m_i,k_{il}}  \prod_{k = 2}^nK_{\h _k}\oint_{P[0,1]}dv_k\;v_k^{\h_k-1}(1-v_k)^{\h_k-1}\frac{(\pt_k-\bar{\pt}_k)^{\h_k}}{(2i)^{1-\h_k}}
\vspace{2.5mm}  
\\
\dps\times
(2i)^{-\h_1}(\bar{\pt}_1-\pt_1)^{\h_1+k_{12}+k_{13}}(s'_{12})^{\sigma_1-k_{12}}
(s'_{13})^{\gamma_1-k_{13}}\prod_{l=2}^{n-1}(v'_{l\,l+1})^{\sigma_l}(v'_{l\,l+2})^{\gamma_l}\,,
\ea
\ee
where the coefficient is introduced 
\be 
\label{Ec1}
{}_{(1)}E_{\h\tilde{\h}}^{m_i,k_{il}} := E_{\h\tilde{\h}}^{m_i}\frac{(-\sigma_1)_{k_{12}}(-\gamma_1)_{k_{13}}(\h_1)_{k_{1 2}+k_{13}}}{k_{12}!k_{13}!(2\h_1)_{k_{12}+k_{13}}}\,.
\ee
To evaluate the integral over $v_2$ one represents it as the Lauricella function $F_D^{(3)}$ \eqref{lauricella_poch}:
\be 
\label{v_2_integral}
\ba{c}
\dps
K_{\h _2}\oint_{P[0,1]}dv_2\,v_2^{\h_2-1}(1-v_2)^{\h_2-1}(s'_{12})^{\sigma_1-k_{12}}(v'_{23})^{\sigma_2}(v'_{24})^{\gamma_2} = (-)^{\h_2}(2i)^{-2\h_2+1}(\bar{\pt}_1-\bar{\pt}_2)^{\sigma_1-k_{12}}
\vspace{2.5mm}  
\\
\dps
\times
(s'_{23})^{\sigma_2}
(s'_{24})^{\gamma_2}F_D^{(3)} \left[\begin{array}{cccc}
     k_{12}-\sigma_1,&-\sigma_2,&-\gamma_2, &\h_2  \\
     &2\h_2 \end{array};\frac{\bar{\pt}_2-\pt_2}{\bar{\pt}_2-\bar{\pt}_1},\frac{\bar{\pt}_2-\pt_2}{s'_{23}},\frac{\bar{\pt}_2-\pt_2}{s'_{24}}\right],
\ea
\ee 
and expands it into series using \eqref{def_lauricella} with $m_1=k_{21}, m_2=k_{23}$ and $m_3=k_{24}$:
\be 
\label{n-point_int}
\ba{c}
\dps
\hspace{-2mm}\cV_{ \h\tilde{\h}}({\bf x})=
\hspace{-2mm}\sum_{\substack{m_1,...,m_{n-3}=0\\k_{12},k_{13},k_{21},k_{23},k_{24}=0}}^{\infty}{}_{(2)}E_{\h\tilde{\h}}^{m_i,k_{il}} \prod_{k = 3}^nK_{\h _k}\oint_{P[0,1]}dv_k\;v_k^{\h_k-1}(1-v_k)^{\h_k-1}\frac{(\pt_k-\bar{\pt}_k)^{\h_k}}{(2i)^{1-\h_k}}
\vspace{2.5mm}  
\\
\dps\times
(2i)^{-\h_1-\h_2} (\bar{\pt}_1-\pt_1)^{\h_1+k_{12}+k_{13}} (\bar{\pt}_2-\pt_2)^{\h_2+k_{21}+k_{23}+k_{24}} (\bar{\pt}_1-\bar{\pt}_2)^{\sigma_1-k_{12}-k_{21}} 
\ea
\ee
$$
\ba{l}
\dps
\times
(s'_{13})^{\gamma_1-k_{13}} 
(s'_{23})^{\sigma_2-k_{23}}
(s'_{24})^{\gamma_2-k_{24}}
\prod_{l=3}^{n-1}(v'_{l\,l+1})^{\sigma_l}(v'_{l\,l+2})^{\gamma_l}\,,
\ea
$$
where 
\be 
\ba{l}
\dps
{}_{(2)}E_{\h\tilde{\h}}^{m_i,k_{il}} := (-)^{k_{21}}{}_{(1)}E_{\h\tilde{\h}}^{m_i,k_{il}}\frac{(\h_2)_{k_{21}+k_{23}+k_{24}}}{k_{21}!k_{23}!k_{24}!(2\h_2)_{k_{21}+k_{23}+k_{24}}}
(k_{12}-\sigma_1)_{k_{21}}(-\sigma_2)_{k_{23}}(-\gamma_2)_{k_{24}}\,,
\ea
\ee 
cf. \eqref{Ec1}.  To proceed further we use the mathematical induction. Suppose that after evaluating $p-1$ integrals the $n$-point AdS vertex function \eqref{start_np} is given by
\be 
\label{n-point_int_2}
\ba{l}
\dps
\hspace{-2mm}\cV_{ \h\tilde{\h}}({\bf x})=
\hspace{-10mm}\sum_{\substack{m_1,...,m_{n-3}=0\\\{k_{i\,i-2},k_{i\,i-1},k_{i\,i+1},k_{i\,i+2}=0\}_{i=1,..., p-1}}}^{\infty}\hspace{-10mm}{}_{(p-1)}E_{\h\tilde{\h}}^{m_i,k_{il}} \prod_{k = p}^nK_{\h _k}\oint_{P[0,1]}dv_k\;v_k^{\h_k-1}(1-v_k)^{\h_k-1}\frac{(\pt_k-\bar{\pt}_k)^{\h_k}}{(2i)^{1-\h_k}}
\vspace{2.5mm}  
\\
\dps\times\prod_{s = 1}^{p-1}\left[
(2i)^{-\h_s} (\bar{\pt}_s-\pt_s)^{\h_s+k_{s\,s-2}+k_{s\,s-1}+k_{s\,s+1}+k_{s\,s+2}}\right] \prod_{s = 1}^{p-2}(\bar{\pt}_s-\bar{\pt}_{s+1})^{\sigma_s-k_{s\,s+1}-k_{s+1\,s}}  
\vspace{2.5mm}  
\\
\times
\prod_{s = 1}^{p-3}\left[(\bar{\pt}_s-\bar{\pt}_{s+2})^{\gamma_s-k_{s\,s+2}-k_{s+2\,s}} \right](s'_{p-2\,p})^{\gamma_{p-2}-k_{p-2\,p}} (s'_{p-1\,p})^{\sigma_{p-1}-k_{p-1\,p}}(s'_{p-1\,p+1})^{\gamma_{p-1}-k_{p-1\,p+1}}
\vspace{2.5mm}  
\\
\dps\times
\prod_{l=p}^{n-1}(v'_{l\,l+1})^{\sigma_l}(v'_{l\,l+2})^{\gamma_l}\,,
\ea
\ee
where the coefficient ${}_{(p-1)}E_{\h\tilde{\h}}^{m_i,k_{il}}$ satisfies the following recurrence relation
\be 
\label{coeff_recursion}
\ba{c}
\dps
{}_{(n)}E_{\h\tilde{\h}}^{m_i,k_{il}} := {}_{(n-1)}E_{\h\tilde{\h}}^{m_i,k_{il}}\frac{(-)^{k_{n\,n-1}+k_{n\,n-2}}(\h_n)_{k_{n\,n-2}+k_{n\,n-1}+k_{n\,n+1}+k_{n\,n+2}}}{k_{n\,n-2}!k_{n\,n-1}!k_{n\,n+1}!k_{n\,n+2}!(2\h_n)_{k_{n\,n-2}+k_{n\,n-1}+k_{n\,n+1}+k_{n\,n+2}}}
\vspace{2.5mm}  
\\
\dps
\times
(k_{n-2\,n}-\gamma_{n-2})_{k_{n\,n-2}}
(k_{n-1\,n}-\sigma_{n-1})_{k_{n\,n-1}}
(-\sigma_n)_{k_{n\,n+1}}
(-\gamma_n)_{k_{n\,n+2}}
\,.
\ea
\ee
To make the expression \eqref{n-point_int_2} more compact one  introduces the new convention: if the indices of  $k_{ij}$ in  \eqref{n-point_int_2} satisfy one of the following four inequalities: $i\leq0$, $j\leq0$, $i>n$, $j>n$, then one sets $k_{ij}=0$ and the sum over this variable is omitted. The base of the induction is already shown in \eqref{n-point_int}. To prove the induction step we show that after evaluating the integral over $v_p$ in \eqref{n-point_int_2} the result can be written in the form \eqref{n-point_int_2} with $p$ being changed to $p+1$. To evaluate the integral over $v_p$ one uses the integral representation of the Lauricella function $F_D^{(4)}$  \eqref{lauricella_poch}:\footnote{Note that the integral over $v_2$ \eqref{v_2_integral} is a special case of this formula with the first parameter $k_{p-2\,p}-\gamma_{p-2}$ in \eqref{v_p_integral}  being taken as zero: in this case the  Lauricella function $F_D^{(4)}$ reduces to $F_D^{(3)}$. The equality $k_{p-2\,p}-\gamma_{p-2} = 0$ at $p = 2$ directly follows from the conventions introduced below \eqref{npoint_aux} and \eqref{coeff_recursion}.}
$$
\ba{c}
\dps
K_{\h _p}\oint_{P[0,1]}dv_p\,v_p^{\h_p-1}(1-v_p)^{\h_p-1}(s'_{p-2\,p})^{\gamma_{p-2}-k_{p-2\,p}}(s'_{p-1\,p})^{\sigma_{p-1}-k_{p-1\,p}}(v'_{p\,p+1})^{\sigma_p}(v'_{p\,p+2})^{\gamma_p} 
\vspace{2.5mm}  
\\
\dps
= (-)^{\h_p}(2i)^{-2\h_p+1}(\bar{\pt}_{p-2}-\bar{\pt}_p)^{\gamma_{p-2}-k_{p-2\,p}}(\bar{\pt}_{p-1}-\bar{\pt}_p)^{\sigma_{p-1}-k_{p-1\,p}}(s'_{p\,p+1})^{\sigma_p}
(s'_{p\,p+2})^{\gamma_p}
\ea
$$
\be 
\label{v_p_integral}
\ba{c}

\times F_D^{(4)} \left[\begin{array}{ccccc}
      k_{p-2\,p}-\gamma_{p-2},&k_{p-1\,p}-\sigma_{p-1},&-\sigma_p,&-\gamma_p, &\h_p  \\
     &2\h_p \end{array};\frac{\bar{\pt}_p-\pt_p}{\bar{\pt}_p-\bar{\pt}_{p-2}},\frac{\bar{\pt}_p-\pt_p}{\bar{\pt}_p-\bar{\pt}_{p-1}},\frac{\bar{\pt}_p-\pt_p}{s'_{p\,p+1}},\frac{\bar{\pt}_p-\pt_p}{s'_{p\,p+2}}\right],
\ea
\ee 
then expands $F_D^{(4)}$ into series \eqref{def_lauricella} with $m_1 = k_{p\,p-2}, m_2 = k_{p\,p-1}, m_3 = k_{p\,p+1}, m_4 = k_{p\,p+2}$ and obtains
$$
\ba{l}
\dps
\hspace{-2mm}\cV_{ \h\tilde{\h}}({\bf x})=
\hspace{-6mm}\sum_{\substack{m_1,...,m_{n-3}=0\\\{k_{i\,i-2},k_{i\,i-1},k_{i\,i+1},k_{i\,i+2}=0\}_{i=1,..., p}}}^{\infty}\hspace{-3mm} \prod_{k = p+1}^nK_{\h _k}\oint_{P[0,1]}dv_k\;v_k^{\h_k-1}(1-v_k)^{\h_k-1}\frac{(\pt_k-\bar{\pt}_k)^{\h_k}}{(2i)^{1-\h_k}}
\vspace{2.5mm}  
\\
\dps\times\prod_{s = 1}^{p-1}\left[
(2i)^{-\h_s} (\bar{\pt}_s-\pt_s)^{\h_s+k_{s\,s-2}+k_{s\,s-1}+k_{s\,s+1}+k_{s\,s+2}}\right] \prod_{s = 1}^{p-2}(\bar{\pt}_s-\bar{\pt}_{s+1})^{\sigma_s-k_{s\,s+1}-k_{s+1\,s}}  
\vspace{2.5mm}  
\\
\dps
\times
\prod_{s = 1}^{p-3}\left[(\bar{\pt}_s-\bar{\pt}_{s+2})^{\gamma_s-k_{s\,s+2}-k_{s+2\,s}} \right](\bar{\pt}_{p-2}-\bar{\pt}_p)^{\gamma_{p-2}-k_{p-2\,p}-k_{p\,p-2}} (\bar{\pt}_{p-1}-\bar{\pt}_p)^{\sigma_{p-1}-k_{p-1\,p}-k_{p\,p-1}}
\ea
$$
\be 
\label{n-point_int_3}
\ba{l}
\dps
\times
(s'_{p-1\,p+1})^{\gamma_{p-1}-k_{p-1\,p+1}}(s'_{p\,p+1})^{\sigma_{p}-k_{p\,p+1}}(s'_{p\,p+2})^{\gamma_{p}-k_{p\,p+2}}
\prod_{l=p+1}^{n-1}(v'_{l\,l+1})^{\sigma_l}(v'_{l\,l+2})^{\gamma_l}
\vspace{2.5mm}  
\\
\dps
\times
{}_{(p-1)}E_{\h\tilde{\h}}^{m_i,k_{il}}\frac{(-)^{k_{p\,p-1}+k_{p\,p-2}}(\h_p)_{k_{p\,p-2}+k_{p\,p-1}+k_{p\,p+1}+k_{p\,p+2}}}{k_{p\,p-2}!k_{p\,p-1}!k_{p\,p+1}!k_{p\,p+2}!(2\h_p)_{k_{p\,p-2}+k_{p\,p-1}+k_{p\,p+1}+k_{p\,p+2}}}
\vspace{3mm}  
\\
\dps
\times
(k_{p-2\,p}-\gamma_{p-2})_{k_{p\,p-2}}
(k_{p-1\,p}-\sigma_{p-1})_{k_{p\,p-1}}
(-\sigma_p)_{k_{p\,p+1}}
(-\gamma_p)_{k_{p\,p+2}}
\,.
\ea
\vspace{1mm}  
\ee
Note that the coefficient in the last two lines can be represented as ${}_{(p)}E_{\h\tilde{\h}}^{m_i,k_{il}}$ which follows from the recursion relation \eqref{coeff_recursion}. Using this observation we conclude that the obtained expression reproduces \eqref{n-point_int_2} with $p$ being changed to $p+1$  that  establishes the induction step. To obtain the series representation of the $n$-point AdS vertex function one substitutes $p=n+1$ in \eqref{n-point_int_2} which corresponds to the case where all $n$ integrals in the HKLL-type representation \eqref{start_np} are evaluated. Then one applies the recurrence relation \eqref{coeff_recursion} to the coefficient ${}_{(n)}E_{\h\tilde{\h}}^{m_i,k_{il}}$ $n$ times and uses the identity  $(a)_k(a+k)_m = (a)_{k+m}$:
\be 
\label{n-point_final}
\ba{l}
\dps
\cV_{ \h\tilde{\h}}({\bf x})=\pref_{\h\tilde{\h}}(2i)^{-\sum_{i=1}^n\h_i}\hspace{-10mm}\sum_{\substack{m_1,...,m_{n-3}=0\\\{k_{i\,i-2},k_{i\,i-1},k_{i\,i+1},k_{i\,i+2}=0\}_{i=1,..., n}}}^{\infty}\prod_{i=1}^{n}(\bar{\pt}_i-\pt_i)^{k_{i\,i-2}+k_{i\,i-1}+k_{i\,i+1}+k_{i\,i+2}+\h_i}
\vspace{2.5mm}  
\\
\dps\times
\prod_{l=1}^{n-1}(-)^{k_{l+1\,l}+k_{l+2\,l}}(\bar{\pt}_l-\bar{\pt}_{l+1})^{\sigma_l-k_{l+1\,l}-k_{l\,l+1}}
(\bar{\pt}_l-\bar{\pt}_{l+2})^{\gamma_l-k_{l+2\,l}-k_{l\,l+2}}
\vspace{2.5mm}  
\\
\dps\times\prod_{s=1}^{n}\frac{(\h_s)_{k_{s\,s-2}+k_{s\,s-1}+k_{s\,s+1}+k_{s\,s+2}}}{k_{s\,s-2}!k_{s\,s-1}!k_{s\,s+1}!k_{s\,s+2}!m_s!(2\h_s)_{k_{s\,s-2}+k_{s\,s-1}+k_{s\,s+1}+k_{s\,s+2}}(2\tilde{\h}_s)_{m_s}}
\vspace{2.5mm}  
\\
\dps\times\prod_{t=1}^{n-1}(-\sigma_t)_{k_{t+1\,t}+k_{t\,t +1}}(-\gamma_t-m_{t-1}-m_t)_{m_{t-1}+m_t+k_{t+2\,t}+k_{t\,t +2}}\,.
\ea
\ee 
Substituting $\sigma_i$ and $\gamma_i$ \eqref{npoint_aux} into \eqref{n-point_final} we obtain the final expression \eqref{n-point_short}. One can see that at $n=3$ the final result \eqref{n-point_final} coincides with the first representation of the  $3$-point AdS vertex function \eqref{3p_first_rep}.

\section{Propagators for free scalar fields in AdS$_2$}
\label{app:bb}

In two dimensions, the bulk-to-bulk propagator in  Poincare coordinates \eqref{metric}  is given by \cite{Fronsdal:1974ew} 
\be 
\label{bulk-to-bulk}
G(x,x'|\h) = \left(\frac{\xi(x,x')}{2}\right)^{\h}{}_2F_1\left(\frac{\h}{2},\frac{\h}{2}+\half; \h+\half|\xi(x,x')^2\right)\,,
\ee
where the AdS invariant distance
\be
\label{AdS_invariant_dist}
\xi(x,x') := \frac{2e^{-\rho-\rho'}}{e^{-2\rho}+e^{-2\rho'}+(z-z')^2}
\ee 
can be expressed in terms of the \ads geodesic distance $\sigma(x_1,x_2)$ as 
\be 
\frac{1}{\xi(x_1,x_2)} = \cosh(\sigma(x_1,x_2))\,.
\ee
The bulk-to-boundary propagator is obtained by placing one of the bulk points on the conformal boundary and regularising the result
\be 
\label{bulk-to-boundary}
K(x,z'|\h) = \chi(x,z')^{\h} = \frac{e^{-\h\rho}}{(e^{-2\rho} + (z-z')^2)^\h}\,,
\ee 
where we introduced a regularized  invariant distance
\be
\label{regul_distance}
\chi(x,z'):=\lim_{\rho'\to\infty}e^{\rho'}\xi(x,x') = \frac{e^{-\rho}}{e^{-2\rho} + (z-z')^2}\,.
\ee

\section{Calculations and proofs}
\label{app:details}

\subsection{Wilson matrix elements in integral form}
\label{app:wils_integ}
At $\forall \h\in\mathbb{R}$ one  applies the Pfaff transformation \eqref{4kummer2f1} twice to the hypergeometric functions in the Wilson matrix elements \eqref{closed_left0}  and uses the integral representation  \eqref{hyper_negative_poch}:
\be 
\label{matrix_integral_rep_interm}
\ba{c}
\dps
\braket{a^-|W^-_\h[0,x]|\h,m} =  \frac{(-)^{-\h}(-\h)!}{(\h-1)!}\left[\frac{1}{(m-\h)!(-\h-m)!(-2\h)!}\right]^\half(-2i)^{-2\h+1}\, 
e^{-\rho m}\,(q+i)^{\h-1}(q-i)^{m}
\vspace{3mm}
\\
\dps
\times\frac{1}{(1-e^{-2\pi i \h})^2}\oint_{P[1,\frac{q-i}{q+i}]}ds \; (1-s)^{\h-1}s^{m-\h}\left(1-\left(\frac{q-i}{q+i}\right)s\right)^{\h-1},
\vspace{8mm}
\\
\dps
\braket{\h,m|W^-_\h[x,0]|a^-} = \frac{ (-)^{-\h-m}(-\h)!}{(\h-1)!}\left[\frac{1}{(m-\h)!(-\h-m)!(-2\h)!}\right]^\half(-2i)^{-2\h+1}\, 
e^{\rho m}\,(q+i)^{-m+\h-1}
\vspace{3mm}
\\
\dps
\times\frac{1}{(1-e^{-2\pi i \h})^2}\oint_{P[1,\frac{q-i}{q+i}]}dt \; (1-t)^{\h-1}t^{m-\h}\left(1-\left(\frac{q-i}{q+i}\right)t\right)^{\h-1},
\ea
\ee 
where $P[1,\frac{q-i}{q+i}]$ is the Pochhammer contour around the points $1$ and $\frac{q-i}{q+i}$, see Fig. \bref{fig:poch}. Changing  $u = -s(q-i)e^{-\rho}$, $v = -\frac{q+i}{t}e^{-\rho}$ and using the  identity $\Gamma(x)\Gamma(1-x) = {\pi}/{\sin(\pi x)}$ one obtains
\be 
\label{integral_rep_inf}
\ba{l}
\dps
\braket{a^-|W^-_\h[0,x]|\h,m} = (-)^{m-\h} K_\h\left[\frac{(-2\h)!}{(m-\h)!(-\h-m)!}\right]^\half\oint_{P[\pt,\bar{\pt}]}du \, ((u-z)^2e^\rho+e^{-\rho})^{\h-1}u^{m-\h}\, ,
\vspace{8mm}
\\
\dps
\braket{\h,m|W^-_\h[x,0]|a^-} = K_\h\left[\frac{(-2\h)!}{(m-\h)!(-\h-m)!}\right]^\half 
\oint_{P[\pt,\bar{\pt}]}dv \,((v-z)^2e^\rho+e^{-\rho})^{\h-1}\,v^{-\h-m}\, ,
\ea
\ee 
where the prefactor and $w, \bar w$ in the Pochhammer contour $P[w, \bar w]$ are given by  
\be 
\label{notation1}
K_\h = \frac{(-\h)!^2}{(-2\h)!}\frac{(-2i)^{-2\h+1}}{\pi(1-e^{-2\pi i \h})} \,,
\qquad  
\pt = z+ie^{-\rho}\,, \quad \bar \pt = z-ie^{-\rho}\,.
\ee 
We separated the coefficient $\left[\frac{(-2\h)!}{(m-\h)!(-\h-m)!}\right]^\half$ from $K_\h$ in the final result to further interpret the factor $\left[\frac{(-2\h)!}{(m-\h)!(-\h-m)!}\right]^\half u^{m-\h}$ as a boundary value of the Wilson matrix element, see Section \bref{sec:matrix}.

\subsection{Calculating the $2$-point AdS vertex function}
\label{app:2p_simpl}
In the $n=2$ case the AdS vertex function \eqref{n-point_short} takes the form 
\be
\ba{c}
\dps
\cV_{\h_1 \h_2}({\bf x})= \pref_{\h_1\h_2}(2i)^{-2\h_1}\sum_{k_{12},k_{21}=0}^{\infty}(\bar{\pt}_{1}-\pt_1)^{k_{12}+\h_1}(\pt_2-\bar{\pt}_{2})^{k_{21}+\h_1}(\bar{\pt}_{1}-\bar{\pt}_{2})^{-2\h_1-k_{12}-k_{21}}
\vspace{3mm}
\\
\dps
\times \frac{(2\h_1)_{k_{12}+k_{21}}(\h_1)_{k_{12}}(\h_1)_{k_{21}}}{k_{12}!k_{21}!(2\h_1)_{k_{12}}(2\h_1)_{k_{21}}}\,.
\ea
\ee
Applying the Pfaff transformation \eqref{pfaff_series} to the sum over $k_{12}$ one obtains 
\be
\ba{c}
\dps
\cV_{\h_1 \h_2}({\bf x})= \pref_{\h_1\h_2}(2i)^{-2\h_1}\sum_{k_{12},k_{21}=0}^{\infty}(\bar{\pt}_{1}-\pt_1)^{k_{12}+\h_1}(\pt_2-\bar{\pt}_{2})^{k_{21}+\h_1}(\bar{\pt}_{1}-\bar{\pt}_{2})^{-\h_1-k_{21}}
\vspace{3mm}
\\
\dps
\times (-)^{k_{12}}(\bar{\pt}_{2}-\pt_{1})^{-\h_1-k_{12}}\frac{(\h_1)_{k_{12}}(\h_1)_{k_{21}}}{k_{12}!(2\h_1)_{k_{12}}(k_{21}-k_{12})!}\,.
\ea
\ee
Making change $k_{21}\to k_{21}' =  k_{21}-k_{12}$, summing over $k_{21}'$, and  using the generalized Newton binomial yields
\be
\cV_{\h_1 \h_2}({\bf x})= \pref_{\h_1\h_2}(2i)^{-2\h_1}\sum_{k_{12}=0}^{\infty}\left(\frac{(\bar{\pt}_{1}-\pt_1)(\bar{\pt}_{2}-\pt_2)}{(\bar{\pt}_{1}-\pt_{2})(\bar{\pt}_{2}-\pt_{1})}\right)^{k_{12}+\h_1}\frac{(\h_1)_{k_{12}}(\h_1)_{k_{12}}}{k_{12}!(2\h_1)_{k_{12}}}\,.
\ee
Then, one sums  over $k_{12}$ by means of the hypergeometric series, applies  the Pfaff transformation \eqref{4kummer2f1}, and uses the following relation between the invariant variables $c_{ij}$ \eqref{integrals_of_motion} and the variables $\pt_i,\pt_j$,
\be 
\label{boundary_points}
c_{ij} = -4\frac{(\pt_i-\pt_{j})(\bar{\pt}_i-\bar{\pt}_{j})}{(\pt_i-\bar{\pt}_i)(\pt_{j}-\bar{\pt}_{j})}\,,
\ee 
to obtain the final result
\be
\cV_{\h_1 \h_2}({\bf x})= \pref_{\h_1\h_2}c_{12}^{-\h_1}\F\left(\h_1,\h_1;2\h_1|-\frac{4}{c_{12}}\right).
\ee
The resulting $2$-point AdS vertex function is further analyzed in Section \bref{sec:2pt}.

\subsection{Second representation of the $3$-point AdS vertex function}
\label{app:3p_second}

To calculate the second representation of the $3$-point AdS vertex function it is convenient to start from the integral representation \eqref{HKLL_vertex}:
\be 
\label{3p_integral}
\ba{c}
\dps
\cV_{\h_1 \h_2 \h_3}({\bf x})=
C_{\h_1 \h_2 \h_3} \prod_{k = 1}^3K_{\h_k} \oint_{P[\pt_k,\bar{\pt}_k]}du_k\;K(x_k,u_k|1-\h_k)
\vspace{3mm}
\\
\dps
\times (u_1-u_2)^{\h_3-\h_1-\h_2}(u_2-u_3)^{\h_1-\h_2-\h_3}(u_1-u_3)^{\h_2-\h_1-\h_3}\,.
\ea
\ee
One can integrate over $u_1$ by using the analytically continued integral representation of the Gauss hypergeometric function \eqref{hyper_integral_contin} and then expanding into  series.  The next steps are the following: (1) integrate over $u_2$ using \eqref{hyper_integral_contin} and expand the resulting hypergeometric function; (2) integrate over $u_3$ using the analytically continued representation of the first Appell function \eqref{appell_poch} and expand into series  according to \eqref{def_appell}. The resulting expression reads
\be
\label{3pt_2form_interm}
\ba{l}
\dps
\cV_{\h_1 \h_2 \h_3}({\bf x})=  \frac{C_{\h_1 \h_2 \h_3}}{(2i)^{\h_3+\h_2+\h_1}}\sum_{\{k_i\geq0;\, i=1,...,4\}}\,  \,(\bar{\pt}_1-\pt_1)^{k_1+\h_1+k_4}(\bar{\pt}_2-\pt_2)^{k_2+\h_2}
\vspace{3mm}
\\
\dps
\times(\pt_3-\bar{\pt}_3)^{k_3+k_4+\h_3}(\bar{\pt}_2-\pt_1)^{-\h_2-k_2}(\pt_2-\pt_1)^{\h_3-\h_1-k_1+k_3}(\pt_3-\pt_1)^{k_2+\h_2}
\vspace{3mm}
\\
\dps
\times(\pt_3-\bar{\pt}_1)^{-\h_1-k_1-k_4}(\pt_3-\pt_2)^{\h_1-\h_2-\h_3+k_1-k_2-k_3}(\bar{\pt}_3-\pt_1)^{-\h_3-k_4-k_3}
\vspace{3mm}
\\
\dps
\times
(-)^{k_3}\frac{(\h_1+\h_2-\h_3)_{k_1}(\h_2+\h_3-\h_1-k_1)_{k_2+k_3}(\h_1)_{k_1+k_4}(\h_2)_{k_2}(\h_3)_{k_3+k_4}}{k_1!k_2!k_3!k_4!(2\h_1)_{k_1}(2\h_2)_{k_2}(2\h_3)_{k_3+k_4}}\,.
\ea
\ee
The rest of the calculation can be done in three steps: (1) make change $k_4\to k_4'=k_4+k_3$ and apply the Pfaff transformation \eqref{pfaff_series} to the sums over $k_3$, $k'_4$, $k_2$; (2) make change $k_3\to s=k_3-k_1$ and change $k_1\to k$. The result is given by
\be
\ba{l}
\dps
\cV_{\h_1 \h_2 \h_3}({\bf x})=  \frac{C_{\h_1 \h_2 \h_3}\, c_{13}^{-\h_3}\,c_{12}^{-\h_2}}{(2i)^{\h_1-\h_2-\h_3}}\sum_{\substack{k,s\geq0\vspace{1mm}\\ k<s}}\, \frac{(\h_1+\h_2-\h_3)_{k}(\h_1-\h_2-\h_3)_{k}(\h_2+\h_3-\h_1)_{s}(\h_1)_{k}}{k!(s-k)!(2\h_1)_{k}}
\vspace{3mm}
\\
\dps
\times y^{\h_3-\h_1+s}
{}_2F_1\left(k+\h_1-s,\h_3;2\h_3\big|\frac{-4}{c_{13}}\right){}_2F_1\left(\h_2+\h_3-\h_1-k+s,\h_2;2\h_2\big|\frac{-4}{c_{12}}\right),
\ea
\ee
where we introduced a new variable $y$ as follows  
\be 
\dps
y:=\frac{(\bar{\pt}_2-\pt_1)(\bar{\pt}_3-\bar{\pt}_1)}{(\bar{\pt}_2-\bar{\pt}_1)(\bar{\pt}_3-\pt_1)} \,,
\ee
which is related to invariant variables by means of \eqref{c_as_y}.

\subsection{$3$-point AdS vertex function with two boundary points}
\label{app:KGK_calc}

Substituting $\rho_2=\rho_3=\rho$ into the first representation of the $3$-point AdS vertex function \eqref{3p_first_rep}  and keeping only the leading term one obtains
\be
\ba{c}
\dps
\cV_{\h_1 \h_2 \h_3}^{\text{reg}}(x_1,z_2,z_3)= \frac{C_{\h_1 \h_2 \h_3}}{(2i)^{\h_1}}(z_2-z_3)^{\h_1-\h_2-\h_3}\sum_{k_1,k_2=0}^{\infty} \, (\pt_1-\bar{\pt}_1)^{k_1+k_2+\h_1}(z_3-\bar{\pt}_1)^{\h_2-\h_3-\h_1-k_2}
\vspace{3mm}
\\
\dps
\times
\,(z_2-\bar{\pt}_1)^{\h_3-\h_2-\h_1-k_1}\frac{(\h_1-\h_2+\h_3)_{k_2}(\h_1+\h_2-\h_3)_{k_1}(\h_1)_{k_1+k_2}}{k_1!k_2!(2\h_1)_{k_1+k_2}}
\,,
\ea
\ee 
where we used the regularized AdS vertex function \eqref{regular}. We used the first representation because the limit $\rho\to\infty$ is taken here more simply than in the second representation. To simplify this expression one applies  the Pfaff transformation \eqref{pfaff_series} to the sum over $k_1$:
\be
\ba{l}
\dps
\cV_{\h_1 \h_2 \h_3}^{\text{reg}}(x_1,z_2,z_3)=  \frac{C_{\h_1 \h_2 \h_3}}{(2i)^{\h_1}}(z_2-z_3)^{\h_1-\h_2-\h_3}\sum_{k_1,k_2=0}^{\infty} \, (\pt_1-\bar{\pt}_1)^{k_1+k_2+\h_1}
\vspace{3mm}
\\
\dps
\times
\,(z_3-\bar{\pt}_1)^{\h_2-\h_3-\h_1-k_2}(z_2-\bar{\pt}_1)^{\h_3-\h_2+k_2}(\pt_1-z_2)^{-\h_1-k_2-k_1}\frac{(\h_1-\h_2+\h_3)_{k_1+k_2}(\h_1)_{k_1+k_2}}{k_1!k_2!(2\h_1)_{k_1+k_2}}\,.
\ea
\ee 
Changing $k_1' = k_1+k_2$ one can sum over $k_2$ using the generalized Newton binomial and represent the sum over $k_1'$ as the Gauss hypergeometric function
\be
\ba{l}
\dps
\cV_{\h_1 \h_2 \h_3}^{\text{reg}}(x_1,z_2,z_3)=  \frac{C_{\h_1 \h_2 \h_3}}{(2i)^{\h_1}} \, (\pt_1-\bar{\pt}_1)^{\h_1}(z_3-z_2)^{\h_1-\h_2-\h_3}(\pt_1-z_2)^{-\h_1}
\vspace{3mm}
\\
\dps
\times
\,(z_3-\bar{\pt}_1)^{\h_2-\h_3-\h_1}(z_2-\bar{\pt}_1)^{\h_3-\h_2}{}_2F_1\left(\h_1+\h_3-\h_2,\h_1; 2\h_1\Big|\frac{(\pt_1-\bar{\pt}_1)(z_3-z_2)}{(\pt_1-z_2)(z_3-\bar{\pt}_1)}\right).
\ea
\ee 
Applying the quadratic \eqref{quadratic} and Pfaff \eqref{4kummer2f1} transformations yields
\be
\label{Wilson_KGK}
\ba{c}
\dps
\cV_{\h_1 \h_2 \h_3}^{\text{reg}}(x_1,z_2,z_3)=  \pref_{\h_1\h_2\h_3} \, z_{23}^{\h_1-\h_2-\h_3}\chi(x_1,z_2)^{\frac{\h_1+\h_2-\h_3}{2}}\chi(x_1,z_3)^{\frac{\h_1+\h_3-\h_2}{2}}
\vspace{3mm}
\\
\dps
\times{}_2F_1\left(\frac{\h_1+\h_3-\h_2}{2},\frac{\h_1+\h_2-\h_3}{2}; \h_1+\half\big|z_{23}^2\chi(x_1,z_2)\chi(x_1,z_3)\right),
\ea
\ee
where we used the regularized invariant distance \eqref{regul_distance}. 

\subsection{$3$-point AdS vertex function with one boundary point}
\label{app:KGG}

Consider the first representation of the 3-point AdS vertex function \eqref{3p_first_rep} and take $\rho_3\to\infty$:
\be
\label{3pt_1boundary}
\ba{l}
\dps
\cV^{\text{reg}}_{\h_1 \h_2 \h_3}(x_1,x_2,z_3) = \frac{C_{\h_1 \h_2 \h_3}}{(2i)^{\h_1+\h_2}}\hspace{-1mm}\sum_{k_1,k_2,k_3,k_4=0}^{\infty} \, (\pt_1-\bar{\pt}_1)^{k_1+k_2+\h_1}
\vspace{3mm}
\\
\dps
\times
\,(\bar{\pt}_2-\bar{\pt}_1)^{\h_3-\h_2-\h_1-k_1-k_4}(\bar{\pt}_2-z_3)^{\h_1-\h_2-\h_3-k_3}(z_3-\bar{\pt}_1)^{\h_2-\h_3-\h_1-k_2}(\bar{\pt}_2-\pt_2)^{k_3+k_4+\h_2}
\vspace{3mm}
\\
\dps
\times
\frac{(\h_1-\h_2+\h_3)_{k_2}(\h_2+\h_3-\h_1)_{k_3}(\h_1+\h_2-\h_3)_{k_4+k_1}(\h_1)_{k_1+k_2}(\h_2)_{k_3+k_4}}{k_1!k_2!k_3!k_4!(2\h_1)_{k_1+k_2}(2\h_2)_{k_3+k_4}}\,.
\ea
\ee 
One  simplifies this  expression by applying  the Pfaff transformation \eqref{pfaff_series} to the sum over $k_2$:
\be
\ba{l}
\dps
\cV^{\text{reg}}_{\h_1 \h_2 \h_3}(x_1,x_2,z_3)=  \frac{C_{\h_1 \h_2 \h_3}}{(2i)^{\h_1+\h_2}}\sum_{k_1,k_2,k_3,k_4=0}^{\infty} \, (\pt_1-\bar{\pt}_1)^{k_1+k_2+\h_1}(\bar{\pt}_2-\bar{\pt}_1)^{\h_3-\h_2-\h_1-k_1-k_4}
\vspace{3mm}
\\
\dps
\times
\,(\bar{\pt}_2-z_3)^{\h_1-\h_2-\h_3-k_3}(z_3-\bar{\pt}_1)^{\h_2-\h_3+k_1}(\bar{\pt}_2-\pt_2)^{k_3+k_4+\h_2}(z_3-\pt_1)^{-\h_1-k_1-k_2}(-)^{k_2}
\vspace{3mm}
\\
\dps
\times
\frac{(\h_1+\h_2-\h_3+k_1)_{k_2}(\h_2+\h_3-\h_1)_{k_3}(\h_1+\h_2-\h_3)_{k_4+k_1}(\h_1)_{k_1+k_2}(\h_2)_{k_3+k_4}}{k_1!k_2!k_3!k_4!(2\h_1)_{k_1+k_2}(2\h_2)_{k_3+k_4}}\,.
\ea
\ee 
After changing $k_2'=k_2+k_1$ one applies the same  Pfaff transformation to the sum over $k_1$ and then to the sum over $k_3$,
\be
\ba{l}
\dps
\cV^{\text{reg}}_{\h_1 \h_2 \h_3}(x_1,x_2,z_3)=  \frac{C_{\h_1 \h_2 \h_3}}{(2i)^{\h_1+\h_2}}\sum_{k_1,k_2',k_3,k_4=0}^{\infty} \, (\bar{\pt}_1-\pt_1)^{k_2'+\h_1}(\bar{\pt}_2-\bar{\pt}_1)^{\h_3-\h_2-\h_1-k_2'-k_4}
\vspace{3mm}
\\
\dps
\times
\,(\bar{\pt}_2-z_3)^{\h_1-\h_3-k_1+k_2'+k_4}(z_3-\bar{\pt}_1)^{\h_2-\h_3+k_1}(\pt_2-\bar{\pt}_2)^{k_3+k_4+\h_2}
(z_3-\pt_1)^{-\h_1-k_2'}
\vspace{3mm}
\\
\dps
\times
(\pt_2-z_3)^{-\h_2-k_4-k_3}(-)^{k_1}\frac{(\h_1+\h_2-\h_3)_{k_2'}(\h_1+\h_2-\h_3)_{k_3+k_4}(\h_1)_{k_2'}(\h_2)_{k_3+k_4}}{k_1!(k_2'-k_1)!k_3!(k_4-k_1)!(2\h_1)_{k_2'}(2\h_2)_{k_3+k_4}(\h_1+\h_2-\h_3)_{k_1}}\,.
\ea
\ee 
Changing $k_4'=k_3+k_4$, summing over $k_3$, using the generalized Newton binomial, and summing over $k_1$ by means of  the identity for the Gauss hypergeometric function  \eqref{2f1_in_1} yields
\be
\label{3pt_1_boundary_app}
\ba{l}
\dps
\cV^{\text{reg}}_{\h_1 \h_2 \h_3}(x_1,x_2,z_3)=  \frac{C_{\h_1 \h_2 \h_3}}{(2i)^{\h_1+\h_2}}\sum_{k_2',k_4'=0}^{\infty} \, (\bar{\pt}_1-\pt_1)^{k_2'+\h_1}(\bar{\pt}_2-\bar{\pt}_1)^{\h_3-\h_2-\h_1-k_2'-k_4'}
\vspace{3mm}
\\
\dps
\times
\,(\bar{\pt}_2-z_3)^{\h_1-\h_3+k_2'}(z_3-\bar{\pt}_1)^{\h_2-\h_3+k_4'}(\pt_2-\bar{\pt}_2)^{k_4'+\h_2}(z_3-\pt_1)^{-\h_1-k_2'}(\pt_2-z_3)^{-\h_2-k_4'}
\vspace{3mm}
\\
\dps
\times
\frac{(\h_1)_{k_2'}(\h_2)_{k_4'}(\h_1+\h_2-\h_3)_{k_2'+k_4'}}{k_2'!k_4'!(2\h_1)_{k_2'}(2\h_2)_{k_4'}}\,.
\ea
\ee 
This expression  can be written as the second Appell function \eqref{def_appell_f2}
\be
\ba{c}
\label{3p_1b_Appell}
\dps
\cV^{\text{reg}}_{\h_1 \h_2 \h_3}(x_1,x_2,z_3)= \frac{C_{\h_1 \h_2 \h_3}}{(2i)^{\h_1+\h_2}}\left[\frac{(z_3-\bar{\pt}_2)(z_3-\bar{\pt}_1)}{\bar{\pt}_2-\bar{\pt}_1}\right]^{-\h_3}
\vspace{3mm}
\\
\dps
\times\,
c_1^{\h_1}c_2^{\h_2} F_2\left[\begin{array}{ccc}
     \h_1,&\h_2, &\h_1+\h_2-\h_3  \\
     2\h_1, &2\h_2 \end{array};c_1,c_2\right],
\ea
\ee 
where we introduced the auxiliary variables
\be 
\label{c_variables}
c_1:=\frac{(\bar{\pt}_1-\pt_1)(\bar{\pt}_2-z_3)}{(z_3-\pt_1)(\bar{\pt}_2-\bar{\pt}_1)}\,;\quad\quad
c_2:=\frac{(\pt_2-\bar{\pt}_2)(z_3-\bar{\pt}_1)}{(\pt_2-z_3)(\bar{\pt}_2-\bar{\pt}_1)}.
\ee 
Using the quadratic transformation \eqref{f_2_to_f_4} of the second Appell function one obtains the final result
\be
\label{3p_KGG}
\ba{c}
\dps
\cV^{\text{reg}}_{\h_1 \h_2 \h_3}(x_1,x_2,z_3)=  \frac{C_{\h_1 \h_2 \h_3}}{(2i)^{\h_1+\h_2}}\left[\frac{(z_3-\bar{\pt}_2)(z_3-\bar{\pt}_1)}{\bar{\pt}_2-\bar{\pt}_1}\right]^{-\h_3}
\left(1-\frac{c_1}{2}-\frac{c_2}{2}\right)^{\h_3-\h_2-\h_1 }
\vspace{3mm}
\\
\dps
\times
c_1^{\h_1}c_2^{\h_2}\, F_4\left[\begin{array}{cc}
 \frac{\h_2+\h_1-\h_3}{2},&\frac{\h_2+\h_1-\h_3}{2}+\half \\
  \h_1+\half, &\h_2 +\half\end{array};\frac{c_1^2}{(2-c_1-c_2)^2},\frac{c_2^2}{(2-c_1-c_2)^2}\right] \,.
\ea
\ee 

The next step  is to consider the $3$-point Witten diagram with one boundary point which was calculated in \cite{Jepsen:2019svc}:
$$
\ba{l}
\dps
\int_{\text{AdS}_2}d^2x\, G(x,x_1|\h_1) G(x,x_2|\h_2)K(x,z_3|\h_3) =  \alpha(\h_1,\h_2,\h_3)\sum_{k_1,k_2=0}^{\infty}c^{\h_1;\h_2;\h_3}_{k_1;k_2}
\vspace{3mm}
\\
\dps
\times
\chi(z_3,x_1)^{\frac{\h_3+\h_1-\h_2}{2}+k_1-k_2}\chi(z_3,x_2)^{\frac{\h_3+\h_2-\h_1}{2}-k_1+k_2}\left(\frac{\xi(x_1,x_2)}{2}\right)^{\frac{\h_1+\h_2-\h_3}{2}+k_1+k_2}
\ea
$$
\be
\ba{l}
\label{KGG}
\dps
\hspace{8mm}+\sum_{k_1,k_2=0}^{\infty}d^{\h_1;\h_2;\h_3}_{k_1;k_2}\chi(z_3,x_1)^{\h_3+k_1}\chi(z_3,x_2)^{-k_1}\left(\frac{\xi(x_1,x_2)}{2}\right)^{\h_2+k_1+2k_2}+(1\leftrightarrow 2)
\vspace{3mm}
\\
\hspace{18mm}\equiv WF^{(1)}(x_1,x_2,z_3) + WF^{(2)}(x_1,x_2,z_3)+ WF^{(3)}(x_1,x_2,z_3)\,,
\ea
\ee 
where $(1\leftrightarrow 2)$ denotes  interchanging the indices $1$ and $2$ in the second sum, and $WF^{(i)}(x_1,x_2,z_3)$ denotes each sum in \eqref{KGG}, the coefficients $c^{\h_1;\h_2;\h_3}_{k_1;k_2}$ and $d^{\h_1;\h_2;\h_3}_{k_1;k_2}$ are  given by

$$
\ba{l}
\dps
c_{k_1 ; k_2}^{\h_1 ; \h_2; \h_3 } = \left(\frac{\h_1-\h_2+\h_3}{2}\right)_{k_1-k_2}\left(\frac{\h_1+\h_2-\h_3}{2}\right)_{k_1+k_2}\left(\frac{\h_2+\h_3-\h_1}{2}\right)_{k_2-k_1}
\vspace{3mm}
\\
\dps
\times
\frac{(-)^{k_1+k_2}}{k_{1}!k_{2}!}F_2\left[\begin{array}{ccc}
     -k_1,&-k_2, &\frac{\h_1+\h_2+\h_3}{2}-\half  \\
     \h_1+\half, &\h_2 +\half\end{array};1,1\right],
\ea
$$
\be 
\ba{c}
\dps
\hspace{13mm}d_{k_1 ; k_2}^{\h_1 ; \h_2; \h_3 } =\frac{\pi^{\half}\Gamma(\frac{\h_1+\h_2+\h_3}{2}-\half)}{2\Gamma(\h_1)}(\h_3)_{k_1}(\h_2)_{2k_2+k_1}\Gamma\left(\frac{\h_1-\h_2-\h_3}{2}-k_1-k_2\right)
\vspace{3mm}
\\
\dps
\times
\frac{(-)^{k_1+k_2}}{k_{1}!k_{2}!}F_2\left[\begin{array}{ccc}
     \frac{\h_1-\h_2-\h_3}{2}-k_1-k_2,&-k_2, &\frac{\h_1+\h_2+\h_3}{2}-\half  \\
     \h_1+\half, &\h_2 +\half\end{array};1,1\right].
\ea
\ee 
In what follows we consider each term $WF^{(1,2,3)}(x_1,x_2,z_3)$ separately. 

\paragraph{Calculating $WF^{(1)}(x_1,x_2,z_3)$.} We observe that the first term $WF^{(1)}(x_1,x_2,z_3)$ is proportional to the $3$-point AdS vertex function. To show this, we use the second Appell series \eqref{def_appell_f2} to obtain
$$
\dps
\hspace{-13mm}
WF^{(1)}(x_1,x_2,z_3) = \alpha(\h_1,\h_2,\h_3)\sum_{k_1,k_2,k_3,k_4=0}^\infty   \left(\frac{\h_1-\h_2+\h_3}{2}\right)_{k_1-k_2}\left(\frac{\h_1+\h_2-\h_3}{2}\right)_{k_1+k_2}
$$
\be 
\label{KGG_start}
\ba{l}
\dps
\times
\left(\frac{\h_2+\h_3-\h_1}{2}\right)_{k_2-k_1}
\frac{(-)^{k_1+k_2}(-k_1)_{k_3}(-k_2)_{k_4}(\frac{\h_1+\h_2+\h_3}{2}-\half)_{k_3+k_4}}{k_1!k_2!k_3!k_4!(\h_1+\half)_{k_3}(\h_2 +\half)_{k_4}}
\vspace{3mm}
\\
\dps
\times
\chi(z_3,x_1)^{\frac{\h_3+\h_1-\h_2}{2}+k_1-k_2}\chi(z_3,x_2)^{\frac{\h_3+\h_2-\h_1}{2}-k_1+k_2}\left(\frac{\xi(x_1,x_2)}{2}\right)^{\frac{\h_1+\h_2-\h_3}{2}+k_1+k_2}\,.
\ea
\ee
Representing the sum over $k_4$ as the Gauss hypergeometric function at unit argument and using the identity \eqref{2f1_in_1} yields
\be 
\ba{l}
\dps
WF^{(1)}(x_1,x_2,z_3) = \alpha(\h_1,\h_2,\h_3)\sum_{k_1,k_2,k_3=0}^\infty \left(\frac{\h_1-\h_2+\h_3}{2}\right)_{k_1-k_2}\left(\frac{\h_1+\h_2-\h_3}{2}\right)_{k_1+k_2}
\vspace{3mm}
\\
\dps
\times
\left(\frac{\h_2+\h_3-\h_1}{2}\right)_{k_2-k_1}
\frac{(-)^{k_1}(-k_1)_{k_3}(\frac{\h_1+\h_2+\h_3}{2}-\half)_{k_3}(\frac{\h_2-\h_3-\h_1}{2}-k_3+1)_{k_2}}{k_1!k_2!k_3!(\h_1+\half)_{k_3}(\h_2 +\half)_{k_2}}
\vspace{3mm}
\\
\dps
\times
\chi(z_3,x_1)^{\frac{\h_3+\h_1-\h_2}{2}+k_1-k_2}\chi(z_3,x_2)^{\frac{\h_3+\h_2-\h_1}{2}-k_1+k_2}\left(\frac{\xi(x_1,x_2)}{2}\right)^{\frac{\h_1+\h_2-\h_3}{2}+k_1+k_2}\,.
\ea
\ee 
Further simplification can be achieved  in four steps: (1) sum over $k_3$ using the hypergeometric series ${}_3F_2$ \eqref{def_hyper_3f2} and apply the identity \eqref{3f2_idenity}; (2) make change $k_1'=k_1-k_3$ and apply the Pfaff transformation \eqref{pfaff_series} to the sum over $k_1'$; (3) make change $k_1=k_1'+k_3$, represent the sum over $k_3$ as the Gauss hypergeometric function at unit argument and use \eqref{2f1_in_1}; (4) apply the Pfaff transformation to the sum over $k_2$. One obtains  
\be 
\label{KGG_inter}
\ba{c}
\dps
WF^{(1)}(x_1,x_2,z_3)= \alpha(\h_1,\h_2,\h_3)\sum_{k_1,k_2=0}^\infty 
\frac{\left(\h_1+\h_2-\h_3\right)_{2k_1+2k_2}}{k_1!k_2!(\h_1+\half)_{k_1}(\h_2 +\half)_{k_2}}\chi(z_3,x_1)^{\h_3-\h_2-2k_2}\chi(z_3,x_2)^{\h_3-\h_1-2k_1}
\vspace{3mm}
\\
\dps
\times
\left(\frac{1}{\chi(z_3,x_2)^2}+\frac{1}{\chi(z_3,x_1)^2} - \frac{2}{\chi(z_3,x_2)\chi(z_3,x_1)\xi(x_1,x_2)}\right)^{\frac{\h_3-\h_1-\h_2}{2}-k_1-k_2}.
\ea
\ee 
Rewriting this expression in terms of the fourth Appell function \eqref{def_appell_f4} and using the following identities
\be 
\ba{l}
\dps
\frac{(2-c_1-c_2)^2}{c_1^2} = 1+\frac{\chi(z_3,x_2)^2}{\chi(z_3,x_1)^2} - \frac{2}{\xi(x_1,x_2)}\frac{\chi(z_3,x_2)}{\chi(z_3,x_1)}\,,
\vspace{3mm}
\\
\dps
\frac{(2-c_1-c_2)^2}{c_2^2} = 1+\frac{\chi(z_3,x_1)^2}{\chi(z_3,x_2)^2} - \frac{2}{\xi(x_1,x_2)}\frac{\chi(z_3,x_1)}{\chi(z_3,x_2)}\,,
\vspace{3mm}
\\
\dps
\frac{(\bar{\pt}_2-\bar{\pt}_1)}{(z_3-\bar{\pt}_2)(z_3-\bar{\pt}_1)}\left(1-\frac{c_1}{2}-\frac{c_2}{2}\right)=\left(2\frac{\chi(z_3,x_2)\chi(z_3,x_1)}{\xi(x_1,x_2)}-\chi(z_3,x_2)^2-\chi(z_3,x_1)^2\right)^{1/2}\,,
\ea
\ee 
one can see that the first term in  \eqref{KGG} is proportional to the regularized $3$-point AdS vertex function with one boundary point \eqref{3pt_1boundary}:
\be 
\label{KGG_first_term}
WF^{(1)}(x_1,x_2,z_3) = \frac{\alpha(\h_1,\h_2,\h_3)}{\pref_{\h_1\h_2\h_3}}\cV^{\text{reg}}_{\h_1 \h_2 \h_3}(x_1,x_2,z_3)\,.
\ee 

\paragraph{Calculating $WF^{(2,3)}(x_1,x_2,z_3)$.}  Moreover, the second and third terms $WF^{(2,3)}(x_1,x_2,z_3)$ in \eqref{KGG} can be represented as linear combination of the $3$-point AdS vertex functions
\be
\label{KGG_2nd_term}
\ba{l}
\dps
WF^{(2)}[x_1,x_2,z_3]=\sum_{n=0}^{\infty}\frac{a(\h_1;\h_2,\h_3;n)}{\pref_{\h_2+\h_3+2n\  \h_2\h_3}}\,\cV^{\text{reg}}_{\h_2+\h_3+2n\  \h_2\h_3}(x_1,x_2,z_3)\,,
\vspace{3mm}
\\
\dps
WF^{(3)}[x_1,x_2,z_3]=\sum_{n=0}^{\infty}\frac{a(\h_2;\h_1,\h_3;n)}{\pref_{\h_1\ \h_1+\h_3+2n\ \h_3}}\,\cV^{\text{reg}}_{\h_1\ \h_1+\h_3+2n\ \h_3}(x_1,x_2,z_3)\,.
\ea
\ee 
To show this, one replaces the $3$-point AdS vertex function with $WF^{(1)}(x_1,x_2,z_3)$ using \eqref{KGG_first_term}, uses the definition of $WF^{(1)}(x_1,x_2,z_3)$ \eqref{KGG} and applies  the identity \eqref{useful_identity}. Combining \eqref{KGG_first_term} with \eqref{KGG_2nd_term} one obtains  Proposition \bref{prop:3pt_one}. We will use the same method to prove Proposition \bref{prop:GGG} in Appendix \bref{app:GGG}.

\subsection{$3$-point AdS vertex functions from $3$-point Witten diagram}
\label{app:GGG}

\paragraph{Proof of Lemma \bref{lem:KG}.} To show the first relation in \eqref{KG_WF} one applies the Klein-Gordon    operator written in the Poincare coordinates \eqref{metric} to the function $WF^{(1)}(x_1,x_2,x_3)$ defined in  \eqref{GGG}: 
$$
\ba{l}
\dps
\hspace{-5mm}\Big(e^{-2\rho_1}\partial^2_{z_1} + \partial^2_{\rho_1}+\partial_{\rho_1}-\h_1(\h_1-1)\Big)WF^{(1)}(x_1,x_2,x_3) = 4\alpha(\h_1,\h_2,\h_3)
\vspace{3mm}
\\
\dps
\times
\sum_{k_1,k_2,k_3=0}^{\infty}\left[\frac{\xi(x_1,x_3)}{2}\right]^{\frac{\h_{13,2}}{2}+k_{13,2}}\left[\frac{\xi(x_2,x_3)}{2}\right]^{\frac{\h_{23,1}}{2}+k_{23,1}}\left[\frac{\xi(x_1,x_2)}{2}\right]^{\frac{\h_{12,3}}{2}+k_{12,3}}
\vspace{3mm}
\\
\dps
\times\left[k_1\left(k_1+\h_1-\half\right)c^{\h_1;\h_2;\h_3}_{k_1;k_2;k_3} -\left(\frac{\h_{12,3}}{2}+k_{12,3}-1\right)\left(\frac{\h_{12,3}}{2}+k_{12,3}-2\right)c^{\h_1;\h_2;\h_3}_{k_1-1;k_2-1;k_3}\right.
\ea
$$
\be
\label{KG_GGG}
\ba{l}
\dps
-\left(\frac{\h_{13,2}}{2}+k_{13,2}-1\right)\left(\frac{\h_{13,2}}{2}+k_{13,2}-2\right)c^{\h_1;\h_2;\h_3}_{k_1-1;k_2;k_3-1}
\vspace{3mm}
\\
\dps
\left.-\left(\frac{\h_{13,2}}{2}+k_{13,2}-1\right)\left(\frac{\h_{12,3}}{2}+k_{12,3}-1\right)c^{\h_1;\h_2;\h_3}_{k_1-1;k_2;k_3}\right],
\ea
\ee
where $a_{ij,k} \equiv a_i+a_j-a_k$. Since the AdS invariant distances $\xi(x_i,x_j)$ are independent variables, then the right-hand side here  equals zero iff the expression in the square brackets (the last three lines) is zero. Substituting the coefficients $c^{\h_1;\h_2;\h_3}_{k_1;k_2;k_3}$ \eqref{GGG_coef} into \eqref{KG_GGG} and expanding the Lauricella functions $F_A^{(3)}$ \eqref{def_lauricella_A} one obtains
$$
\ba{l}
\dps
\Big(e^{2\rho_1}\partial^2_{z_1} + \partial^2_{\rho_1}+\partial_{\rho_1}-\h_1(\h_1-1)\Big)WF^{(1)}(x_1,x_2,x_3) = 4\alpha(\h_1,\h_2,\h_3)
\vspace{3mm}
\\
\dps
\times
\sum_{k_1,k_2,k_3=0}^{\infty}\left[\frac{\xi(x_1,x_3)}{2}\right]^{\frac{\h_{13,2}}{2}+k_{13,2}}\left[\frac{\xi(x_2,x_3)}{2}\right]^{\frac{\h_{23,1}}{2}+k_{23,1}}\left[\frac{\xi(x_1,x_2)}{2}\right]^{\frac{\h_{12,3}}{2}+k_{12,3}}\frac{(-)^{k_1+k_2+k_3}}{k_{1}!k_{2}!k_3!}
\ea
$$
\be 
\label{KG_GGG_2}
\ba{l}
\dps
\times
\left(\frac{\h_{13,2}}{2}\right)_{k_{13,2}}\left(\frac{\h_{12,3}}{2}\right)_{k_{12,3}}\left(\frac{\h_{23,1}}{2}\right)_{k_{23,1}}\sum_{l_1,l_2,l_3=0}^{\infty}\frac{(\h_1+\h_2+\h_3-\half)_{l_1+l_2+l_3}(-k_1)_{l_1}(-k_2)_{l_2}(-k_3)_{l_3}}{l_1!l_2!l_3!(\h_1 +\half)_{l_1}(\h_2 +\half)_{l_2}(\h_3 +\half)_{l_3}}
\vspace{3mm}
\\
\dps
\times
\left[(\h_1+\h_2+\h_3-\half+l_1+l_2+l_3)(k_1-l_1)+l_1(\h_1 +\half+l_1)\right].
\ea
\ee
The sum over $l_1$ is equal to zero. To show this, one uses the following identity relations
\be
\label{D34} 
\ba{l}
\dps
\sum_{l_1=0}^{\infty}\frac{(\h_1+\h_2+\h_3-\half)_{l_1+l_2+l_3}(-k_1)_{l_1}}{l_1!(\h_1 +\half)_{l_1}}\left[l_1(\h_1 +\half+l_1)\right]
\vspace{3mm}
\\
\dps
=\sum_{l_1=0}^{\infty}\frac{(\h_1+\h_2+\h_3-\half)_{l_1+l_2+l_3+1}(-k_1)_{l_1+1}}{l_1!(\h_1 +\half)_{l_1}}
\vspace{3mm}
\\
\dps
=-\sum_{l_1=0}^{\infty}\frac{(\h_1+\h_2+\h_3-\half)_{l_1+l_2+l_3}(-k_1)_{l_1}}{l_1!(\h_1 +\half)_{l_1}}\left[(\h_1+\h_2+\h_3-\half+l_1+l_2+l_3)(k_1-l_1)\right].
\ea
\ee 
Adding the first and third lines in  \eqref{D34}  one notes that sum in the square brackets in \eqref{KG_GGG_2}  equals zero, i.e. the first term of the $3$-point Witten diagram $WF^{(1)}(x_1,x_2,x_3)$ satisfies  the homogeneous  Klein-Gordon equation over the variable $x_1$ with the mass $m_1^2 = \h_1(\h_1-1)$. Since the expression for the $3$-point Witten diagram \eqref{GGG} is symmetric with respect to the permutation of  indices (up to an overall sign) the first term $WF^{(1)}(x_1,x_2,x_3)$ also satisfies the homogeneous Klein-Gordon equation over  $x_2$ and $x_3$ with  masses $m_2^2 = \h_2(\h_2-1)$ and $m_3^2 = \h_3(\h_3-1)$, respectively.

In order to prove that the boundary condition in \eqref{KG_WF} is satisfied we consider the first term $WF^{(1)}(x_1,x_2,x_3)$ and take the limit $\rho_3\to\infty$ with the regularization factor $e^{\rho_3\h_3}$:
\be 
\ba{l}
\dps
\lim_{\rho_3\to\infty}e^{\rho_3\h_3}WF^{(1)}(x_1,x_2,x_3) = \alpha(\h_1,\h_2,\h_3)\sum_{k_1,k_2=0}^{\infty}\left[\frac{\h_1-\h_2+\h_3}{2}\right]_{k_1-k_2}\left[\frac{\h_1+\h_2-\h_3}{2}\right]_{k_1+k_2}
\vspace{3mm}
\\
\dps
\times
\left[\frac{\h_2+\h_3-\h_1}{2}\right]_{k_2-k_1}\frac{(-)^{k_1+k_2}}{k_{1}!k_{2}!}F_2\left[\begin{array}{ccc}
     -k_1,&-k_2, &\frac{\h_1+\h_2+\h_3}{2}-\half  \\
     \h_1+\half, &\h_2 +\half\end{array};1,1\right]

\vspace{3mm}
\\
\dps
\times
\left[\frac{\chi(x_1,x_3)}{2}\right]^{\frac{\h_3+\h_1-\h_2}{2}+k_1-k_2}\left[\frac{\chi(x_2,x_3)}{2}\right]^{\frac{\h_3+\h_2-\h_1}{2}-k_1+k_2}\left[\frac{\xi(x_1,x_2)}{2}\right]^{\frac{\h_1+\h_2-\h_3}{2}+k_1+k_2}
\vspace{3mm}
\\
\dps
=WF^{(1)}(x_1,x_2,z_3)
\,,
\ea
\ee 
where $WF^{(1)}(x_1,x_2,z_3)$ is the first term of the $3$-point Witten diagram with one boundary point \eqref{KGG}. In the last step we use the relation between $WF^{(1)}(x_1,x_2,z_3)$ and the regularized  $3$-point AdS vertex function \eqref{KGG_first_term} to obtain the following relation 
\be 
\label{GGG_boundary}
\lim_{\rho_3\to\infty}e^{\rho_3\h_3}WF^{(1)}(x_1,x_2,x_3) = \frac{\alpha(\h_1,\h_2,\h_3)}{\pref_{\h_1\h_2\h_3}}\cV^{\text{reg}}_{\h_1 \h_2 \h_3}(x_1,x_2,z_3)\,,
\ee 
which proves Lemma \bref{lem:KG}.
\paragraph{Proof of Lemma \bref{lem:Cauchy}.}  Lemma \bref{lem:KG} claims that the boundary asymptotics of the first term of $3$-point Witten diagram and the $3$-point AdS vertex function are proportional to each other as $\rho_3\to\infty$ \eqref{GGG_boundary}. In order to prove that these functions coincide in the bulk ($\rho_3\in\mathbb{R}$) we show that the following Cauchy problem
\be
\label{Cauchy_problem}
\ba{l}
\dps
\left(\Box_{3}- m_3^2\right) WF^{(1)}(x_1,x_2,x_3) = 0\,,
\qquad
m_3^2 = h_3(h_3-1)\,.
\vspace{3mm}
\\
\dps
\lim_{\rho_3\to\infty}e^{\rho_3\h_3}WF^{(1)}(x_1,x_2,x_3) = \frac{\alpha(\h_1,\h_2,\h_3)}{\pref_{\h_1\h_2\h_3}}\lim_{\rho_3\to\infty}e^{\rho_3\h_3}\cV_{\h_1\h_2\h_3}(x_1,x_2,x_3)\,,
\ea
\ee
has a unique solution.\footnote{One can consider other Cauchy problems where the boundary condition is defined by placing two or three points on the boundary, i.e. $\rho_1\to \infty$ and/or $\rho_2\to \infty$. On one hand, such boundary conditions are easier to prove, but on the other hand, it takes a lot more effort to show the uniqueness of solutions to these Cauchy problems since a  single Klein-Gordon equation in \eqref{Cauchy_problem} is replaced with a system of two or three Klein-Gordon equations, respectively. Therefore, we consider the simplest case with a boundary condition for a single point on the boundary since we already proved that the $3$-point AdS vertex function and the first term of $3$-point Witten diagram with one boundary point  are proportional \eqref{GGG_boundary}.} Obviously, the proof boils down to solving the Klein-Gordon equation and analyzing boundary asymptotics. Despite the fact that such a consideration is standard in AdS/CFT (for review see  e.g.  \cite{Ramallo:2013bua}), here we repeat the main steps and highlight some subtleties.  

Consider the Klein-Gordon equation which is  supplemented  with a particular  condition  on the conformal boundary: 
\be 
\label{KG_equation}
\ba{l}
\dps
\Big(u^2\partial^2_{z} + u^2\partial^2_u-h(h-1)\Big)f(z,u)=0\,,
\vspace{3mm}
\\
\dps
u^{-h}f(z,u)\Big|_{u\to0} = g(z)\,.
\ea
\ee 
Here we changed $u = e^{-\rho}$ to simplify the Klein-Gordon  equation; the conformal boundary in this parametrization lies at $u=0$; $g(z)$ is a given function. To solve the equation one makes the Fourier transform over $z$ to obtain
\be 
\label{KG_fourier}
\ba{l}
\dps
\left(\partial^2_u+k^2-\frac{h(h-1)}{u^2}\right)F(k,u)=0\,,
\vspace{3mm}
\\
\dps
u^{-h}F(k,u)\Big|_{u\to0} = \cF_z[g(z)](k)\,,
\ea
\ee 
where $k$ is the dual variable,  $F(k,u) := \cF_{z}[f(z,u)](k,u)$ and $\cF_z$ stands for the Fourier image. The resulting equation is the Bessel equation which  solution is given in terms of the Bessel functions of the first kind \eqref{bes}\footnote{In the case $h+\half\in\mathbb{Z}$ the second solution is replaced with the Bessel function of the second kind but it is irrelevant for our analysis as the boundary behavior remains the same.}
\be 
\label{KG_solution}
\dps
F(k,u)=C_1(k)u^\half J_{h-\half}(|k|u) + C_2(k)u^\half J_{\half-h}(|k|u)\,,
\ee 
where the coefficient functions $C_i(k)$ are to be fixed by the boundary condition. Expanding the Bessel functions near $u=0$ one finds that the boundary asymptotics of the left-hand side of the boundary condition \eqref{KG_fourier} reads
\be 
\label{boundary_expanded}
\ba{l}
\dps
u^{-h}F(k,u)\Big|_{u\to0} = \frac{C_1(k)|k|^{h-\half}}{2^{h-\half}\Gamma(h+\half)} + \left(u^{1-2h}\frac{C_2(k)|k|^{\half-h}}{2^{\half-h}\Gamma(\frac{3}{2}-h)}+O(u^{2-2h})\right)\Bigg|_{u\to0}\,.
\ea
\ee 
Note that in the case $h>\half$ the boundary condition fixes $C_{2}(k)=0$, otherwise the right-hand side of \eqref{boundary_expanded} is not regular. Then, the solution \eqref{KG_solution}  takes the form
\be 
\label{KG_solution_with_bndry}
\dps
F(k,u)=2^{h-\half}\Gamma(h+\half)|k|^{\half-h}u^\half J_{h-\half}(|k|u)\cF_z[g(z)](k)\,.
\ee 
As one can see, in the case $h>\half$ the coefficients $C_1(k)$ and $C_2(k)$ are fixed by the function $g(z)$. In other words, the  Cauchy problem \eqref{KG_equation} is solved uniquely  provided that  $g(z)$ is absolutely integrable and continuous over  $z$ in order to apply the inverse Fourier transform in \eqref{KG_solution_with_bndry} \cite{vretblad2006fourier}. 

In our case, $g(z)$ is the $3$-point AdS vertex function with one boundary point \eqref{Cauchy_problem}. One can show that it is absolutely integrable and continuous for a particular domain of conformal dimensions. It follows that the first term of the $3$-point Witten diagram and  the $3$-point AdS vertex function coincide in this case. Since  both functions are defined for general weights, then any  restrictions on the weights can be removed by analytic continuation.

\paragraph{Proof of Proposition \bref{prop:GGG}.} To obtain the relation between the $3$-point AdS vertex function and the $3$-point Witten diagram consider the explicit expression for the latter \eqref{GGG}. The first term $WF^{(1)}(x_1,x_2,x_3)$ therein is proportional to the 3-point AdS vertex function while the other terms are expressed as linear combinations of the first term with different weights. E.g., the second term of the $3$-point Witten diagram $WF^{(2)}(x_1,x_2,x_3)$ can be written in the following form
\be 
WF^{(2)}(x_1,x_2,x_3) = \sum_{n=0}^{\infty}\frac{a(\h_1;\h_2,\h_3;n)}{\alpha(\h_2+\h_3+2n,\h_2,\h_3)}\left(WF^{(1)}(x_1,x_2,x_3)\Big|_{\h_1=\h_2+\h_3+2n}\right),
\ee 
where the coefficients in the sum are defined in \eqref{alpha_beta} and \eqref{a_coef}. To show this, one substitutes the coefficients and the first term $WF^{(1)}(x_1,x_2,x_3)$ from \eqref{GGG} and obtains
\be 
\ba{l}
\dps
WF^{(2)}(x_1,x_2,x_3) = 2\pi^{\half}\frac{\Gamma(\h_1+\half)}{\Gamma(\h_1)}\sum_{\substack{k_1,k_2,k_3,n=0\\l_1,l_2,l_3=0}}^{\infty}\frac{1}{k_{1}!k_{2}!k_3!l_1!l_2!l_3!} 
\vspace{3mm}  
\\
\dps
\times\frac{(-)^{n}(\h_2+\h_3+2n-\half)\left(\h_3\right)_{k_1-k_2+k_3+n}\left(\h_2\right)_{k_1+k_2-k_3+n}}{[\h_1(\h_1-1)-(\h_2+\h_3+2n)(\h_2+\h_3+2n-1)](n+k_1-k_2-k_3)!}
\vspace{3mm}  
\\
\dps
\times
\frac{}{}\frac{(-k_1)_{l_1}(-k_2)_{l_2}(-k_3)_{l_3}(\h_2+\h_3+n-\frac{3}{2}+l_1+l_2+l_3)!}{(\h_2+\h_3+2n-\half+l_1)!(\h_2 +\half)_{l_2}(\h_3 +\half)_{l_3}}\left[\frac{\xi(x_1,x_3)}{2}\right]^{\h_3+n+k_1-k_2+k_3}
\vspace{3mm}
\\
\dps
\times
\left[\frac{\xi(x_2,x_3)}{2}\right]^{k_2+k_3-k_1-n}\left[\frac{\xi(x_1,x_2)}{2}\right]^{\h_2+n+k_1+k_2-k_3}\,,
\ea
\ee 
where the Lauricella function $F_A^{(3)}$ in the coefficient \eqref{GGG_coef} was expanded into series. To prove this relation: (1) sum over  $l_1$  by using the Gauss hypergeometric series and its value at unit argument \eqref{2f1_in_1}; (2)  change $k_1\to s = k_1-k_2-k_3+n$ and then $n\to p = n-l_2-l_3$; (3) use the identity \eqref{useful_identity} with $a=\h_2+\h_3+2l_2+2l_3$, $b = \h_1$, $k=s+k_2+k_3-l_2-l_3$ to sum over $p$. Finally, renaming $s=k_1$ one obtains 
\be 
\ba{l}
\dps
WF^{(2)}(x_1,x_2,x_3)= \frac{\pi^{\half}}{2}\frac{\Gamma(\h_1+\half)}{\Gamma(\h_1)}\sum_{\substack{k_1,k_2,k_3=0\\l_2,l_3=0}}^{\infty}\frac{1}{k_1!k_{2}!k_3!l_2!l_3!} 
\vspace{2.5mm}  
\\
\dps
\times
\frac{(\frac{\h_1-\h_2-\h_3}{2}-k_1-k_2-k_3-1)!(k_1+k_2+k_3-l_2-l_3)!(\frac{\h_2+\h_3+\h_1}{2}+l_2+l_3-\frac{3}{2})!}{(\frac{\h_2+\h_3+\h_1}{2}+k_1+k_2+k_3-\half)!(\frac{\h_1-\h_2-\h_3}{2}-l_2-l_3)!}
\vspace{3mm}
\\
\dps
\times
\frac{(-)^{k_1+k_2+k_3}\left(\h_3\right)_{k_1+2k_3}\left(\h_2\right)_{k_1+2k_2}(-k_2)_{l_2}(-k_3)_{l_3}}{(\h_2 +\half)_{l_2}(\h_3 +\half)_{l_3}}\left[\frac{\xi(x_1,x_3)}{2}\right]^{\h_3+k_1+2k_3}\left[\frac{\xi(x_2,x_3)}{2}\right]^{-k_1}
\vspace{3mm}
\\
\dps
\times
\left[\frac{\xi(x_1,x_2)}{2}\right]^{\h_2+k_1+2k_2}\,.
\ea
\ee 
Representing the second line here  as the Gauss hypergeometric function at unit argument \eqref{2f1_in_1} and expanding into series yields the second term in \eqref{GGG}:
$$
\ba{c}
\dps
WF^{(2)}(x_1,x_2,x_3) = \frac{\pi^{\half}}{2}\frac{\Gamma(\frac{\h_1+\h_2+\h_3}{2}-\half)}{\Gamma(\h_1)}\sum_{k_1,k_2,k_3=0}^{\infty}\frac{(-)^{k_1+k_2+k_3}}{k_1!k_{2}!k_3!} 
\vspace{2.5mm}  
\\
\dps
\times
\left(\h_3\right)_{k_1+2k_3}\left(\h_2\right)_{k_1+2k_2}\Gamma\left(\frac{\h_1-\h_2-\h_3}{2}-k_1-k_2-k_3\right)
\ea
$$
\be 
\ba{c}
\dps
\times
\sum_{l_1,l_2,l_3=0}^{\infty}\frac{\left(\frac{\h_2+\h_3+\h_1}{2}-\half\right)_{l_1+l_2+l_3}\left(\frac{\h_1-\h_2-\h_3}{2}-k_1-k_2-k_3\right)_{l_1}(-k_2)_{l_2}(-k_3)_{l_3}}{l_1!l_2!l_3!(\h_1 +\half)_{l_1}(\h_2 +\half)_{l_2}(\h_3 +\half)_{l_3}}
\vspace{3mm}
\\
\dps
\times
\left[\frac{\xi(x_1,x_3)}{2}\right]^{\h_3+k_1+2k_3}
\left[\frac{\xi(x_2,x_3)}{2}\right]^{-k_1}
\left[\frac{\xi(x_1,x_2)}{2}\right]^{\h_2+k_1+2k_2}\,.
\ea
\ee 

The  similar relations hold for other terms of the $3$-point Witten integral, $WF^{(3)}(x_1,x_2,x_3)$ and $WF^{(4)}(x_1,x_2,x_3)$, as can be proven by applying the same technique. The resulting relation is given by \eqref{GGG_rel}.

\section{AdS vertex functions in terms of  special functions}
\label{app:special}

Here, we summarize various representations for the lower-point  AdS vertex functions which we managed to express in terms of particular special functions. 
\be 
\label{E1}
\cV_{\h_1 \h_2}(x_1,x_2)= \pref_{\h_1\h_2}\left[\frac{\xi(x_1,x_2)}{2}\right]^{\h_1}\F\left(\frac{\h_1}{2},\frac{\h_1}{2}+\half;\h_1+\half\, \Big|\, \xi(x_1,x_2)^2\right).
\ee 
\be
\label{E2}
\ba{c}
\dps
\cV^{\text{reg}}_{\h_1 \h_2 \h_3}(x_1,x_2,z_3)= \frac{\pref_{\h_1\h_2\h_3} c_1^{\h_1}c_2^{\h_2}}{(2i)^{\h_1+\h_2}}\left[\frac{(z_3-\bar{\pt}_2)(z_3-\bar{\pt}_1)}{\bar{\pt}_2-\bar{\pt}_1}\right]^{-\h_3}
 F_2\left[\begin{array}{ccc}
     \h_1,&\h_2, &-\h_3+\h_1+\h_2  \\
     2\h_1, &2\h_2 \end{array};c_1,c_2\right],
\ea
\ee 
\be
\label{E3}
\ba{l}
\dps
\cV^{\text{reg}}_{\h_1 \h_2 \h_3}(x_1,x_2,z_3)=  \frac{C_{\h_1 \h_2 \h_3}}{(2i)^{\h_1+\h_2}}\left(\frac{(z_3-\bar{\pt}_2)(z_3-\bar{\pt}_1)}{\bar{\pt}_2-\bar{\pt}_1}\right)^{-\h_3}c_1^{\h_1}c_2^{\h_2}
\vspace{3mm}
\\
\dps
\hspace{10mm}\times
\left(1-\frac{c_1}{2}-\frac{c_2}{2}\right)^{\h_3-\h_2-\h_1 } F_4\left[\begin{array}{cc}
 \frac{\h_2+\h_1-\h_3}{2},&\frac{\h_2+\h_1-\h_3}{2}+\half \\
  \h_1+\half, &\h_2 +\half\end{array};\frac{c_1^2}{(2-c_1-c_2)^2},\frac{c_2^2}{(2-c_1-c_2)^2}\right].
\ea
\ee 
\be
\label{E4}
\ba{l}
\dps
\cV_{\h_1 \h_2 \h_3}^{\text{reg}}(x_1,z_2,z_3)=  \pref_{\h_1\h_2\h_3} \, z_{23}^{\h_1-\h_2-\h_3}\chi(x_1,z_2)^{\frac{\h_1+\h_2-\h_3}{2}}\chi(x_1,z_3)^{\frac{\h_1+\h_3-\h_2}{2}}
\vspace{3mm}
\\
\dps
\hspace{40mm}\times{}_2F_1\left(\frac{\h_1+\h_3-\h_2}{2},\frac{\h_1+\h_2-\h_3}{2}; \h_1+\half\,\Big|\, z_{23}^2\,\chi(x_1,z_2)\,\chi(x_1,z_3)\right),
\ea
\ee
Definitions of variables can be found in the main text. Two different forms of $\cV^{\text{reg}}_{\h_1 \h_2 \h_3}(x_1,x_2,z_3)$ in \eqref{E2} and \eqref{E3} can be shown to be related by identical transformations.

\bibliographystyle{JHEP}
\bibliography{refs}

\providecommand{\href}[2]{#2}\begingroup\raggedright\begin{thebibliography}{10}

\bibitem{Achucarro:1987vz}
A.~Achucarro and P.~K. Townsend, {\it {A Chern-Simons Action for
  Three-Dimensional anti-De Sitter Supergravity Theories}},  {\em Phys. Lett.}
  {\bf B180} (1986) 89.

\bibitem{Witten:1988hc}
E.~Witten, {\it {(2+1)-Dimensional Gravity as an Exactly Soluble System}},
  {\em Nucl. Phys. B} {\bf 311} (1988) 46.

\bibitem{Fukuyama:1985gg}
T.~Fukuyama and K.~Kamimura, {\it {Gauge Theory of Two-dimensional Gravity}},
  {\em Phys. Lett. B} {\bf 160} (1985) 259--262.

\bibitem{Penrose}
R.~Penrose, {\em {Angular momentum; an approach to combinatorial space time}}.
\newblock Quantum Theory and Beyond. Cambridge University Press, Cambridge,
  1971.

\bibitem{Moore:1988qv}
G.~W. Moore and N.~Seiberg, {\it {Classical and Quantum Conformal Field
  Theory}},  {\em Commun. Math. Phys.} {\bf 123} (1989) 177.

\bibitem{Ammon:2013hba}
M.~Ammon, A.~Castro, and N.~Iqbal, {\it {Wilson Lines and Entanglement Entropy
  in Higher Spin Gravity}},  {\em JHEP} {\bf 10} (2013) 110,
  [\href{http://arxiv.org/abs/1306.4338}{{\tt arXiv:1306.4338}}].

\bibitem{deBoer:2013vca}
J.~de~Boer and J.~I. Jottar, {\it {Entanglement Entropy and Higher Spin
  Holography in AdS$_3$}},  {\em JHEP} {\bf 04} (2014) 089,
  [\href{http://arxiv.org/abs/1306.4347}{{\tt arXiv:1306.4347}}].

\bibitem{deBoer:2014sna}
J.~de~Boer, A.~Castro, E.~Hijano, J.~I. Jottar, and P.~Kraus, {\it {Higher spin
  entanglement and $ {\mathcal{W}}_{\mathrm{N}} $ conformal blocks}},  {\em
  JHEP} {\bf 07} (2015) 168, [\href{http://arxiv.org/abs/1412.7520}{{\tt
  arXiv:1412.7520}}].

\bibitem{Hegde:2015dqh}
A.~Hegde, P.~Kraus, and E.~Perlmutter, {\it {General Results for Higher Spin
  Wilson Lines and Entanglement in Vasiliev Theory}},  {\em JHEP} {\bf 01}
  (2016) 176, [\href{http://arxiv.org/abs/1511.05555}{{\tt arXiv:1511.05555}}].

\bibitem{Bhatta:2016hpz}
A.~Bhatta, P.~Raman, and N.~V. Suryanarayana, {\it {Holographic Conformal
  Partial Waves as Gravitational Open Wilson Networks}},  {\em JHEP} {\bf 06}
  (2016) 119, [\href{http://arxiv.org/abs/1602.02962}{{\tt arXiv:1602.02962}}].

\bibitem{Besken:2016ooo}
M.~Besken, A.~Hegde, E.~Hijano, and P.~Kraus, {\it {Holographic conformal
  blocks from interacting Wilson lines}},  {\em JHEP} {\bf 08} (2016) 099,
  [\href{http://arxiv.org/abs/1603.07317}{{\tt arXiv:1603.07317}}].

\bibitem{Besken:2017fsj}
M.~Besken, A.~Hegde, and P.~Kraus, {\it {Anomalous dimensions from quantum
  Wilson lines}},  \href{http://arxiv.org/abs/1702.06640}{{\tt
  arXiv:1702.06640}}.

\bibitem{Hikida:2017ehf}
Y.~Hikida and T.~Uetoko, {\it {Correlators in higher-spin AdS$_3$ holography
  from Wilson lines with loop corrections}},  {\em PTEP} {\bf 2017} (2017),
  no.~11 113B03, [\href{http://arxiv.org/abs/1708.08657}{{\tt
  arXiv:1708.08657}}].

\bibitem{Anand:2017dav}
N.~Anand, H.~Chen, A.~L. Fitzpatrick, J.~Kaplan, and D.~Li, {\it {An Exact
  Operator That Knows Its Location}},  {\em JHEP} {\bf 02} (2018) 012,
  [\href{http://arxiv.org/abs/1708.04246}{{\tt arXiv:1708.04246}}].

\bibitem{Hikida:2018eih}
Y.~Hikida and T.~Uetoko, {\it {Superconformal blocks from Wilson lines with
  loop corrections}},  {\em JHEP} {\bf 08} (2018) 101,
  [\href{http://arxiv.org/abs/1806.05836}{{\tt arXiv:1806.05836}}].

\bibitem{Hikida:2018dxe}
Y.~Hikida and T.~Uetoko, {\it {Conformal blocks from Wilson lines with loop
  corrections}},  {\em Phys.\ Rev.\ D} {\bf 97} (2018), no.~8 086014,
  [\href{http://arxiv.org/abs/1801.08549}{{\tt arXiv:1801.08549}}].

\bibitem{Besken:2018zro}
M.~Besken, E.~D'Hoker, A.~Hegde, and P.~Kraus, {\it {Renormalization of
  gravitational Wilson lines}},  {\em JHEP} {\bf 06} (2019) 020,
  [\href{http://arxiv.org/abs/1810.00766}{{\tt arXiv:1810.00766}}].

\bibitem{Bhatta:2018gjb}
A.~Bhatta, P.~Raman, and N.~V. Suryanarayana, {\it {Scalar Blocks as
  Gravitational Wilson Networks}},  {\em JHEP} {\bf 12} (2018) 125,
  [\href{http://arxiv.org/abs/1806.05475}{{\tt arXiv:1806.05475}}].

\bibitem{DHoker:2019clx}
E.~D'Hoker and P.~Kraus, {\it {Gravitational Wilson lines in AdS$_{\bf 3}$}},
  \href{http://arxiv.org/abs/1912.02750}{{\tt arXiv:1912.02750}}.

\bibitem{Castro:2018srf}
A.~Castro, N.~Iqbal, and E.~Llabrés, {\it {Wilson lines and Ishibashi states
  in AdS$_{3}$/CFT$_{2}$}},  {\em JHEP} {\bf 09} (2018) 066,
  [\href{http://arxiv.org/abs/1805.05398}{{\tt arXiv:1805.05398}}].

\bibitem{Kraus:2018zrn}
P.~Kraus, A.~Sivaramakrishnan, and R.~Snively, {\it {Late time Wilson lines}},
  {\em JHEP} {\bf 04} (2019) 026, [\href{http://arxiv.org/abs/1810.01439}{{\tt
  arXiv:1810.01439}}].

\bibitem{Blommaert:2018oro}
A.~Blommaert, T.~G. Mertens, and H.~Verschelde, {\it {The Schwarzian Theory - A
  Wilson Line Perspective}},  {\em JHEP} {\bf 12} (2018) 022,
  [\href{http://arxiv.org/abs/1806.07765}{{\tt arXiv:1806.07765}}].

\bibitem{Hulik:2018dpl}
O.~Hulik, J.~Raeymaekers, and O.~Vasilakis, {\it {Multi-centered higher spin
  solutions from $ {\mathcal{W}}_N $ conformal blocks}},  {\em JHEP} {\bf 11}
  (2018) 101, [\href{http://arxiv.org/abs/1809.01387}{{\tt arXiv:1809.01387}}].

\bibitem{Hung:2018mcn}
L.-Y. Hung, W.~Li, and C.~M. Melby-Thompson, {\it {Wilson line networks in
  $p$-adic AdS/CFT}},  {\em JHEP} {\bf 05} (2019) 118,
  [\href{http://arxiv.org/abs/1812.06059}{{\tt arXiv:1812.06059}}].

\bibitem{Castro:2020smu}
A.~Castro, P.~Sabella-Garnier, and C.~Zukowski, {\it {Gravitational Wilson
  Lines in 3D de Sitter}},  {\em JHEP} {\bf 07} (2020) 202,
  [\href{http://arxiv.org/abs/2001.09998}{{\tt arXiv:2001.09998}}].

\bibitem{Alkalaev:2020yvq}
K.~Alkalaev and V.~Belavin, {\it {More on Wilson toroidal networks and torus
  blocks}},  {\em JHEP} {\bf 11} (2020) 121,
  [\href{http://arxiv.org/abs/2007.10494}{{\tt arXiv:2007.10494}}].

\bibitem{Belavin:2022bib}
V.~Belavin and J.~R. Cabezas, {\it {Wilson lines construction of osp(1$|$2)
  conformal blocks}},  {\em Nucl. Phys. B} {\bf 985} (2022) 115981,
  [\href{http://arxiv.org/abs/2204.12149}{{\tt arXiv:2204.12149}}].

\bibitem{Belavin:2023orw}
V.~Belavin, P.~Oreglia, and J.~R. Cabezas, {\it {Wilson lines construction of
  sl3 toroidal conformal blocks}},  {\em Nucl. Phys. B} {\bf 990} (2023)
  116186, [\href{http://arxiv.org/abs/2301.04575}{{\tt arXiv:2301.04575}}].

\bibitem{Castro:2023bvo}
A.~Castro, I.~Coman, J.~R. Fliss, and C.~Zukowski, {\it {Coupling Fields to 3D
  Quantum Gravity via Chern-Simons Theory}},  {\em Phys. Rev. Lett.} {\bf 131}
  (2023), no.~17 171602, [\href{http://arxiv.org/abs/2304.02668}{{\tt
  arXiv:2304.02668}}].

\bibitem{Alkalaev:2023axo}
K.~Alkalaev, A.~Kanoda, and V.~Khiteev, {\it {Wilson networks in AdS and global
  conformal blocks}},  {\em Nucl. Phys. B} {\bf 998} (2024) 116413,
  [\href{http://arxiv.org/abs/2307.08395}{{\tt arXiv:2307.08395}}].

\bibitem{Hamilton:2005ju}
A.~Hamilton, D.~N. Kabat, G.~Lifschytz, and D.~A. Lowe, {\it {Local bulk
  operators in AdS/CFT: A Boundary view of horizons and locality}},  {\em Phys.
  Rev. D} {\bf 73} (2006) 086003,
  [\href{http://arxiv.org/abs/hep-th/0506118}{{\tt hep-th/0506118}}].

\bibitem{Alkalaev:2015fbw}
K.~B. Alkalaev and V.~A. Belavin, {\it {From global to heavy-light: 5-point
  conformal blocks}},  {\em JHEP} {\bf 03} (2016) 184,
  [\href{http://arxiv.org/abs/1512.07627}{{\tt arXiv:1512.07627}}].

\bibitem{Rosenhaus:2018zqn}
V.~Rosenhaus, {\it {Multipoint Conformal Blocks in the Comb Channel}},  {\em
  JHEP} {\bf 02} (2019) 142, [\href{http://arxiv.org/abs/1810.03244}{{\tt
  arXiv:1810.03244}}].

\bibitem{Fortin:2020zxw}
J.-F. Fortin, W.-J. Ma, and W.~Skiba, {\it {All Global One- and Two-Dimensional
  Higher-Point Conformal Blocks}},  \href{http://arxiv.org/abs/2009.07674}{{\tt
  arXiv:2009.07674}}.

\bibitem{Fortin:2023xqq}
J.-F. Fortin, W.-J. Ma, S.~Parikh, L.~Quintavalle, and W.~Skiba, {\it {One- and
  two-dimensional higher-point conformal blocks as free-particle wavefunctions
  in $ {\textrm{AdS}}_3^{\otimes m} $}},  {\em JHEP} {\bf 01} (2024) 031,
  [\href{http://arxiv.org/abs/2310.08632}{{\tt arXiv:2310.08632}}].

\bibitem{Fitzpatrick:2016mtp}
A.~L. Fitzpatrick, J.~Kaplan, D.~Li, and J.~Wang, {\it {Exact Virasoro Blocks
  from Wilson Lines and Background-Independent Operators}},  {\em JHEP} {\bf
  07} (2017) 092, [\href{http://arxiv.org/abs/1612.06385}{{\tt
  arXiv:1612.06385}}].

\bibitem{Nakayama:2015mva}
Y.~Nakayama and H.~Ooguri, {\it {Bulk Locality and Boundary Creating
  Operators}},  {\em JHEP} {\bf 10} (2015) 114,
  [\href{http://arxiv.org/abs/1507.04130}{{\tt arXiv:1507.04130}}].

\bibitem{Nakayama:2016xvw}
Y.~Nakayama and H.~Ooguri, {\it {Bulk Local States and Crosscaps in Holographic
  CFT}},  {\em JHEP} {\bf 10} (2016) 085,
  [\href{http://arxiv.org/abs/1605.00334}{{\tt arXiv:1605.00334}}].

\bibitem{Jepsen:2019svc}
C.~B. Jepsen and S.~Parikh, {\it {Propagator identities, holographic conformal
  blocks, and higher-point AdS diagrams}},  {\em JHEP} {\bf 10} (2019) 268,
  [\href{http://arxiv.org/abs/1906.08405}{{\tt arXiv:1906.08405}}].

\bibitem{Zhou:2018sfz}
X.~Zhou, {\it {Recursion Relations in Witten Diagrams and Conformal Partial
  Waves}},  {\em JHEP} {\bf 05} (2019) 006,
  [\href{http://arxiv.org/abs/1812.01006}{{\tt arXiv:1812.01006}}].

\bibitem{Giombi:2020xah}
S.~Giombi, H.~Khanchandani, and X.~Zhou, {\it {Aspects of CFTs on Real
  Projective Space}},  {\em J. Phys. A} {\bf 54} (2021), no.~2 024003,
  [\href{http://arxiv.org/abs/2009.03290}{{\tt arXiv:2009.03290}}].

\bibitem{DHoker:1999mqo}
E.~D'Hoker, D.~Z. Freedman, and L.~Rastelli, {\it {AdS / CFT four point
  functions: How to succeed at z integrals without really trying}},  {\em Nucl.
  Phys. B} {\bf 562} (1999) 395--411,
  [\href{http://arxiv.org/abs/hep-th/9905049}{{\tt hep-th/9905049}}].

\bibitem{Hijano:2015zsa}
E.~Hijano, P.~Kraus, E.~Perlmutter, and R.~Snively, {\it {Witten Diagrams
  Revisited: The AdS Geometry of Conformal Blocks}},  {\em JHEP} {\bf 01}
  (2016) 146, [\href{http://arxiv.org/abs/1508.00501}{{\tt arXiv:1508.00501}}].

\bibitem{Chamseddine:1989wn}
A.~H. Chamseddine and D.~Wyler, {\it {Topological Gravity in
  (1+1)-dimensions}},  {\em Nucl. Phys. B} {\bf 340} (1990) 595--616.

\bibitem{Isler:1989hq}
K.~Isler and C.~A. Trugenberger, {\it {A Gauge Theory of Two-dimensional
  Quantum Gravity}},  {\em Phys. Rev. Lett.} {\bf 63} (1989) 834.

\bibitem{Teitelboim:1983ux}
C.~Teitelboim, {\it {Gravitation and Hamiltonian Structure in Two Space-Time
  Dimensions}},  {\em Phys. Lett. B} {\bf 126} (1983) 41--45.

\bibitem{Jackiw:1984je}
R.~Jackiw, {\it {Lower Dimensional Gravity}},  {\em Nucl. Phys. B} {\bf 252}
  (1985) 343--356.

\bibitem{HOLMAN19661}
W.~J. Holman and L.~C. Biedenharn, {\it Complex angular momenta and the groups
  su(1, 1) and su(2)},  {\em Annals of Physics} {\bf 39} (1966), no.~1 1 -- 42.

\bibitem{Banados:1994tn}
M.~Banados, {\it {Global charges in Chern-Simons field theory and the (2+1)
  black hole}},  {\em Phys. Rev. D} {\bf 52} (1996) 5816--5825,
  [\href{http://arxiv.org/abs/hep-th/9405171}{{\tt hep-th/9405171}}].

\bibitem{Ishibashi:1988kg}
N.~Ishibashi, {\it {The Boundary and Crosscap States in Conformal Field
  Theories}},  {\em Mod. Phys. Lett. A} {\bf 4} (1989) 251.

\bibitem{10.1007/BF02392525}
S.~Saran, {\it {Transformations of certain hypergeometric functions of three
  variables}},  {\em Acta Mathematica} {\bf 93} (1955), no.~none 293 -- 312.

\bibitem{Witten:1998qj}
E.~Witten, {\it Anti-de {S}itter space and holography},  {\em Adv. Theor. Math.
  Phys.} {\bf 2} (1998) 253--291,
  [\href{http://arxiv.org/abs/hep-th/9802150}{{\tt hep-th/9802150}}].

\bibitem{Kabat:2012av}
D.~Kabat and G.~Lifschytz, {\it {CFT representation of interacting bulk gauge
  fields in AdS}},  {\em Phys. Rev. D} {\bf 87} (2013), no.~8 086004,
  [\href{http://arxiv.org/abs/1212.3788}{{\tt arXiv:1212.3788}}].

\bibitem{Castro:2024cmf}
A.~Castro and P.~J. Martinez, {\it {Revisiting Extremal Couplings in AdS/CFT}},
   \href{http://arxiv.org/abs/2409.15410}{{\tt arXiv:2409.15410}}.

\bibitem{Nishida:2016vds}
M.~Nishida and K.~Tamaoka, {\it {Geodesic Witten diagrams with an external
  spinning field}},  {\em PTEP} {\bf 2017} (2017), no.~5 053B06,
  [\href{http://arxiv.org/abs/1609.04563}{{\tt arXiv:1609.04563}}].

\bibitem{Dyer:2017zef}
E.~Dyer, D.~Z. Freedman, and J.~Sully, {\it {Spinning Geodesic Witten
  Diagrams}},  {\em JHEP} {\bf 11} (2017) 060,
  [\href{http://arxiv.org/abs/1702.06139}{{\tt arXiv:1702.06139}}].

\bibitem{Sleight:2017fpc}
C.~Sleight and M.~Taronna, {\it {Spinning Witten Diagrams}},  {\em JHEP} {\bf
  06} (2017) 100, [\href{http://arxiv.org/abs/1702.08619}{{\tt
  arXiv:1702.08619}}].

\bibitem{Blencowe:1988gj}
M.~P. Blencowe, {\it {A Consistent Interacting Massless Higher Spin Field
  Theory in $D$ = (2+1)}},  {\em Class. Quant. Grav.} {\bf 6} (1989) 443.

\bibitem{Bateman:100233}
H.~Bateman and A.~Erdélyi, {\em {Higher transcendental functions}}.
\newblock California Institute of technology. Bateman Manuscript project.
  McGraw-Hill, New York, NY, 1953.

\bibitem{Bailey:1938}
W.~N. Bailey, {\it The generating function of jacobi polynomials},  {\em
  Journal of the London Mathematical Society} {\bf s1-13} (1938), no.~1 8--12,
  [\href{http://arxiv.org/abs/https://londmathsoc.onlinelibrary.wiley.com/doi/pdf/10.1112/jlms/s1-13.1.8}{{\tt
  https://londmathsoc.onlinelibrary.wiley.com/doi/pdf/10.1112/jlms/s1-13.1.8}}].

\bibitem{CARLSON1963452}
B.~Carlson, {\it Lauricella's hypergeometric function fd},  {\em Journal of
  Mathematical Analysis and Applications} {\bf 7} (1963), no.~3 452--470.

\bibitem{Fronsdal:1974ew}
C.~Fronsdal, {\it {Elementary particles in a curved space. ii}},  {\em Phys.
  Rev. D} {\bf 10} (1974) 589--598.

\bibitem{Ramallo:2013bua}
A.~V. Ramallo, {\it {Introduction to the AdS/CFT correspondence}},  {\em
  Springer Proc. Phys.} {\bf 161} (2015) 411--474,
  [\href{http://arxiv.org/abs/1310.4319}{{\tt arXiv:1310.4319}}].

\bibitem{vretblad2006fourier}
A.~Vretblad, {\em Fourier Analysis and Its Applications}.
\newblock Graduate Texts in Mathematics. Springer New York, 2006.

\end{thebibliography}\endgroup

\end{document}